\documentclass[a4]{article}
\usepackage[a4paper]{geometry}
\usepackage{cite}
\usepackage[pdftex]{graphicx}
\usepackage[cmex10]{amsmath}

\usepackage{times}
\usepackage{array}
\usepackage[tight,footnotesize]{subfigure}
\usepackage[font=footnotesize]{subfig}
\usepackage{graphicx}
\usepackage{mathtools}
\usepackage{amsmath}
\usepackage{enumitem}
\def\union{{\cup}}
\def\intersect{{\cap}}

\usepackage{alltt}

\def\boxx{{\vcenter{\vbox{\hrule height.3pt
          \hbox{\vrule width.3pt height6pt
          \kern6pt\vrule width.3pt}\hrule height.3pt}}\;}}

\def\impos{{\;\vcenter{\hbox{\rule{5mm}{0.2mm}}} \vcenter{\hbox{\rule{1.5mm}{1.5mm}}} \;}}

\def\lrarrow{\leftrightarrow \kern-8pt \rightarrow}

\def\Unspec{A_{\rm ?}}

\def\2{\frac{1}{2}}

\def\adj{{\rm adj}}    

\def\beq{\begin{eqnarray}}
\def\eeq{\end{eqnarray}}
\def\2{\frac{1}{2}}

\newtheorem{assumption}{Assumption}

\newtheorem{example}{Example}
\newtheorem{lemma}{Lemma}

\newtheorem{definition}{Definition}
\newtheorem{law}{Law}

\newtheorem{hype}{Hypothesis}

\def\valence{{\bf Valence}}
\def\Coarse{{\bf G}}
\def\dim{{\rm dim}}

\def\lrarrow{\leftrightarrow \kern-8pt \rightarrow}

\def\frightarrow{\rightarrow \kern-11pt /~~}
\def\reducesto{\simeq \kern -3pt >}
\def\intersection{\cap}

\def\host{{\rm Host}}
\def\tenant{{\rm Tenant}}

\def\occupier{{\rm Occupier}}

\begin{document}
\newcommand{\strust}[1]{\stackrel{\tau:#1}{\longrightarrow}}
\newcommand{\trust}[1]{\stackrel{#1}{{\rm\bf ~Trusts~}}}
\newcommand{\promise}[1]{\xrightarrow{#1}}
\newcommand{\revpromise}[1]{\xleftarrow{#1} }
\newcommand{\assoc}[1]{{\xrightharpoondown{#1}} }
\newcommand{\imposition}[1]{\stackrel{#1}{\impos}}
\newcommand{\scopepromise}[2]{\xrightarrow[#2]{#1}}
\newcommand{\handshake}[1]{\xleftrightarrow{#1} \kern-8pt \xrightarrow{} }
\newcommand{\cpromise}[1]{\stackrel{#1}{\frightarrow}}
\newcommand{\policy}{\stackrel{P}{\equiv}}
\newcommand{\field}[1]{\mathbf{#1}}
\newcommand{\bundle}[1]{\stackrel{#1}{\Longrightarrow}}

\title{Spacetimes with Semantics (II)\\\small Scaling of agency, semantics, and tenancy\\~\\\Large Notes (draft v1.2)}

\author{Mark Burgess}

\maketitle

\begin{abstract}
  Using Promise Theory as a calculus, I review how to define agency in
  a scalable way, for the purpose of understanding semantic
  spacetimes. By following simple scaling rules, replacing individual
  agents with `super-agents' (sub-spaces), it is shown how agency can
  be scaled both dynamically and semantically.

  The notion of occupancy and tenancy, or how space is used and filled
  in different ways, is also defined, showing how spacetime can be
  shared between independent parties, both by remote association and
  local encapsulation. I describe how to build up dynamic and semantic
  continuity, by joining discrete individual atoms and molecules of
  space into quasi-continuous lattices.
\end{abstract}

\tableofcontents


\section{Introduction} 

This is the second part of a series of papers on semantic spacetimes,
exploring the idea of {\em scaling} of the structure and tenancy of
space, i.e.  describing how properties (i.e. information) reside
within a space and occupy it.  These notes contain technical ideas
to show an approach unifying semantics with dynamics, for later
refinement.

The laying out of patterns in space is the basis for the encoding of
information, as well as of functional behaviours in systems.
Semantics add strong constraints to the nature of spacetime that
dynamics alone cannot provide in a natural way, e.g. how do we explain
intentional inhomogeneity is space? Does topology have meaning?  Is
there a smallest grain size in spacetime?  The role of functional
asymmetry is another of the key features of semantically rich systems
that requires a simple explanation.

Examples of scale and
tenancy in spacetimes with semantics include everything from the
mundane, where people occupy housing, offices, or hotels rooms, to
computer programs that live in `the cloud', where data live on a
network, and how we fill storage resources, or search for items in
storage warehouses.  The structure of spacetime is key to answering
all of these issues.

The theoretical calculus for this paper is promise
theory\cite{spacetime1, promisebook}.  Elements of spacetime are
described by autonomous agents that cooperate through the keeping of
promises. Promises describe observer semantics and shared dynamics of
agents in a unified way. The scope of promise information is local.

\section{Properties of spacetime agency}

The properties of spacetime refer both to the notions of location,
adjacency, and to the possibly inhomogeneous properties exhibited by
both.  In paper I\cite{spacetime1}, it was seen that one could
construct a model of spacetime in which the elements of space were
{\em agents}, or units of agency, representing locations, and the
topology or connectivity was made through promises, as well as the
functional semantics. The purpose of this construction was to
attribute a necessary but sufficient degree of capability to space
with the aim of understanding and utilizing it
operationally\footnote{I am grateful to Susan Bennet for making this
  crisp observation.}. What is elegant about this is the small number
of concepts required in order to build an entire picture of functional
spaces. A semantic spacetime, then, is more or less simply:
\begin{itemize}
\item A collection of autonomous agents, usually denoted $A_i$.
\item A set of promises made between these agents, usually denoted $\pi$ (or as a matrix $\Pi$).
\end{itemize}
This is the starting point.

\subsection{Agency by attachment, encapsulation, and association}

Agency is that property which ascribes intentionality and semantics to
an object.  In the natural sciences, the idea that space and time
could have semantics has been skirted and avoided. Indeed, one
normally suppresses this issue altogether, by making spacetime a
neutral backdrop to what goes on\footnote{The exception here is
  Einstein's theory of general relativity, which ascribes semantics of
  gravitational force to spacetime curvature, or vertex functions in
  the Regge calculus.}. However, once one embraces the notion that
space and time can have semantics, it is possible to describe all
kinds of structures from cell biology to databases as special cases of
a generalized idea of semantic spacetimes, while embracing concepts of
adjacency, regularity, symmetry, and relativity, normally associated with their
physics.

\begin{figure}[ht]
\begin{center}
\includegraphics[width=7.5cm]{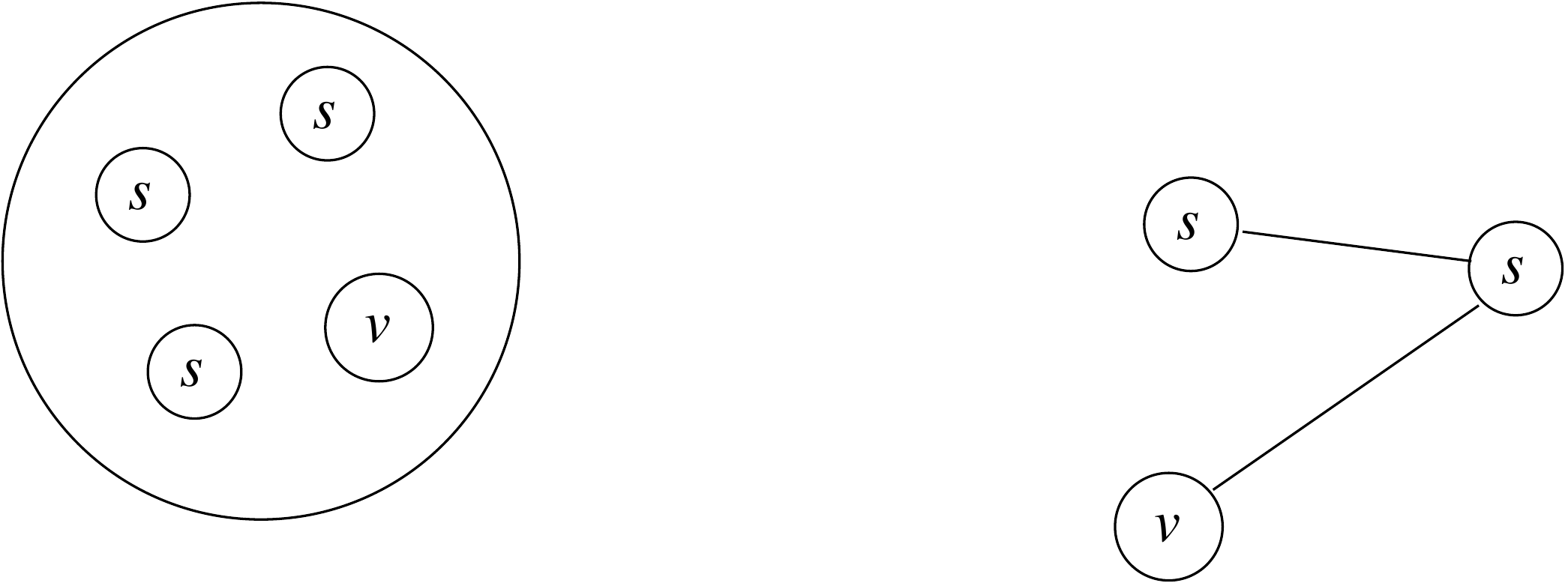}
\caption{\small Local versus non-local scaling of agency, by encapsulation and by
association.\label{one}}
\end{center}
\end{figure}

Agency, i.e. intentionality and meaning, may be attributed to any kind
of process, smart or dumb, by intelligent, thinking observers, but this
does not make its perception less important, or less real. 
Abstraction informs us that there are two ways by which agency can be wielded:
\begin{itemize}
\item Locally, by encapsulation of intentions within a spacetime element (what we call autonomy).
\item Non-locally (remotely), by association of a spacetime element with a source of agency which is
elsewhere (what we call subordination).
\end{itemize}
These two modes will be studied in the remainder of these notes in the guise of the more prosaic
terminology:
\begin{itemize}
\item The scaling of local agency (by encapsulation, absorption, clique and membership, etc).
\item The scaling of remote agency (intentionality at a distance, by attachment, correlation, entanglement, etc).
\end{itemize}
These two notions are complementary, and yet they also become intertwined as we scale up.

\subsection{Recap of the propagation model, and observation}

\begin{figure}[ht]
\begin{center}
\includegraphics[width=7.5cm]{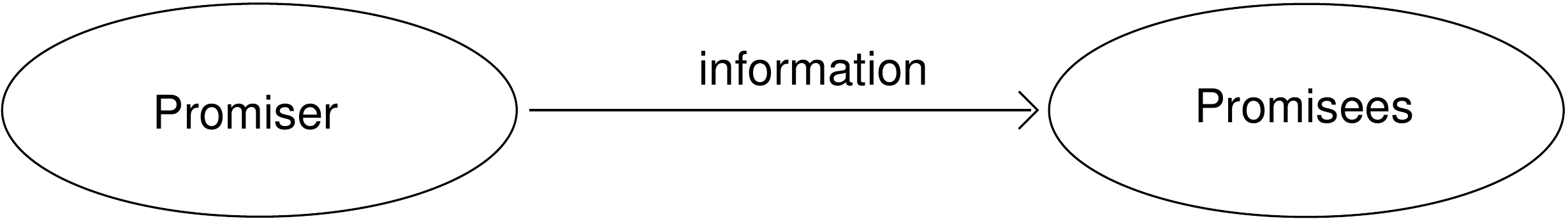}\\
\includegraphics[width=4.5cm]{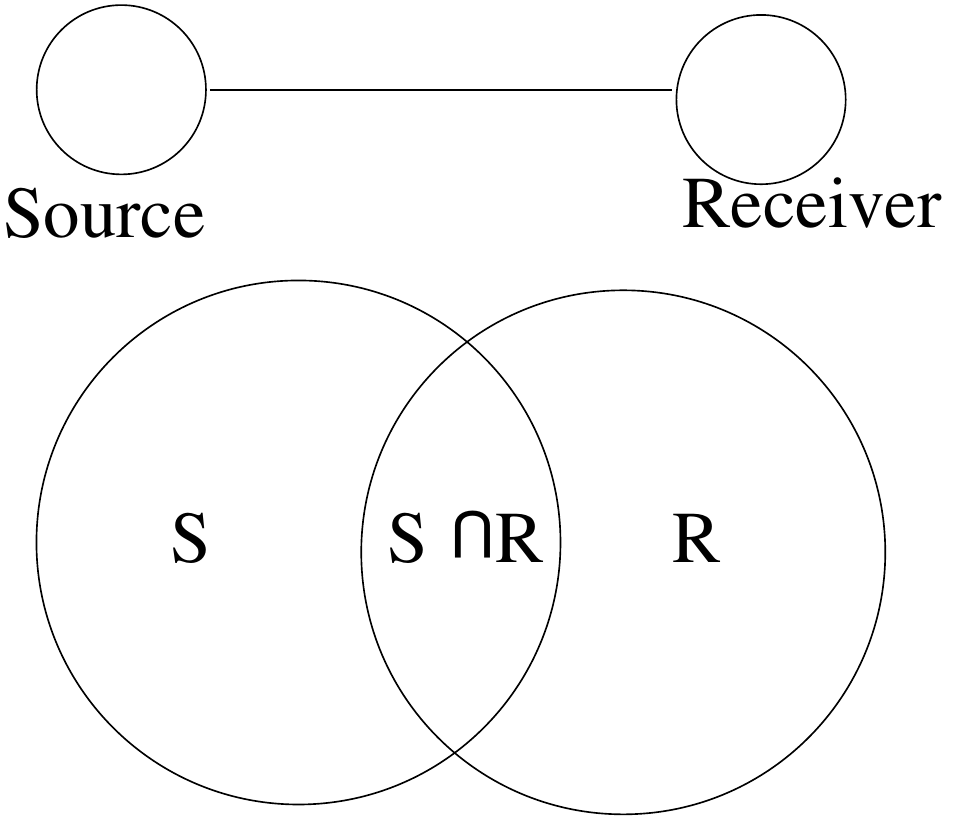}
\caption{\small Shannon's basic source-sink view of information transmission forms
the basis of a promise-promisee interaction, and all transmission of influence
between locations. This model encapsulates the way in which observation
breaks spacetime symmetries.\label{shannon}}
\end{center}
\end{figure}

The basic model underlying promise theory is one of transmission of
influence, through signalling of intent and by measurement, that
guides causation.  There is a basic asymmetry already encoded into
this elementary intentional process: the asymmetry of `have' and `have not'.
For influence to be transmitted from an agent $A$ to an agent $A'$, it
requires a promise to be communicated by $A$, and accepted by $A'$:
\beq
A &\promise{+b}& A'\\
A' &\promise{-b}& A
\eeq
How we describe causation within a promise model
depends on how we understand the agencies involved.  However,
Shannon's basic model of transmitter and receiver has to be at the
core of all other considerations\cite{shannon1} (see fig
\ref{shannon}). It enshrines the way in which source and observer
break the translational symmetries of the system in order to make a
scale for measurements.

The recipient of the source's communicated intent (the semantics),
might not be the same as the recipients of the impact of the intent (dynamics).
Thus transmitted information has these two aspects, which must be understood together.
This is why promise theory is a harder set of constraints to satisfy
than a theory in which we take the semantics as trivial.
Causation is transferred through {\em vector promises} (see section
\ref{tensor}), especially so-called `use-promises'. These, in turn,
may be interpreted as special cases of the following general
semantics:
\begin{enumerate}
\item $A_1$ can {\em influence} $A_2$ (change/causation)
\item $A_1$ is {\em connected to} $A_2$ (topology)
\item $A_1$ is {\em part of} $A_2$ (containment)
\end{enumerate}
\begin{example}
Examples include the following quasi-transitive and quasi-causative relations: 
\beq
A_1 &\promise{\rm Instructions}& A_2\\
A_1 &\promise{\rm Precedes/Follows}& A_2\\
A_1 &\promise{\rm Affects}& A_2\\
A_1 &\promise{\rm Is~a~special~case~of}& A_2\\
A_1 &\promise{\rm Generalizes}& A_2\\
A_1 &\promise{\rm Neighbour}& A_2 
\eeq 
\end{example}
These examples lead to what we
call quasi-transitivity, which in turn allows the propagation of
intent\footnote{This is similar to the way in which the equations of motion of a dynamical
variable allow the propagation of state from an initial configuration
to a final one, over a `free' interval.  One expects to derive other
kinds of measurements from this basic evolutionary structure.}.

\subsection{The tensor structure of a promise body}\label{tensor}

The spatial relationships encoded by promises do not necessarily
follow the graph of promiser-promisee bindings. It is useful to have a
language by which to discuss these.  Tensors are well-known structures
that interrelate constellations of points in a space.  Promises bind
agents (the elements in a semantic space) in two ways:
\begin{itemize}
\item Binding together agents that receive information about intent (scope, promisee).
\item Binding agents who must cooperate in keeping the promise (promiser and body)
\end{itemize}
We write a promise from $A$ to $A'$ in the form 
\beq A
\scopepromise{b(A_1,A_2,\ldots)}{\sigma} A', 
\eeq 
where $b$ is the body of the
promise, possibly depending on specific agencies, 
and $\sigma$ is a collection of agents known as the scope of
the promise.  

\begin{figure}[ht]
\begin{center}
\includegraphics[width=9.5cm]{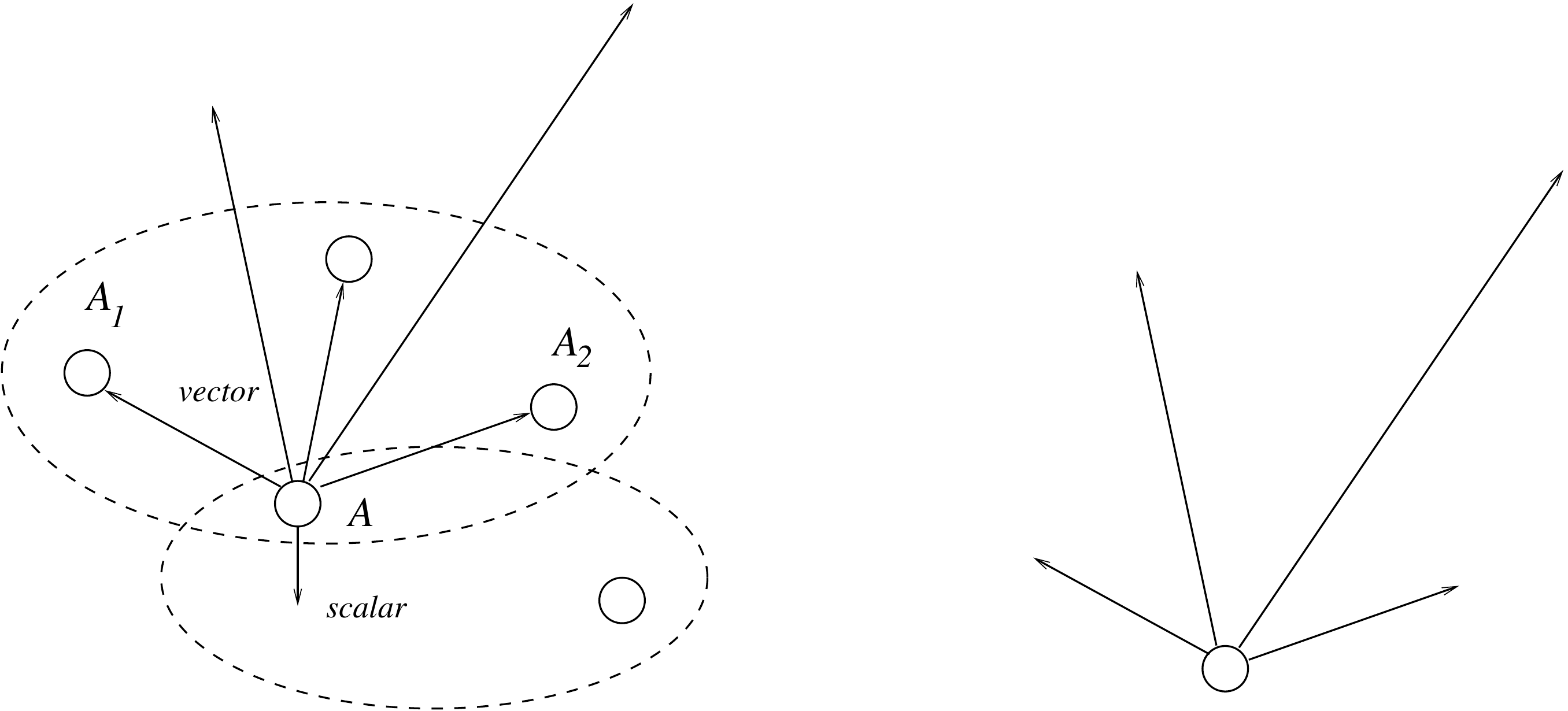}
\caption{\small The promise arrows emanating from a point (agent) in space
mark out a spanning basis for influence, while the scope regions
are aware of the promise, including those for whom the promise is intended. Scalar promises
can pass simple semaphores to their scope, while vector promises
entangle agents in an exchange with a specific location.\label{span}}
\end{center}
\end{figure}

If no other agent is involved in keeping a promise, i.e. no agent appears
mentioned in the promise body $b$, then we call it a
scalar promise.  If a single agent is involved in defining the
promise, we call it a vector promise.  Promises of higher dependence
can also exist, but might be decomposable into multiple vector
promises. There is thus a notion of promise {\em rank}, in the same way as for
tensors (see paper I, section 3.3.4).  This distinction between scalar
and vector promises becomes increasingly important as we apply
promises to tenancy, so let's begin by examining this in more detail (see fig \ref{span}).

\begin{lemma}[Use promises acquire tensor rank 1 implicitly]
A use-promise such as
\beq
\pi: S \promise{-b} \{ R\} ~ {\rm also\,written} ~ S \promise{U(b)} \{R\}
\eeq
is a special case in which there is an implicit reference to the
acceptance of a promise in a binding initiated by a remote promiser.
Hence the promisees $\{R\}$ are implicitly involved parties in the promise
body. Since all promises are separable into promises between
pairs of agents, the rank of a use-promise
\beq
R_i \promise{+b} S,
\eeq 
is always 1 (a vector).
\end{lemma}

A promise, {\em conditional} on a promise from an external agent,
is also a special case, in which the promise body refers to a
promise quenched by an external agent through the dependency. This
does not refer to the agent quenching the promise, however. The
conditional promise law (see \cite{promisebook}) implies it will be
accompanied by a use-promise, however, which will lead to a vector
relationship. Thus conditional promises lead implicitly to vector
relationships also.

\begin{definition}[Tensor promises]
Promise rank is determined by the number of agencies related by a promise.
\begin{enumerate}
\setcounter{enumi}{-1}
\item A scalar promise refers to no other agency in its body, i.e. expresses a property of just the promiser.
\item A vector promise refers to one other agency in its body, i.e. a property in relation to one other independent agent.
\item A rank $n$ tensor promise, at point $A$, refers to $n$ other 
independent agencies in its body. In all cases, the number of
promisees in undefined.
\end{enumerate}
\end{definition}

\begin{example}
A scalar promise might be: I promise to brush my teeth every day.
A vector promise might be: I promise to read a message from agent $X$.
A 2-tensor promise might be: I promise to relay information from $X$ to $Y$.
An $n$-tensor promise might be: I promise to select give my money to one of
$A_1, \ldots, A_n$ that has the cheapest goods.
\end{example}
A promise body has the form: 
\beq 
b = b(A_1,\ldots, A_n) 
\eeq 
A set of vector promises that correlates the promiser with the points
$A_n$ appearing in the body form a matroid, or a vector basis, for a
local region of the space by choosing to make the adjacent agencies
$A_1, A_2, \ldots$ (see figure \ref{span}). Intuitively, the set of
paths radiating out from a point to promisees, describe its adjacencies (or
virtual adjacencies), and represents the possible different directions by
which information can be passed to neighbouring agents.
Those directions, which are local (adjacent) to the promiser, acquire the role of a
spanning set of component directions, i.e. promise vectors at the
promiser's location. If the promise body refers to any remote
agencies, i.e. which are not a direct neighbour, one may
resolve `virtual adjacencies' (such as tunnels or overlaid spatial
structures) in terms of the the actual vectors of this local set.

In a general promise graph, each agent location has no knowledge of
other agents, i.e. it has only local information. This is analogous to
the inhomogeneous constraints that explain most influential fields of
force in fundamental physics.  We thus rediscover a natural
representation of {\em gauge theories} in promises, where scalar and vector properties
depend on location, due to the
inhomogeneous symmetry properties\footnote{In the original paper\cite{burgessdsom2005}, I suggested
  that the appearance of other agent identities in promise bodies was
  undesirable as it let to potential inconsistency. This was a
  potential issue when dealing with non-scalar promises. However, this
  was before the role of impositions was properly understood in
  Promise Theory.  The inconsistencies arise when trying to make
  promises on behalf of non-local agencies.}.

\subsection{Completeness of promise information}

The bindings between promiser and promisee relate to the two aspects of behaviour:
\begin{itemize}
\item {\em Dynamics}: (change and measure) agencies mentioned in the promise body involved in the keeping of the promise.
\item {\em Semantics}: (purpose) the promisees to whom intent is communicated.
\end{itemize}
Promisees are the intended recipients of the outcome, but there is
also the scope of the promise by which agents receive the pure
kinematic or dynamic information content about the promiser's intent.
Influence can thus also occur between pairs of agents via indirection.

\begin{example}
  Intent and execution can be communicated separately, with different
  consequences.Imagine promising to follow someone without informing
  them of that promise. Then the promise body would refer to the
  person, but the promisee would not include the person.

The agent being followed might still observe a correlation with the
promiser, without understanding the information of the promise to
signal the possible origin of the behaviour, and could
even speculate about the existence of a promise, as emergent
behaviour, but still has formally incomplete information.
\end{example}
We may define the completeness of information to another agent by saying that
\begin{definition}[Promise with complete information]
In a tensor of rank $n$, made by agent $A$, referring either
explicitly or implicitly to $n$ other agents, the promise provides
complete information iff all $n$ agents referred to in the promise body are in the scope
of the promise, i.e. are promisees or informees of $A$.
\end{definition}
A promise with complete information thus makes other involved agents aware
of promises made about them.

\subsection{Defining promise matrices}

It is useful to define two matrices:

\begin{definition}[Promise matrix]
Let $i,j$ label a collection of $n$ agents, where $i=1,\ldots, n$.
The promise matrix $\Pi_{ij}$ is the matrix of all promises between
$A_i$, $A_j$, stripped of the agent labels, i.e. in which the agents are implicit. Hence, the
complete set of promises between the agents may be written:
\beq
\bigcup_{i,j=1}^n \; A_i\,\pi_{ij}\,A_j = \sum_{i,j=1}^n \; A_i\,\pi_{ij}\,A_j, 
\eeq
Note: this notation defines the meaning of $\sum$ and $+$ for the rest
of the paper.
\end{definition}

\begin{definition}[Promise adjacency matrix]
The directed graph adjacency matrix which records a link if there is a promise
of any type between the labelled agents.
\beq
\Pi_{ij} = \left.
\begin{array}{cc}
1& ~ {\rm iff}~ A_i \promise{b_{ij}} A_j, \\
0&
\end{array}\right\}~~~~\forall b_{ij} \not=\emptyset
\eeq
\end{definition}

These may be further decomposed into useful subsets, for instance, the rank decomposition
of the promise matrix, into matrices of promises of rank $r$, will be useful:
\beq
\Pi_{ij} = \sum_{r=0} \; \Pi^{(r)}_{ij}.
\eeq
The concept of an agent's interior and exterior promises will also be defined below.

\subsection{Differentiating spatial role: agent types and labels}

We may benefit from differentiating {\em types} of agents,
when applying promise theory to functional spaces.  A priori, agents
have no type: they are homogeneous, structureless, universal elements
(analogous to biological `stem cells') that may only be differentiated
via the promises they make. A {\em type} may thus be defined either by
identification of a role\cite{promisebook}, or by explicitly making a
scalar promise.

Consider a universal (typeless) agent $A_\emptyset$, that makes no initial promises, or empty promises
to everyone: 
\beq
A_\emptyset \promise{\emptyset} *.
\eeq
We may add a scalar promise, with additional scope $\sigma$:
\beq
A_\emptyset \scopepromise{+\rm bla}{\sigma} *
\eeq
Any agent in the scope $\{*,\sigma\}$ may now identify the former agent as being equivalent
to an agent of type $b$, making no promise. In other words, within the scope, the agent effectively
has a new name:
\beq
A_{\rm bla} \scopepromise{+\emptyset}{\sigma} *,
\eeq
thence
\beq
{\rm bla} \scopepromise{+\emptyset}{\sigma} *.
\eeq
i.e., we can drop the promiser's agent symbol `$A$' and label agents simply by their names.
I'll use this notion from here on to write agent types implicitly, e.g.
\beq
T_1 \promise{\emptyset} A &\equiv& A_1 \promise{+T} A\\
R_2 \promise{\emptyset} A &\equiv& A_2 \promise{+R} A\\
H_3 \promise{\emptyset} A &\equiv& A_3 \promise{+H} A
\eeq
and so on.
In this way, we can move scalar labels from the promise body to the 
agent's identifier at will. This is in keeping with the idea that
the agent's name is the basic promise that identifies it.

\subsection{Promise valency and saturation}

To develop the formal chemistry of intent, we need to clarify how many
agents a promise can support. In other words, how many `slots',
`binding sites' or occupyable resources does an agent have, to share
between promisees?

Declaring a finite number of such slots, explicitly allows for a
simple discussion of resource exclusivity around promises.  The
concept is basically analogous to the valences (oxidation numbers) of
electrons in physical chemistry.  Think also of the binding sites for
receptors, viruses and major histocompatability proteins in biology.

\begin{definition}[Valence of an agent promise, and overcommitting]
  A promise which provides $+b$ to a number of other agents may
  specify how many agents $n$ for which the promise will be kept exclusively. 
The valency number of an exclusive promise is a
  positive integer $n$, written
\beq
A& \promise{+b\#n}& \{ A_1, \ldots A_p \}.
\eeq
A promise body may be called over-promised (or over-committed) if $p > n$.
\end{definition}
\begin{example}
A reserved parking area promises 10 spaces, to 20 employees. The parking promise
is over-committed, since it cannot keep all of its promises simultaneously.
\end{example}
Over-promising is not a problem unless all of the promisees accept the promise,
and promise to use it.
Thus a separate concept of saturation arises by using up all of the
valence slots:
\begin{definition}[Use-promise saturation]
Suppose we have
\beq
A& \promise{+b\#n}& \{ A_1, \ldots A_p \}\\
\{ A_1, \ldots A_m \} &\promise{-b\#m} &A
\eeq
is saturated if $m \ge n$.
\end{definition}
It is useful to define a function whose value is the net valence of
a particular type of promise body.
\begin{definition}[Net valence of a promise graph and utilization]
 $\pm b$, for a collection of agents $\{ A_i\}$:
\beq
\valence( b; \{A_i\}) &=& \sum_i \valence( b; A_i) \\
&=& n - m
\eeq
Hence we may assign an integer value to the level of usage,
or a rational fraction $m/n$ for utilization of the resource.
If this fraction exceeds unity, or the net valency is negative, the keeping of the promise
effectively becomes a queue of length $|m-n|$, requiring the agent to
multiplex its resources in time to keep its promise.
\end{definition}
\begin{example}
Consider the two agents $A_1, A_2$:
\beq
A_1 &\promise{+b\#2}& A_2\\
A_2 &\promise{-b\#3}& A_1
\eeq
$A_1$ offers two possible slots for its promise of $+b$, while $A_2$ requests three units
of it, leaving a net deficit:
\beq
\valence( b; A_1, A_2) = -1
\eeq
\end{example}
This notation allows us to simplify the discussion of occupancy and tenancy
in later sections.

\begin{example}
Consider the following promises made by a network switching device:
\beq
{\rm switch} &\promise{+(10Gb)\#48}& {\rm client}\\
{\rm client} &\promise{+(1Gb)\#1}&  {\rm switch}.
\eeq
The switch makes 48 promises offering 10Gb `bandwidth' to the clients.
The client accepts one valency slot (leaving 47 more), and promises
to consume only 1Gb of the maximum possible 10Gb.
\end{example}

\subsection{The language structure of promise bodies}

In order to even comprehend one another's promises, agents need a
common language, with which to express body content.  The problem of
how agents come to develop a mutually acceptable language for
information exchange between independent agents has been studied in
connection with linguistics both of the traditional variety and in the
biology of the genetic code\cite{durbin1}. It is not a simple problem,
and I shall not try to address it here in full; however, there are some
simple things we can say about it.

In all cases, what we derive from these studies is that, regardless of
whether language is executed continuously or
discretely\footnote{Messages may be sent with or without words, with
  body postures, or coded by melody (frequency division multiplexing)
  rather than representations discrete in time (time division
  multiplexing). These are well known in information theory.}, the
possible intended meanings form a discrete alphabet of symbols,
representing capabilities, intentions, and so on. Thus, semantics
constrain promise bodies to a set of linguistic atoms (morphemes)
which are discrete.  In nature, we see this in everything from gene
codons, to cells and organisms, to Chinese ideograms\footnote{One can
  speculate about the reason for the size of discrete patterns used to
  convey meaning.  Dynamical scales will ultimately place limits of
  the comprehension of an agent. If agents could have infinite
  resolution, there would be no limit to what information could be
  conveyed in an arbitrary promise. But a recipient has to be able to
  parse this information in a finite time, shorter than that which is
  needed to keep its promise. This suggests that information density
  must be finite, whatever the nature of the agents.  Even a concept
  like `happiness' cannot have an infinite number of shades of grey!
  Protein size limits the size of a gene; variety of length and time
  scales limit the complexity of a key used by a human or a computer,
  and even the density of musical notes in the scale is limited to
  approximately quarter tones by the size of the human ear.}.

\begin{figure}[ht]
\begin{center}
\includegraphics[width=7.5cm]{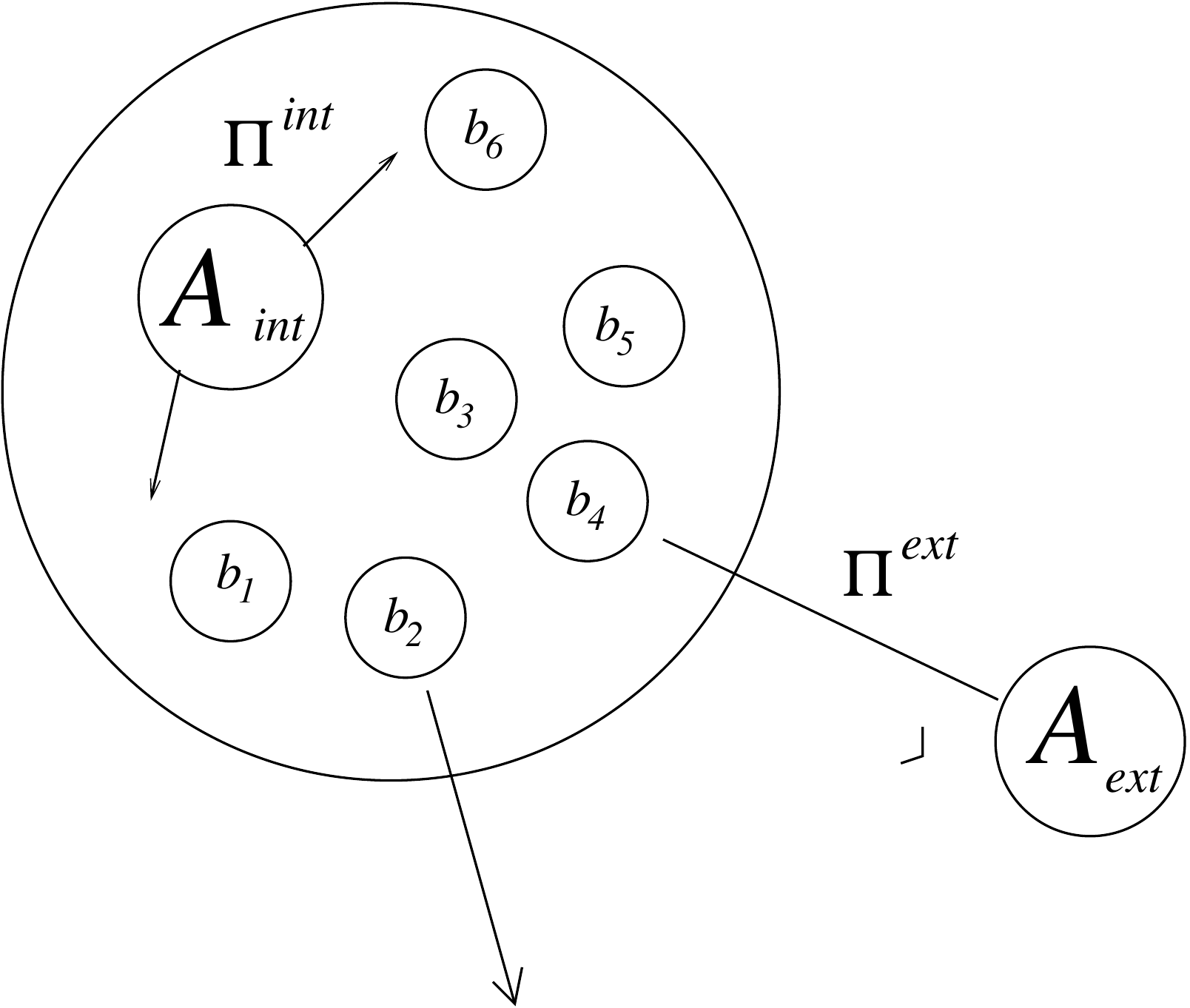}
\caption{\small Body parts or linguistic atoms of intent may be used
  as a spanning set for the body of any promise in a
  region. This is like a coordinate system of intent.
For $b_4$ to be promised to $A_{\rm ext}$, $b_4$ must be
understood by both promiser and promisee agents.\label{agentstructure1}}
\end{center}
\end{figure}

The representation of promises in arbitrary natural language is unlikely to be simple, since
the metaphoric basis of natural languages do not lend themselves to direct mapping. 
However, we can imagine formulating a set of restricted {\em Domain
Specific Languages}, $\ell^{(\alpha)} = \{ \beta_a^{(\alpha)}\}$, 
consisting of symbols $\beta^{(\alpha)}_a$,
whose purpose is the represent
specific intentional behaviours, and which map to a set of concepts that can be
represented in a natural language $N^{(\alpha)}$ of all agencies in a coordinate patch:
\beq
\ell^{(\alpha)} \rightarrow N^{(\alpha)}.
\eeq

\begin{assumption}[Discreteness of promise body encodings]
The language of promise bodies is assumed to be a discrete language pattern,
of fixed, but unspecified, alphabet $\beta$, comprising basis symbols $\beta_a$.
\beq
b(A_i) = \sum_{a=1}^C c_a\beta_a,
\eeq
for some coefficients $c_a$, and $C=\dim(\beta(A_i))$ for agent $A_i$.
\end{assumption}
Based on this decomposition of intent, we may now consider a local
observer view on the measurement of agent structure, assuming
finiteness of information.  Consider the syntax space of alphabetic
strings in a set of languages $\ell^{(\alpha)} \in \cal L$, where
$\alpha$ labels different languages, each of which might have its own
alphabet.

In order for agencies with different languages to be able to
communicate intent, these languages must be mutually comprehensible,
i.e. we must be able to map parts of them to one another.
The existence of a discrete alphabet of symbolic words allows us to think of the
alphabet of intentions as a matroid or spanning set $\beta_a$ over a vector space.
We may assume that the language of agent $A_i$ has $\dim(\beta(A_i))$ dimensions or classes of intention,
so that $a=1,\ldots, \dim(\beta)$ (see figure \ref{agentstructure1}).
This alphabet may or may not necessarily be shared between all agents, but
needs to be partially translatable for agents to make promises
to one another.

Suppose we have two alphabets with symbols $\beta$ and $\beta'$.
In order for a promise body to be encodable in either language, we must have:
\beq
b = \sum_{a=1}^{\dim(\beta)} c_a\beta_a =\sum_{a'=1}^{\dim(\beta')} c'_a\beta'_a
\eeq
i.e. both languages have to be able to span the promise body, or represent
it in their own spanning sets or words.
We need not require a single common spanning set of body parts to span
every message, the language of agents does not have to be a global
symmetry across spacetime, but we do require local continuity in the
couplings, in order for information to be passed on.

\subsection{Agent homogeneity, and transmission of intent}\label{disting}

A related issue concerns the ability for an observtional arbiter to distinguish
between different agents. This depends on the ability of the agent to
comprehend the language being promised.
Suppose an observer suspects that an agency is non-atomic, i.e. it contains
internal structure.
\begin{itemize}
\item Multiple agencies within a single agent might be identified if:
\begin{itemize}
\item If an agent seems to make independent partial-promises to different
  promisees, it could be natural to formally resolve it into
  separate sub-agents for the independent promises (`disaggregation').
\end{itemize}
\item A compound promise could be resolved into several simpler promises if:
\begin{itemize}
\item The details of the promise body can be expressed as a number
of independent items that can be made (+) or consumed (-) independently.
\item If, by emergent agreement, a set of primitives (like a table
of elements) can be seen to form a spanning set for the promises made
by an ensemble of one or more indistinguishable agents\footnote{The latter case is like the Millikan experiment for measuring electric charge. If
you look for differences, then the smallest difference may be assumed to be
an elementary.}.
\end{itemize}
\end{itemize}
To know this information for sure, it would have to be promised. For this,
we can define the concept of an agent {\em directory} (see section \ref{pcg}).

\begin{definition}[Homogeneity of agent languages, and transmission of intent]
Agents $A_1$ and $A_2$ may be said to have distinguishable promises 
that can be resolved by a receiver $A_r$ iff the measures of the promises,
which overlap with the receiver, are unequal.
Suppose $A_1, A_2$ each make a promise to an observing receiver $A_r$:
\beq
A_1 \promise{+b_1} A_r\\
A_2 \promise{+b_2} A_r
\eeq
$A_r$ might judge these two agents identical if $b_1 \intersection b_2 = b_r \not= \emptyset$,
i.e. if both promises contain a common part (the intersection $b_1 \intersection b_2$),
which the promisee promises to see:
\beq
A_r \promise{-(b_1\intersection b_2)} A_1, A_2.
\eeq
\end{definition}
In other words, if the receiver filters its perceptions
according to what is common to all agents, then it is
unable to distinguish them. 

We can break this into two cases: if sources $A_1$ and $A_2$
share common components (e.g. share common genes), i.e. $b_1 \intersection
b_2 \not= \emptyset$, then the receiver can observe a similarity
between the agents. 
If, further, the receiver only perceives the influence of what is common
between them, i.e.  it promises to accept $-b_r$, then the agents will 
perceive an elementary unit of promise equal to:

\beq b_1 \intersection b_2 \intersection b_r
\not= \emptyset.  \label{overlap}
\eeq 

\begin{example}
In genetics, the body elements correspond to genes. A gene can be passed on $(+)$
from a  parent to a child, but whether or not it is `expressed' or activated
depends on the proteins use the gene $(-)$ during morphogenesis. Thus, simply passing
genese from generation to generation need not result in transmission of
attributes (promises kept). Similarly, environmental conditions can play a role
in activating or de-activating particular gene promises in different circumstances. 
\end{example}

\begin{example}
  For instance, imagine one agent believes it is promising to deliver
  a letter to a recipient.  The agent receiving what the promiser
  considers a letter might, in fact, be promising to evidence in an investigation as a DNA
  sample on the letter. The rest of the letter vehicle has no semantic value.
  A second agent then promises to deliver a blood sample to the same
  recipient. This also qualifies as an evidential DNA sample to the recipient.
  Since the agency of DNA is encapsulated by both the letter and the
  blood sample: $DNA \subset Letter$ and $DNA \subset Blood$, an agent
  that can only measure DNA would see the letter and the blood sample
  as being equivalent sources of DNA.

DNA, itself, is a vehicle (agency) for genes that are embedded within it.
Exactly the same argument now applies at the level of DNA as a container.
The presence or absence of an allele (gene flavour) within a strand of
DNA indicates a similarity of intent.
\end{example}

The impact of a promise is defined through its binding strength, or effective coupling constant.
In earlier work, I defined the notion of a trajectory for
an agent, and the corresponding notion of a generalized force, obeying
Newtonian semantics\cite{sirimace2007,organization}. Intuitively, one expects a force
to be something that impacts an agent's trajectory
\beq
F: b \rightarrow b + \delta b
\eeq
Though, readers should note that promise trajectories are rarely Newtonian.
From this, one may construct a generalized force, which with the help of assessment 
function $\alpha(\pi)$ takes on a familiar form of a field-charge like interaction:
\beq
F \simeq \alpha \left( \underbrace{S \promise{+b} R}_{\text{Field}}, \underbrace{R\promise{-b} S}_{\text{Charge}} \right).
\eeq
See \cite{sirimace2007,organization} for the details. This gives us a simple notion of a coupling
strength by which to define such a measure of impact. The analogies to physics are attractive, but
we should beware that the trajectories are `rough walks' not smooth curves, in spite of the
analogy to differential notation.

\subsection{Continuity and spatial homogeneity of promise semantics}

Promises comprise information transmitted between agents. The
effective transmission of information requires the existence of a
common language\cite{shannon1,shannon2}.  If each agent is an
autonomous entity, how may agents learn a common language, or
equilibrate different languages, in order to understand one anothers'
promises?

Consider the existence of a language operation that transforms a body
string $b_1$ by agent $A_1$ into a body string $b_2$ for agent $A_2$.
\beq
b^{(a)} &=& L_{ab}(b^{(b)})\label{l1}\\
b^{(b)} &=& L_{ba}(b^{(a)})\label{l2}.
\eeq
In order for $L(\cdot)$ to be faithful and express transitive properties
such as long-range order, we need piecewise reversibility. Substituting (\ref{l2})
into (\ref{l1})
\beq
b^{(b)} = L_{ba}(L_{ab}(b^{(b)}))
\eeq
which implies that
\beq
L_{ab}(L_{ba}) = 1
\eeq
or the inverse relationship is the transpose:
\beq
L_{ab} = L_{ba}^{-1}.
\eeq
In a general matrix representation, this implies that universal representation of the matrices
representing $L$ belong to the {\em unitary group} over language space $a,b$:
\beq
L^\dagger L = I.
\eeq
The full unitary symmetry (if we take the general solution to this seriously, in the absence
of other constraints) allows for general rotations of symbols. Thus so-called entangled
states are, in principle, allowed for in this observation.

The set of transformations represented by $L$ does not have to be assumed a global
symmetry. The index labels gloss over the piecewise locality of the assumption that
$L_{ij}(A_i)$ is a transformation that takes place at the location $A_i$, on its
way from $A_j$. Similarly $L_{ji}(A_j)$ takes place at $A_j$ om its way 
from $A_i$\footnote{There is an analogy here with local gauge symmetries, as imagined by Weyl.}. 
The limit of locality is thus the adjacency length between $A_i$ and $A_j$
\beq
L_{ab}(A_i^{(b)})\cdot L_{ba}(A_j^{(a)}) = 1.
\eeq
If $A_i$ and $A_j$ are nearest neighbours, this is straightforward. However, if we
regard the transmission of a promise through intermediate proxy agents, then
comprehension and message integrity depend on the existence of non-local correlations,
somewhat analogous to entanglement in quantum mechanics.

This symmetry is closely related to the observation that, even with a common
language, in any promise relationship between agents: 
\beq
U(U(+b)) &\not=& +b\\
--b &\not=& +b 
\eeq
(see section 3.10.2 in \cite{promisebook}). Both relations imply a
kind of long-range cooperation between the agents. These are analogous
to the global symmetries of particle physics.

On seeing this familiar symmetry of the physical world, it is tempting
to look for a conserved quantity, or a conservation law for the
alphabets $\beta$, but it cannot be the case that the alphabets are
preserved. A conservation law would make the transfer of alphabetic
messages into a zero sum game: what was passed on to a neighbour would
be lost by the sender. This is not how evolution works.  Instead, the
process of equilibration is more like an epidemic duplication\footnote{Why, for
  example, would genes be preserved in number and type across species?
  If that were the case, all species would eventually equilibrate into
  one, and what was gained by one species would be lost by another.}.
The transformations of language a location are more likely to be
non-conserved, in general, and depend on the proper time (evolutionary change).
One expects transmission of symbols in both time and space, but without
conservation. Thus one could imagine dividing the inter-lingual transformations
into two parts:
\beq
L = \underbrace{L(A)}_{\text{Conserved}} + \underbrace{L(A,\tau)}_{\text{Non-conserved}}
\eeq
The first part could lead to a zero-sum conserved current of symbols from one agent to another,
allowing migration without preservation, while the latter part allows symbols
to be duplicated and spread. It might be fruitful, in the future, to consider how this
process takes place, and compute Kubo relations for the transmission of promises\cite{forster1}.

\begin{figure}[ht]
\begin{center}
\includegraphics[width=7.5cm]{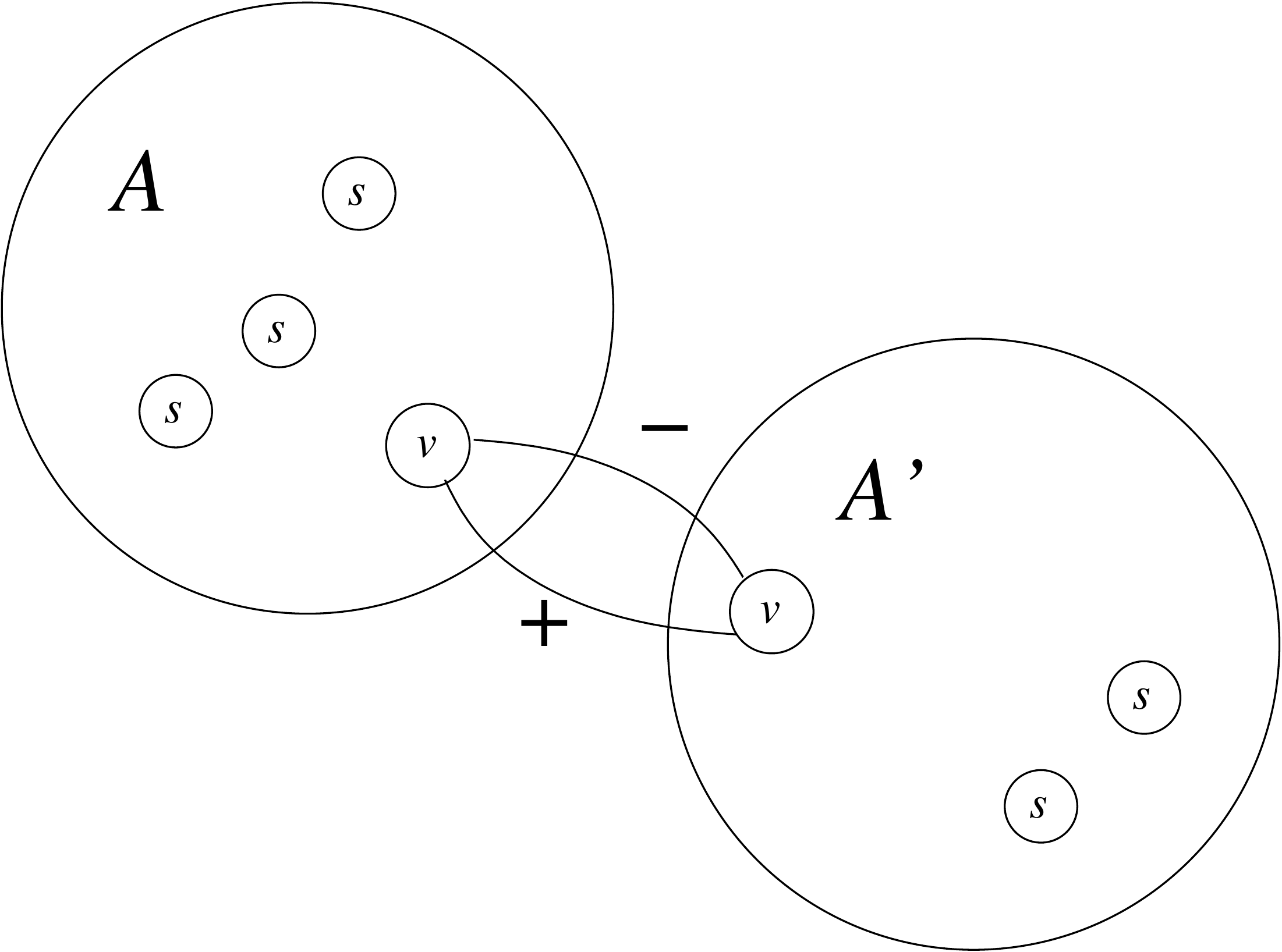}
\caption{\small Agents need to have similar structure to make promises
across agent boundaries, referring to internals of the other.
The promise bodies do not have to be identical as long
as each agent recognizes its own version of the other's promise.\label{agentstructure2}}
\end{center}
\end{figure}

In order to cooperate, agents interacting at a distance need to have a
sufficient level of similarity to local agents in order to be able to
make sense of what they are promising to one another (see figure
\ref{agentstructure2}). This is true regardless of whether they directly
adjacent or not. This must be a semantic equilibrium, mediated by
a dynamic exchange process, in order to this cooperative behaviour to emerge.
Unlike elementary physics, where locality is more obvious, cooperation between
agencies could be long range, as long as there is adjacency over a long range.
Semantics tend to follow humans, companies, organizations, races, countries, etc,
and humans form multiple outposts with geographic separation.

\subsection{Inter-agency language translations}

It is possible, in principle, to construct a linear transformation of one language
into another:
\beq
\beta_a' = L_{a'a}( \beta_a).
\eeq
Then, from the linearity, we may use the distributive law to say that
a body $b$
\beq
b = \sum_{a=1}^{\dim(\beta)} c_a\beta_a =\sum_{a=1}^{\dim(\beta)} c_a L_{aa'}(\beta'_a)
\eeq
Thus, as long as the matric $L$ exists, the languages will be translatable.
If the dimensions of the languages are not the same, only a subset of
meanings will be translatable from one to the other. This might be
asymmetric, allowing one agent to understand another, but not vice
versa.  At the level of atomic intentions, we can introduce coding
transformations a transition matrix for mapping
\beq
\ell^{(\alpha)}(A_i) \in L_{\alpha\beta}( \ell^{(\beta)}(A_j))
\eeq
for some invertible matrix-set of maps $L_{\alpha\beta}$.

\begin{example}
Consider a body language with alphabet $\beta = \{ SEND, RECEIVE, SEEK, FORWARD, BACK \}$,
and  $\beta' = \{ PUT, GET, APPEND \}$, then we can translate these:

\beq
c_1'\beta_1' = PUT &=& c_1\beta_1 = SEND\\
c_2'\beta_2' = GET &=&  c_2\beta_2 = RECEIVE\\
c_3'\beta_3' =APPEND &=& c_2'\beta_2'+c_4'\beta_4'+c_1'\beta_1' = SEEK+FORWARD+SEND
\eeq
Hence there is a translation matrix:
\beq
L_{a'a} = \left( 
\begin{array}{ccccc}
1 & 0 & 0 & 0 & 0\\
0 & 1 & 0 & 0 & 0\\
1 & 0 & 1 & 1 & 0\\
\end{array}
\right)
\eeq
which, in this case, is not invertible. Hence the language is translatable in one direction
only.
\end{example}
In principle, however, it should be possible to restrict $\ell^{(\alpha)}$, such that translation may be
performed faithfully as a bijection, by postulating a `table of elements' for the chemistry
of all promises in a semantic spacetime:
\beq
\ell^{(a)} \leftrightarrow \ell^{(b)} ~~~ \forall \;a,b \in {\cal L}.
\eeq

\begin{figure}[ht]
\begin{center}
\includegraphics[width=7.5cm]{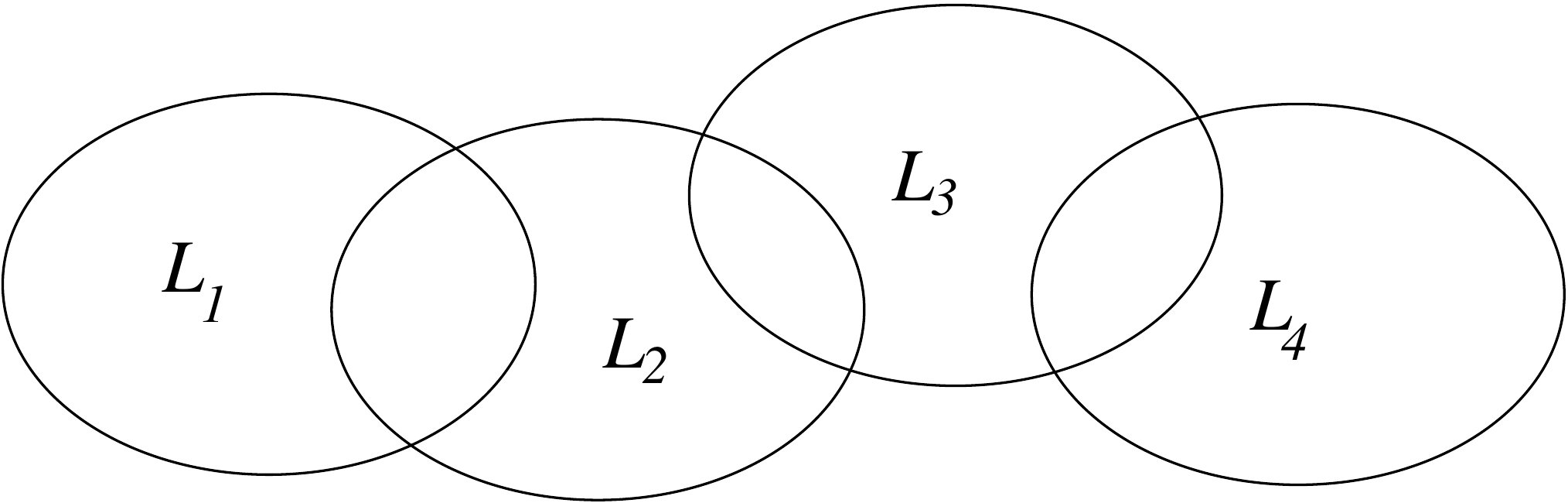}
\caption{\small Overlapping patches of language require all agents in a patch
to use a compatible language, and for (at least some) agents in each patch
to comprehend a faithful translation of a neighbouring language, in order to
bring long range order.\label{languages}}
\end{center}
\end{figure}

From here, the requirement for transmission of intent is that there be a piecewise continuity by
 partial overlap between neighbouring languages (see figure
\ref{languages}) becomes: 

\beq \ell^{(a)} \intersection L_{a,a\pm
  1}(\ell^{(a+1)}) \not= \emptyset~~~\forall\; a\label{continuity}
\eeq so that for $a\pm 1$ a neighbouring language of
$\ell^{(a)}(A_i)$, i.e. $\Pi^{\rm ext}_{ij}(A_i)=1$, a promise
binding, of the following kind, may be mutually and equivalently
comprehended: \beq \left\lbrace
\begin{array}{c}
A_i^{(a)} \promise{+b} A_j^{(a)}\\
A_j^{(a)} \promise{(-b)} A_i^{(a)}
\end{array}
\right.
\stackrel{a\rightarrow b}{ =}
\left\lbrace
\begin{array}{c}
A_j^{(a)} \promise{L_{ab}(+b)} A_i^{(b)}\\
A_i^{(b)} \promise{L_{ba}(-b^{(b)})} A_j^{(a)}
\end{array}
\right.
\eeq
This assumes that both $L_{ab}$ and $L_{ba}$ exist over the relevant bodies.
The solution of the continuity relation (\ref{continuity}) would then take the form:
\beq 
\ell^{(a)} = \lbrace \beta^{(a)}\rbrace \; \union\; \lbrace L_{a,a\pm 1}( \beta^{(a\pm 1)}) \rbrace ~~~\forall\; \alpha
\eeq
i.e. the language of an agent in a coordinate patch $\alpha$ should consist of all interior
body symbols, together with native translations of neighbouring languages.

\section{The scaling of agency}

To describe the spatial structure of agents, in which
the composition of elements is consistent, we need to describe how
agency scales collectively, through aggregation and reduction of elementary agents.
The concepts of scaling are familiar to physicists for dynamics, but here
we also want to extend them to incorporate semantics. This might be motivated
by questions of the following kind:
\begin{itemize}
\item How does a team promise something as a unit? 
\item How does an organization appear as a coherent entity? 
\item How does a collection of components promise to be a car?
\end{itemize}

Since we can aggregate promises into a single promise, and aggregate
agents into a single agent, the ability to detect or resolve parts
within a whole depends on the observer's capabilities. Similarly, the
ability for an agent to perceive a collection of individual agents
with a collective identity (i.e. a super-agent) depends on the
capabilities of the observer agent.  Elementarity and composability of
agency thus go together with a hierarchy of observable agency, which
needs to be elucidated.

\subsection{Subspaces}

A partial region of a semantic space, at any scale, may be called {\em subspace}.
We may distinguish the boundaries of such a region in any way convenient.
A subspace is assumed to be connected, but not necessarily homogeneous
of isotopic.
\begin{definition}[Subspace]
  A subspace is a collection of agents, in which every agent is
  adjacent to at least one other. The agents may be:
\begin{enumerate}
\item Identified and associated by an external observer by their (-) promises, or
\item Intentionally labelled and coordinated by its members with a (+) promise.
\end{enumerate}
\end{definition}
Subspaces can be defined by partitioning a space, or by constructing
a space agent by agent. There is good reason to consider both of these
points of view, so let's describe them below.

\subsection{Independence of agents under aggregation}

We begin by considering how to identify discrete, elementary
components within a system of autonomous agents, making promises.
If we assume that sufficient knowledge of agents in available, then an
observer can assess the independence of agents by the absence of mutual
information, i.e. zero overlap.

\begin{definition}[Independent agent]
Two agents $A_1$ and $A_2$ are independent iff the following overlap relation holds:
\beq
A_1 \intersect A_2 = \emptyset.
\eeq
\end{definition}
An agent that it independent of another agent may be said to be {\em outside} or {\em exterior} to
the agent. An agent that overlaps with another agency may be said to be {\em inside} or {\em interior} to it.

\subsection{Composition of agents (sub- and super-agency)}

\begin{figure}[ht]
\begin{center}
\includegraphics[width=5cm]{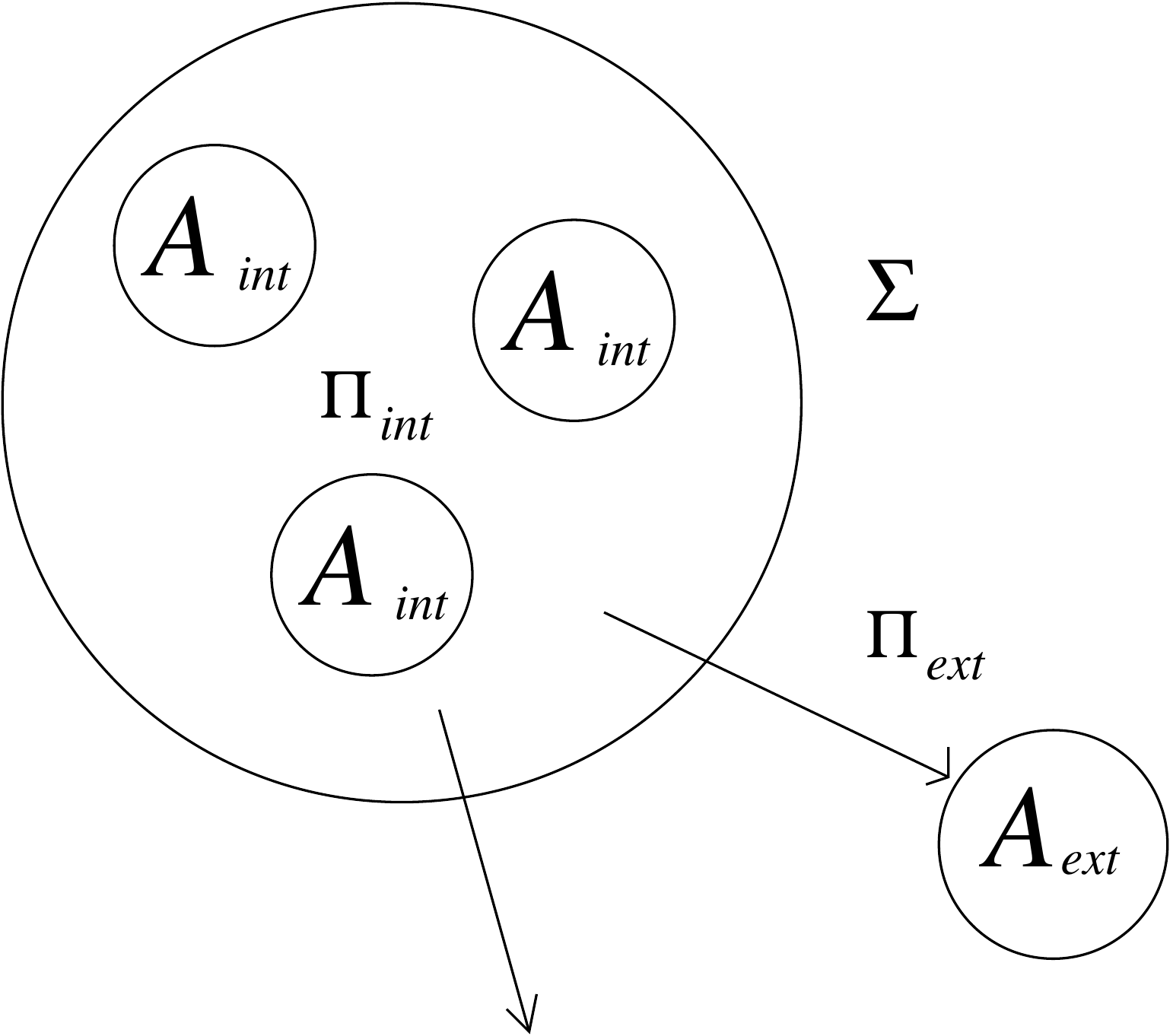}
\caption{\small Agent structure consists of an element that makes a number of
exterior promises, some of which are scalar, some vector, etc.
Interior promises are invisible from the outside.\label{agentstructure}}
\end{center}
\end{figure}

The treatment of a collection of agents as a single entity is a choice
made by any observer. It can be made with or without
promises from the composite agents themselves (see figure \ref{agentstructure}).
Agency can be defined recursively to build up hierarchies of component
parts.  In paper I, I showed how spatial boundaries can be defined
by membership to a group or role. We still have to explicate the
relationship between the internal members and the structure of the
whole, as perceived by an observer.

We define a collective super-agent as a spacetime structure that has
collective agency, i.e. its intended semantics relate to a collection
of agents surrounded by a logical boundary, with collective semantics (see figure
\ref{agentstructure}).

\begin{definition}[Bare super-agent]
  A super-agent of size $S$ is any bounded agency composed of individually
  separable agencies, partially or completely linked by internal vector promises. The bare
  super-agent is defined by the closed graph, without any external
  adjacencies.  It is a doublet: 
\beq 
A_{\rm super} = \langle \{A_i\},
  \Pi^{\rm int}_{ij}\rangle, ~~~~~ i = 1, 2 \ldots S.   \label{sa}
\eeq 
where $A_i$ is an internal agent of $A_{\rm super}$, and $\Pi_{ij}$ is
the promise adjacency matrix between the $S$ internal constituents.
\end{definition}

\begin{definition}[Dressed super-agent]
A dressed super-agent is the bare super-agent together with its set of
exterior promises.
  It is a triplet: 
\beq 
A_{\rm super} = \langle \{A_i\},
  \Pi^{\rm int}_{ij},
  \Pi^{\rm ext}_{s\epsilon_1}
\rangle, ~~~~~ i = 1, 2 \ldots S.   \label{dsa}
\eeq 
\end{definition}

Super-agency allows promises to exist within and without a super-agent
boundary.  We call these interior and exterior promises, respectively.

\begin{definition}[Exterior promises]
Exterior promises are made by agencies within the super-agent boundary,
to agents outside. They
represent inputs and outputs of the super-agent,
i.e. how it interacts with an external space. 
\end{definition}
\begin{definition}[Interior promises]
Internal promises are made by agents inside the super-agent boundary to
other agents inside the boundary. They
represent the bindings that make the super-agent behave as a single
cohesive entity.
\end{definition}
In principle an observer could draw a line around any collection of
agents and call it a cell or composite super-agent. This is an
assessment any agent can make, as part of its definition of an agency
scale. However, it might still be of interest to distinguish special
criteria by which such an arbitration might occur.  In component
design, for instance, the choice of boundary has often to do with the
a choice interface an agent wants to interact with.

\begin{figure}[ht]
\begin{center}
\includegraphics[width=10.5cm]{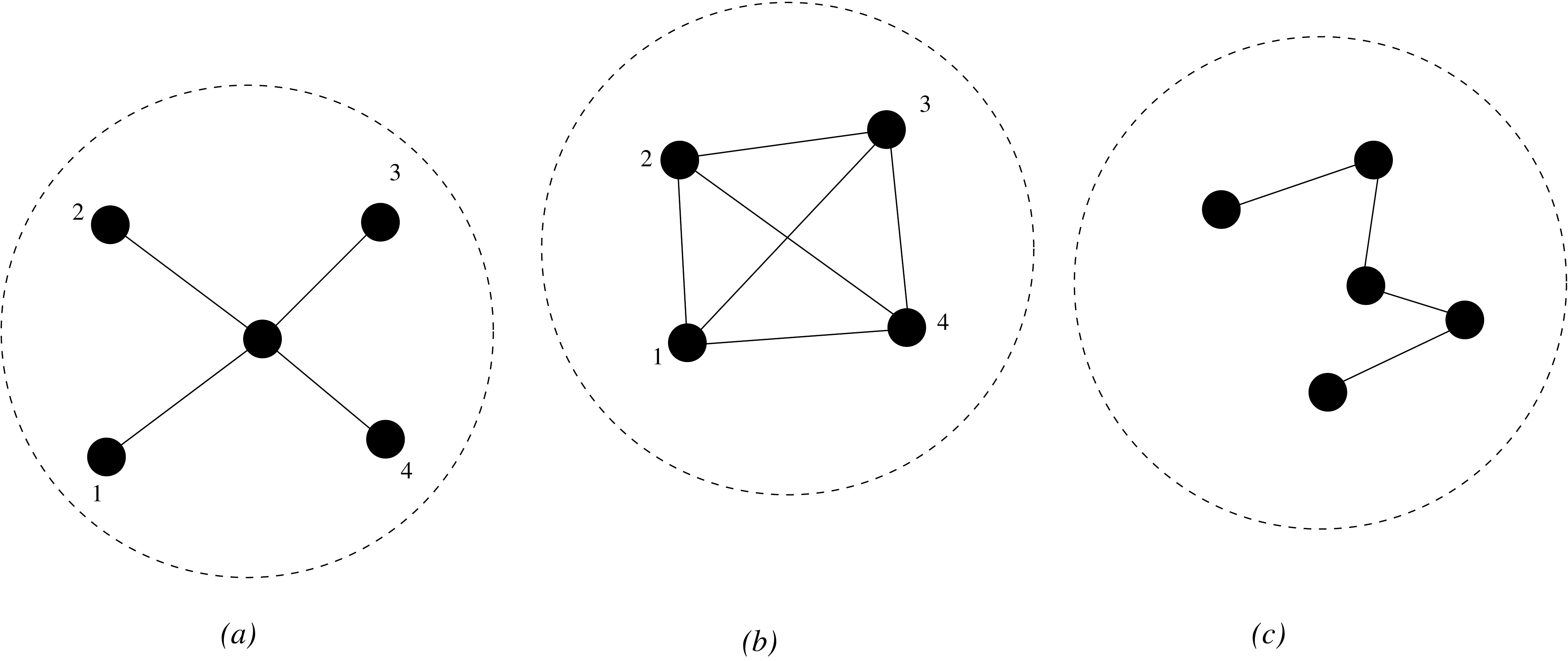}
\caption{\small Three ways of binding collective agency. 
Another way is to simply make an arbitrary collection.\label{superagent}}
\end{center}
\end{figure}
The alternatives fall into three basic categories (see figure \ref{superagent}):
\begin{enumerate}
\item[(a)] A membership in a group or associative role, where the central membership authority
may be either inside or outside the boundary. In this case, we are identifying a group of symmetrical
agents.

\beq
A_{\rm host} &\promise{+{\rm membership}}& \{A_{\rm tenant}\}\\
\{A_{\rm tenant}\} &\promise{-{\rm membership}}& A_{\rm host}
\eeq

\item[(b)] A total graph or collaborative role. In this case, we are identifying agents with coordinated
behaviours.

\beq
A_i &\promise{\pm{\rm membership}}& A_j ~~~\forall A_i,A_j \in A_{\rm super}.
\eeq

\item[(c)] A dependency graph, path or story. In this case we are identifying dependency bindings.
\end{enumerate}
Let us now define the converse properties:
\begin{definition}[Sub-agent]
  A sub-agent is an agent assessed to be a resident of the internal
  structure forming a composite (super) agent.
\end{definition}

\begin{definition}[Residency]
A sub-agent $A$ is resident at a location $L$ iff it is defined to be within the
boundary of the agent:
\beq
A \intersection L \not= \emptyset.
\eeq
\end{definition}

Since an observer can form their own judgement about super-agent
boundaries, we cannot say that residence is the same as a promise of
adjacency.

\begin{figure}[ht]
\begin{center}
\includegraphics[width=10.5cm]{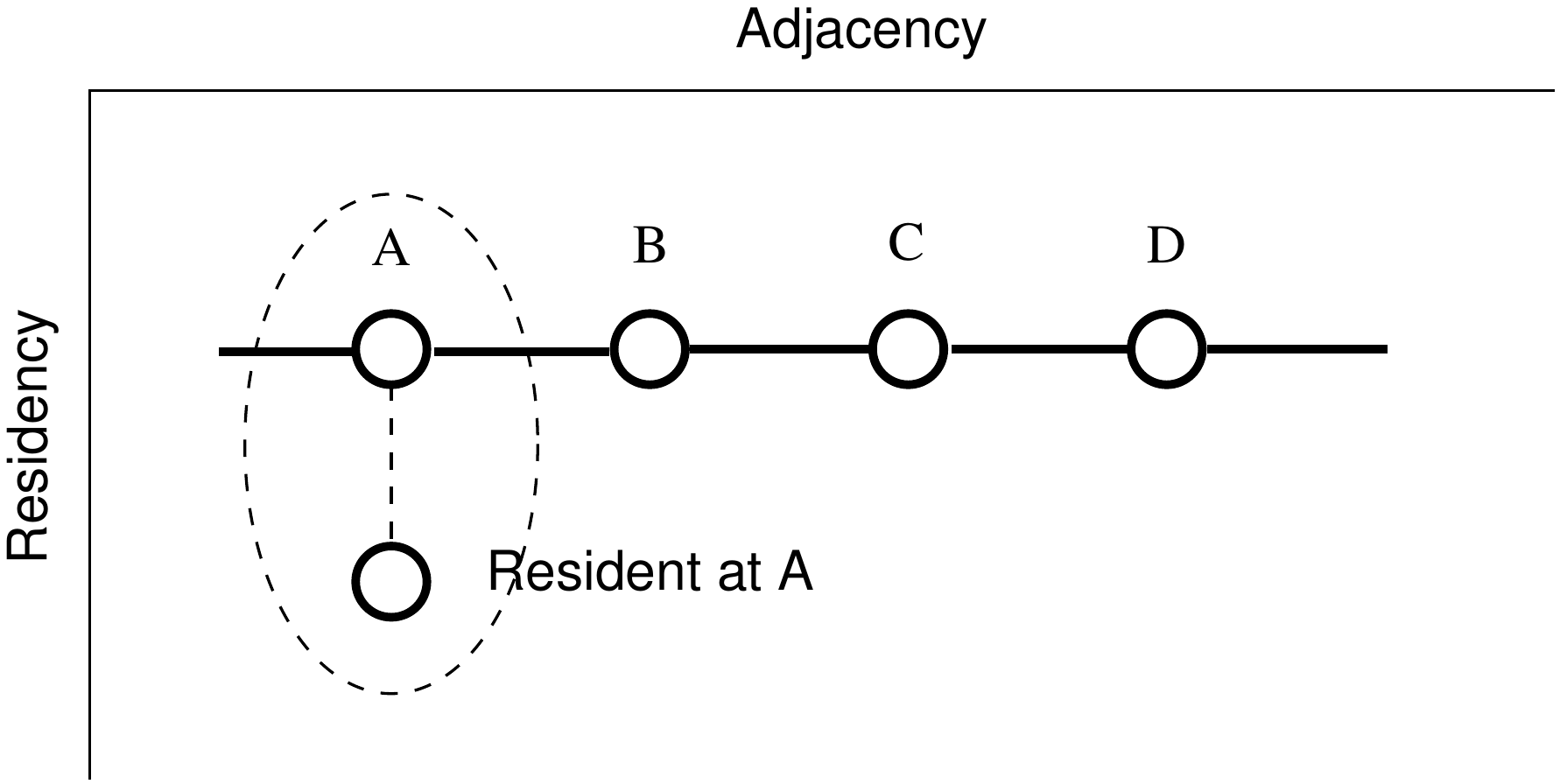}
\caption{\small A resident adjacency forms a super-agent by accreting to a seed agency that represents the
location anchor.\label{resident}}
\end{center}
\end{figure}

There are two types of adjacency, somewhat spacelike, which may or may
not be interchangeable (see figure \ref{resident}).  Normal `physical'
adjacency promises, and resident adjacency, which might link agents
virtually even though they are not physically adjacent. I'll return to
this topic when discussing tenancy below.

Lesser agents can become satellites of other agents.
This leads to a hierarchy or `planetary' structure,
by accretion into super-agents.

\subsection{Super-agent surface boundary}

Super-agent boundaries may be formed with different structural biases.
\begin{itemize}
\item Simple aggregations of agents, related through membership to a single leader (leader may be inside or outside the super-agent), and the connections are made through the leader as a proxy-hub.
\item A cluster of agents linked by cooperative vector promises.
\item Strongly cooperative agents which are inseparable without breaking an external promise.
E.g. an organism made of components that are all different and non-redundant to the functioning of the whole.
\end{itemize}

Interior promises are those entirely within the surface boundary of a super-agent.
We may define interior and exterior promise matrices for any agency,
using a matrix analogous to the adjacency matrix:
\beq
\Pi^{\rm int}_{ij} = \left.
\begin{array}{cc}
1& ~ {\rm iff}~ A_i \promise{*} A_j\\
0&
\end{array}\right\} &~~&  A_i, A_j \in A_{\rm super} \\
\Pi^{\rm ext}_{i\epsilon} = \left.
\begin{array}{cc}
1& ~ {\rm iff}~ A_i \promise{*} A_\epsilon\\
0&
\end{array}\right\} &~~& A_i \in A_{\rm super} , A_\epsilon \not\in A_{\rm super} 
\eeq

\begin{definition}[Surface of a super-agent]
  The exposed {\em surface} $\Sigma$ of the agent is the subset of interior/internal agents
  that have adjacencies to agencies outside the super-agent.
\beq
\Sigma \equiv \{ A_i \} \subset A_{\rm super} \;\Big|\; \Pi^{\rm ext}_{i\epsilon} \not= 0 
\eeq
\end{definition}
A super-agent surface may also make new explicit promises that are not identifiable with
a single component agency (see section \ref{irreducible}).
\begin{example}
In the molecular example above, the super-agent makes interior promises\footnote{These ionic properties
may reach outside the super-agent boundary too in some circumstances, allowing oxidation, etc.}
\beq
A_1 &\promise{+H}& *\\
A_2 &\promise{+H}& *\\
A_3 & \promise{+O}& *
\eeq
from internal agents, and collectively $M = \{ A_1, A_2, A_3\}$
\beq
M \promise{+ H_2O} *
\eeq
This promise implies some interior structure, and it does not emanate from any
smaller agent. Thus if an external observer is able to resolve the component
agents within $M$, the promise of $H_2O$ is not longer a promise.
\end{example}

As noted in section \ref{disting}, an observing agent may or may not
be able to discern an internal elementary agent within a super-agent, i.e.
whether the agent has internal structure, or whether it is atomic.
This depends on whether the agent promises transparency across its
surface.
\begin{definition}[Super-agent transparency]
  All promises whose scope extends beyond the boundary pass
  transparently through the surface of the super-agent. Scope
includes the list of promisees.
\end{definition}

\subsection{Sub-agents as the subject of a promise: emission and absorption}

An agent may itself be the subject of a promise body.  Agents can
conceivably promise to spawn new agencies: cells multiply, particles
transmute into clusters, humans give birth, organizations spin off
departments into new organizations, etc.

\begin{assumption}[Emission and absorption, parent-child relationships]
  It is within the allowed behaviours of an agent to emit (send) or
  absorb and incorporate (receive) a child sub-agent, which has a formally
  distinguishable identity to the parent.
\end{assumption}

The exchange of a sub-agent as an independent promise implies the exchange of the
promises made by it too. However, there is no assumption about the
promises having been kept before or after the keeping of the promise
to emit or absorb the sub-agent. This might be reified later.  The
promising of an agent is thus somewhat like the passing of a point
reference, with possibly late binding.

\begin{lemma}[Emission and residency]
In order to not violate the autonomy of agents, an agent $A_{\rm
  parent}$ could only make a promise to produce an agent $A_{\rm
  child}$ if, at the outset $A_{\rm child} \subset A_{\rm parent}$ i.e. resident.
\end{lemma}
The understanding that agents embody strings of
information (with an arbitrary physical realization) makes all of the arguments simple.  An agent might spawn
another, such as a device $A_{\rm device}$ emitting a network packet
$A_{\rm packet}$:
\beq
A_{\rm device} \promise{+A_{\rm packet}} A_{\rm network}.
\eeq
The agent referred to in the body can, in turn, promise another agent in its body, e.g. a verifier
of the checksum authenticator:
\beq
A_{\rm packet} \promise{+A_{\rm checksum}} A_{\rm verifier}.
\eeq
In order to be an intermediary, the exchanged package
should promise to both sender and receiver:
\beq
A_{\rm checksum} \promise{+\rm structure} \{ A_{\rm packet},\; A_{\rm verifier}\}.
\eeq
This does not necessarily imply the existence of a promise about another promise.
A promise of the existence of an agent has to result in an autonomous agent, by the
rules of promise theory.
\begin{figure}[ht]
\begin{center}
\includegraphics[width=5.5cm]{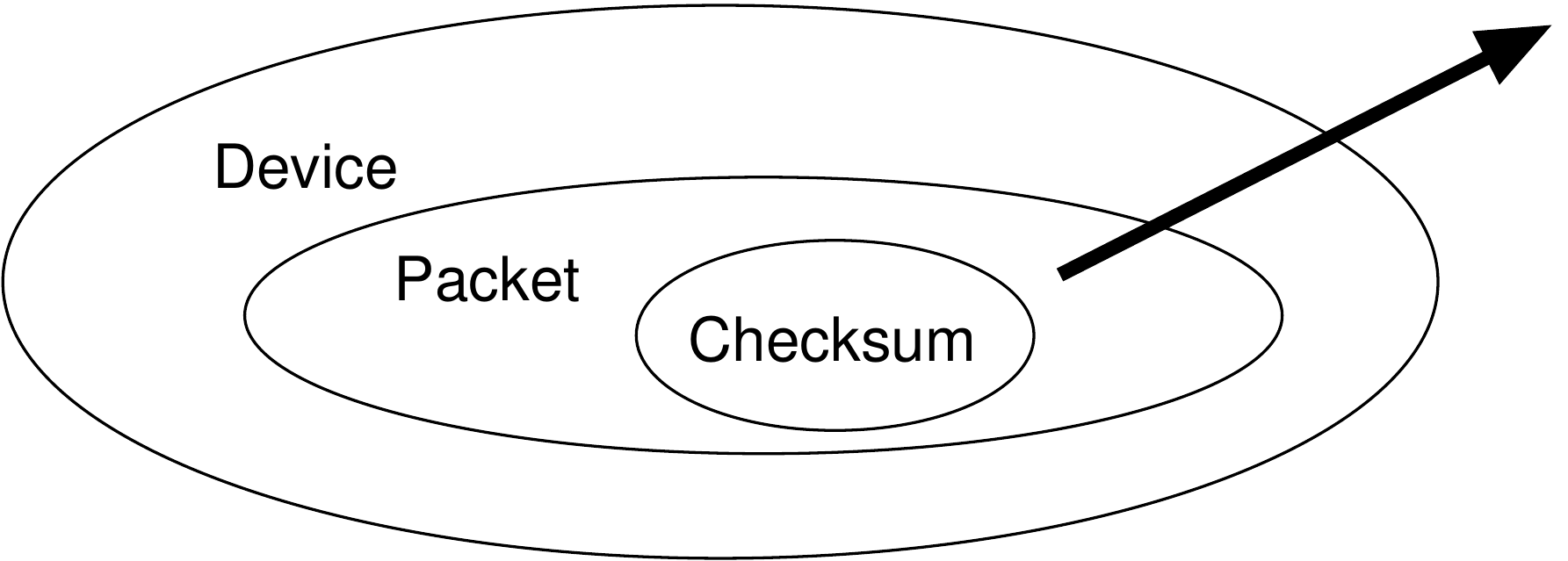}
\caption{\small Agents may be emitted if they start as residents.\label{packet}}
\end{center}
\end{figure}

\begin{definition}[Emission of a body part]
\beq
A_{\rm super} &\promise{+A_{\rm sub}}& A_{\rm recipient}\\
A_{\rm super} &\promise{A_{\rm super} \rightarrow \{A_{\rm super} - A_{\rm sub}\}}& A_{\rm recipient}
\eeq
\end{definition}

\begin{definition}[Absorption of a body part]
\beq
A_{\rm recipient} &\promise{-A_{\rm sub}}& A_{\rm sender}\\
A_{\rm recipient} &\promise{A_{\rm super} \rightarrow \{A_{\rm super} + A_{\rm sub}\}}& A_{\rm sender}
\eeq
\end{definition}
The signs inside the set braces imply union and complement removal, i.e. set difference.

\subsection{Is there an empty space?  The existence of a ground state.}

Is there a discrete lower bound on agency? Can we infer the existence of a state
of empty space in any system?
\beq
A \promise{\emptyset} \Unspec
\eeq
Suppose there is a lowest level to the hierarchy of agency at which
point no new intention can be inserted. The promises made by the agent
are purely names, since a name is a promise that identifies the
presence of the agent. A name or label cannot be subdivided without simply
resulting in more than one name, i.e. without increasing the number of
promises.

\begin{hype}[Lowest level of hierarchy]
  There is a level at which names cannot be subdivided without losing
  the ability to function and be understood (or connected to). The
  ground state is a gaseous state of maximal symmetry.  At this level,
  the only promise that an object can make is its name (this is a
  tautology).

\end{hype}

Does an agent that makes no promises even exist? As mentioned in paper
I, it is completely disconnected from the rest of a space, so we may
consider it either to be non-existent relative to an existing
spacetime, or be an entirely separate spacetime\footnote{If one can
  imagine a disconnected agent that makes no promises at all, then
  from there, one can imagine postulating the existence of an
  infinite number of empty spacetimes. These matters become
  technical, and we have no way of deciding whether they are real or
  fictional, assuming that there is a distinction.}.  If we do not
allow a complete absence of promises, then this suggests that empty
spacetime must, at least, promise adjacency. That is, empty space
requires at least a promise of adjacency. The promise might not be kept all the time,
however, as is the case in a gaseoues phase. This issue need further study.

\subsection{Agency scales}\label{scales}

Super-agency is a scaling transformation, in the sense of a dynamical
system.  It is therefore an important bridge between dynamics and
semantics, with consequences for both.  Because promises incorporate
semantics, which may be arbitrarily applied to a collection of agents,
boundaries for super-agency may be defined around any collection of
agents.

\subsubsection{Definition of semantic agent scales}

\begin{definition}[Agency scale]\label{semscale}
  A named collection of agencies considered to be
  the irreducible entities of space, i.e. it defines a set of
  possibly aggregate atomic agencies that are to be considered the
  set of addressable agents at the scale concerned.
\end{definition}
 Unlike dynamical scales, there is no {\em a priori}
identifiable measuring stick for semantic scales: they are 
non-ordered symbolic quantities, and must be promised
independently by an observer, by promise types and body constraints.
An agency scale is a thing in between a semantic scale and a dynamical
scale. The identification of agency is a semantic issue, but the 
scale is a well-defined dynamical unit.
\begin{example}
Suppose we have atomic agents $A_1, A_2, A_3$, which make exterior promises:
\beq
A_1 &\promise{+H}& *\\
A_2 &\promise{+H}& *\\
A_3 & \promise{+O}& *
\eeq
and interior promises
\beq
A_1 &\promise{+e}& *\\
A_2 &\promise{+e}& *\\
A_3 & \promise{-e, -e}& * ~ \text{(valency~ 2)}
\eeq

We may combine these agents into a super-agent by defining a scale:
\beq
{\rm Molecular} \equiv \{ M \},
\eeq
where $M \equiv \{A_1, A_2, A_3\}$. At the molecular scale, we now have a single 
super-agent $A$, instead of three resolvable atomic agents.
\end{example}

\subsubsection{Explicit schematic example of agent scaling}
\begin{example}
Consider figure \ref{agentstructure3}.
\begin{figure}[ht]
\begin{center}
\includegraphics[width=9.5cm]{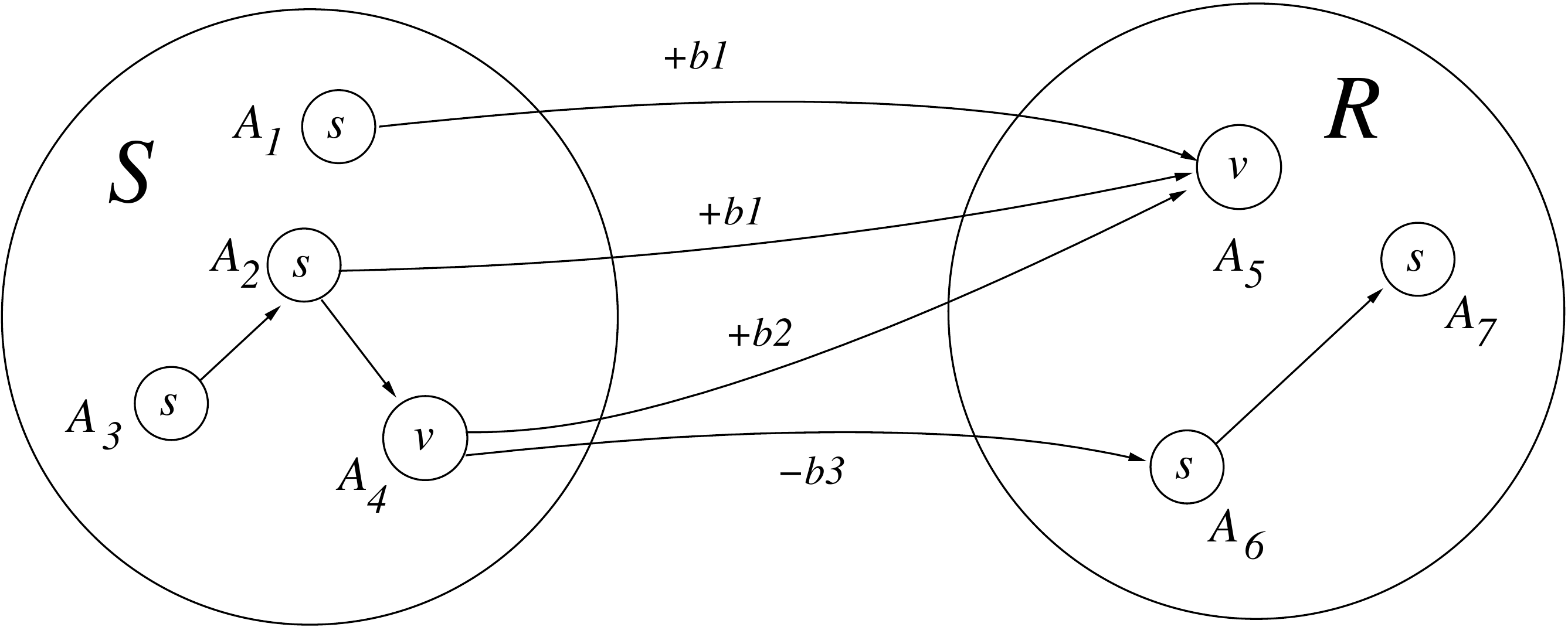}
\caption{\small The transformation of promises under a coarse-graining
  transformation. How promises appear to emanate from the super-agent
  surface.\label{agentstructure3}}
\end{center}
\end{figure}

At the level of atomic agents, we have exterior promises:
\beq
\pi_1: A_1 \promise{+b_1} A_5\label{eqstart}\\
\pi_2: A_2 \promise{+b_1} A_5\\
\pi_3: A_4 \promise{+b_2} A_5\\
\pi_4: A_4 \promise{-b_3} A_6
\eeq
and interior promises
\beq
\pi_5: A_3 \promise{+b_4} A_2\\
\pi_6: A_2 \promise{+b_5} A_4\\
\pi_7: A_6 \promise{+b_6} A_7\label{eqend}
\eeq
The super-agents are defined by:
\beq
S &=& \langle \{ A_1, A_2, A_3, A_4\}, \{ \pi_5, \pi_6\} \rangle\\
R &=& \langle \{ A_5, A_6, A_7 \}, \{ \pi_7 \} \rangle
\eeq
Scales have to be defined semantically in a promise model, since the interpretation of
an aggregate boundary is arbitrary; thus,
if we define the following scales:
\beq
{\rm Atomic} &=& \{ A_1, A_2, A_3, A_4, A_5, A_6, A_7 \}\\
{\rm Hybrid} &=& \{ S, A_5, A_6, A_7 \}\\
{\rm Super} &=& \{ S, R \}
\eeq
Then, at hybrid scale, the promises (\ref{eqstart}-\ref{eqend}) above collapse to:
\beq
S &\promise{+(b_1 \union b_2)}& A_5\\
S &\promise{-b3}& A_6
\eeq
and at super scale:
\beq
S \promise{+(b_1 \union b_2), -b_3} R
\eeq
\end{example}

\begin{example}
  The human eye is an example of a coarse grained receptor, i.e. a
  use-promise to accept radiation in certain frequency range. Light
  emitted at all frequencies in the range may be promised by emitters;
  however, only three coarse receptors (for red, green, and blue)
  cover this range. Thus colour vision in humans is limited to interpretation
through these three channels.
\end{example}

\subsubsection{Gauss' law for coarse-grained agencies and their external promises}

The divergence theorem (Gauss' theorem) is a simple universal identity that applies to vector fields
enclosed by a boundary. We may apply it to vector promises too. In its well-known form, if may be written:
\beq
\int_\sigma \vec V \cdot d\sigma = \int_v (\vec \nabla\cdot \vec V) dv\label{gauss}
\eeq
i.e. the integral of vector flux emanating from a surface $\sigma$ is
the result of the divergence of the vector field generated from the
volume enclosed by it.

Using definitions for unadorned graph theory, we can show that a
coarse graining is a simple application of this result. In other
words, the promises that come out of any volume or grain of space are
only a result of what agencies are inside it.
Consider a grain consisting of a number of agents with a surface boundary, and let
\beq
\pi_{ij} = \pi^{\rm int}_{ij} + \pi^{\rm ext}_{ij}.
\eeq
We observe that $\pi_{ij}$ is the $j$-th component of a local basis vector $\hat e_j$,
surrounding any agent $A_i$. We can define a vector field in this basis by
\beq
\vec \pi(A_i) = \sum_{j \in } b_{ij} \hat e_j  = \sum_{j \in grain} b_{ij} \pi_{ij},
\eeq
where $b_{ij}$ is the body (semantics) of agent $A_i$'s promise to agent $A_j$,
where $i,j$ run over all agents (or just the nearest neighbours of $A_i$) that are picked out
by $\pi_{ij}$. The components of the $\vec \pi(A_i)$ as a row vector are thus easily constructed:
\beq
\vec \pi(A_i) = (b_{i1},b_{i2},\ldots).
\eeq
The values are not really important, as long as they can be defined, since they will cancel
out of the sums. The derivative of this vector is anti-symmetric in $i,j$:
\beq
\vec d_j \vec \pi(A_i) = (b_{(i+j)1} - b_{i1}, \ldots )
\eeq
and thus the divergence is the sum of these where $i=j$.
It is thus easy to see that the sum over the volume enclosed by any grain surface cancels
everywhere except for those vector components that protrude through the surface:
\beq
\int d({\rm vol}) \sum_i (\vec d_i\cdot \vec \pi(A_i)) &=& 0 + {\rm exterior\; contributions}\\
&=& \sum_{i \in \rm surface} (b_{i\epsilon_i}, \ldots)\cdot \vec \sigma\\
&=& \sum_{i \in \rm surface} \vec \pi^{\rm ext}(A_i) \cdot  \vec \sigma
\eeq
where $\vec \sigma$ is the unit surface vector. This is exactly (\ref{gauss})
for $\vec V = \vec \pi$, summed over the grain. i.e.
\beq
\sum_{i\in \rm surface} \vec \pi(A_i) = \sum_{\epsilon \in \rm neighbours}  \vec \pi^{\rm ext}(A_\epsilon)
\eeq

\subsection{Coarse-graining (agent aggregation)}

Coarse-graining is what happens when one invokes a change of scale
from a high level of detail to a low level of detail, i.e. from a
large number of smaller agents to a smaller number of larger agents,
by aggregation.  This is usually done for systems assumed to have
infinite detail, in the so-called continuum limit.

In a discrete system, it is straightforward to define a coarse
graining by aggregation of autonomous agents into collections. In
physics (dynamically), one does this by defining characteristic
lengths (see the discussion for pseudo-continuous information in
\cite{burgesstheory}). In a semantically labelled theory, there are no
such easily defined lengths, and we are forced to define granular scales
explicitly as sets, see definition \ref{semscale} (section \ref{scales}).

\begin{definition}[Coarse-graining]
A transformation in which the collection of agencies representing spacetime 
are changed for a smaller set, with a corresponding reduction in detail.
Coarse graining involves:
\begin{itemize}
\item The summation of bulk properties, and relabelling of composite super-agents by a single collective label.
\item The elimination of all interior promises, which are not visible between the agents 
at the coarser scale.
\item Loss of information to exterior from the above.
\end{itemize}
\end{definition}

In a spacetime without boundaries, we identify self-similarity by 
the idea that a system is functionally and dynamically
similar in its promises and behaviours, before and after coarse-graining
by the coarse graining function:
\begin{definition}[Coarse-graining function]
A many-to-one function whose domain is the semantic spacetime $\langle \{ A_i\}, \pi^M_{ij}\rangle$,
composed of a collection of agents $\{A_i\}$ at scale $M$, along with
its promise matrix $\pi^M_{ij}$, and whose co-domain is the image doublet $\langle \{ S_k\}, \pi^{M'}_{kl}\rangle$, at scale $M'$:
\beq
\langle \{ A_i\}, \pi^M_{ij}\rangle \rightarrow \Coarse(\langle \{ A_i\}, \pi^M_{ij}\rangle) 
= \langle \{ S_k\}, \pi^{M'}a_{kl}\rangle
\eeq
where $i,j = 1\ldots \dim M$, and $k,l \ldots \dim M'$, where $\dim M' \le \dim M$
\end{definition}
The function $\Coarse()$ is a dimensionless renormalization
transformation.  Scale-free behaviour can not usually
apply to a bounded space, as boundaries pin the system at a fixed
scale.  It might still be possible to compute effective equations for
dynamical systems however\cite{scale1,scale2}.
Two spacetimes may be considered equivalent iff:
\begin{definition}[Spacetime equivalence function $\simeq$]
The equivalence of two semantic spacetimes designated by
$\langle \{ A_i\}, \pi^M_{ij}\rangle$, and  $\langle \{ S_k\}, \pi^{M'}_{kl}\rangle$
may be written 
\beq
\langle \{ A_i\}, \pi^M_{ij}\rangle \simeq \langle \{ S_k\}, \pi^{M'}_{kl}\rangle.
\eeq
where there exists a permutation group relabelling of $i,j \rightarrow k,l$
$\pi^{M}_{ij} \equiv \pi^{M'}_{kl}$ and \dim\, M = \dim\, M'.
\end{definition}

The transmission of influence by promise exchange is a dynamical
property that must also be affected by coarse graining. This is the
essence of the Renormalization Group in scaling systems\cite{scale1}.
When a coarser grain size influences a smaller grain from the top
down, it takes the form of effective boundary conditions on the
behaviour of the smaller scale.  This tends to introduce
non-linearity, as it modifies the propagation law for behavioural
trajectories.  If a smaller scale interacts with a larger scale, from the 
bottom up, it takes the form of a perturbation, which probes the stability 
of the larger grains.

There is no {\em a priori} connection between scale and influence; this depends,
as usual, on what exterior promises are made by the agencies regardless of grain-size.
Multi-grain-size models are perfectly possible. Indeed, they are the natural state of
materials like steel and plastics.

\begin{figure}[ht]
\begin{center}
\includegraphics[width=3.5cm]{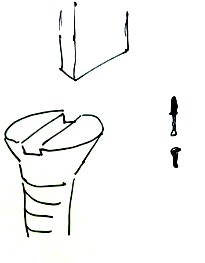}
\caption{\small Agency coarse graining. If there is a mismatch in scale between promiser and
receptor, the promise might not make sense. In some cases, it might be possible to
build a large scale solution from smaller grains, but not vice versa. \label{coarsegrain1}}
\end{center}
\end{figure}

\subsubsection{Coarse graining of interior and exterior promises}

Which promises are lost through coarse graining?  Effective external
promises, made by a super-agent, have their origin in the internal
constituent agents, but they do not preserve all of the internal
details, and they might add new ones (see section \ref{irreducible}).
In order to scale agency, we therefore need to understand the scaling
of both interior and exterior promises, and how they preserve the
semantics of a system during a scaling transformation.

\begin{example}
The chemistry promised by an atom does not depend strongly on the nature
of the isotope of a chemical element, only on the electron structure.
Thus electron shells made exterior promises from the surface of the
agent.
\end{example}

\subsubsection{Promiser coarse graining, and coarse-graining directories}\label{pcg}

Promises that remain exterior to a given grain, under a coarse
graining process, may be combined under the idempotency rule for
promises, and the combinatorics between common end-points. Hence
promises of the same type will merge into a single maximal promise,
defined below.

It remains possible for a single agent (coarse or fine grained) to
make different promises to multiple recipients. These cannot be
combined. However, promising from different sub-agents to the same
recipient becomes an instance for the idempotence rule once the
difference between the promisers is coarse-grained away. Promises with
uniform promisees are idempotent, or cumulative by virtue of the set
union.  The general scaling rule for promisers is thus:

\begin{definition}[Scaling rule for promiser]
Let $b_i$ be a set of promise bodies of the same type, i.e. $\tau(b_i) = \tau, \; \forall i$,
originating from a set of distinct agents $A_i$.
Then we may define the coarse graining of the set of agents, over constant $\tau, \varepsilon$, by:
\beq
\Coarse_n\left(
A_1 \promise{b_1} A_\varepsilon, \ldots, A_n \promise{b_n} A_\varepsilon
\right) &=&
\Coarse_n\left( A_1,\ldots, A_n \right) \promise{\max_i(\{b_i\})} A_\varepsilon\\
&=& A_{\rm super} \promise{\max_i(\{b_i\})} A_\varepsilon\label{cg}
\eeq
where $\max(x_j)$ represents the union of the sets pertaining to set $x_j$ intended for agent $A_j$:
\beq
\max(x_i) \equiv \bigcup_i x_i.
\eeq
Not this rule does not depend on the sign of the promise body $b_i$.
\end{definition}
In other words, the coarse-graining of a set of promises is equivalent
to a possibly smaller set of aggregate agents, which promise the
composite effect of the individual promises.  The total number of
promises is reduced from $n$ to 1 in this coarse graining.

A coarse-graining procedure is not a bijection, i.e. it is not an
invertible map or function. When a coarse graining of promisers is
undertaken, information about the specific fine-grain promises is
lost.  In order to reverse a coarse-graining process, we would have to
preserve the information lost.  
\begin{definition}[Coarse-graining directory]
The information lost during coarse-graining in equation (\ref{cg}), takes the form
  \beq 
  \pi_{\rm directory}(\tau) = \left\{ A_1 \promise{b_1 \in \beta_1}
    A_\varepsilon, \ldots, A_n \promise{b_n\in\beta_n} A_\varepsilon \right\}
  \eeq 
for promises of type $\tau$, expressed in language $\beta$.
Hence, whenever we group agents under
a common umbrella, by coarse-graining, there will be associated map, of the form,
\beq
\langle \tau, \pi_{\rm directory}(\tau), \beta \rangle, ~~~~~\forall \tau,
\eeq
that preserves the lost information, and could restore the detailed
view of which sub-agent is able to keep a promise perceived at the
level of the whole.  This is the information that becomes unavailable
at the coarser scale\footnote{Note, this information is related to the
  entropy of the system, but it is not related to disorder, merely a
  loss of information about order.}.  We may call such a map the
super-agent {\em directory} or {\em index} map, and write its promise:
\beq
A \promise{+{\rm directory}(A)} A'
\eeq
\end{definition}

Coarse-graining replaces $\pi_{\rm directory}(\tau)$ with a single
promise of type $\tau$, and an effective promise body.  So, for each
external promise of type $\tau(b_i) = \tau, \; \forall i$,
coarse-graining leads to a surjective map, between the collection of
promises made by sub-agents and the single effective promise made by
the super-agent. This map cannot be inverted, however it can be
resolved to direct the external agent to an appropriate microscopic
part of the super-agent.

\begin{lemma}[Resolvability of super-agent detail]\label{resolver}
  Let $A_{\rm super}$ be a super-agent at scale $M_1$, formed by
  coarse graining promises at scale $M_0$. From in equation
  (\ref{cg}), we see that the information lost during coarse-graining $M_0 \rightarrow M_1$,
takes the form of a collection of promises called a directory or index $\pi_{\rm directory}(\tau)$.
By preserving, this directory we may expose and resolve the scale $M_0$ under $M_1$:
\beq
\pi_{M_1} + \pi_{\rm directory}(\tau) \ge \pi_{M_0}.
\eeq
\end{lemma}
How an external agent resolves a microscopic inspection of a super-agent will be
discussed in section \ref{distributive}.

\subsubsection{Promisee coarse graining}\label{cgpromisee}

For scaling of the promisees, the scaling rule is simpler than for the promiser. 
It basically amounts only to a redefinition of scope. Information about the scope
within the promisee is lost to coarse-graining, but can be restored by using
the coarse-graining directory (section \ref{pcg}), through scale transduction (section \ref{transduce}).

\begin{definition}[Scaling rule for promisee]

\beq
\Coarse_n\left( A_1 \scopepromise{b_1}{\sigma} A_{\varepsilon_1},
\ldots,
A_n \scopepromise{b_n}{\sigma} A_{\varepsilon_n}\right)
&=& A_1 \scopepromise{b_1}{\Coarse_n(\sigma)} \Coarse_n\left( A_{\varepsilon_1},\ldots,
A_{\varepsilon_n} \right), \ldots\nonumber\\
&~& ~~~ \;~~~\ldots,
 A_n \scopepromise{b_1}{\Coarse_n(\sigma)} \Coarse_n\left( A_{\varepsilon_1},\ldots,
A_{\varepsilon_n} \right),\nonumber\\
&=&
 A_1 \scopepromise{b_1}{\Coarse_n(\sigma)}  A_{\rm super},\ldots,
 A_n \scopepromise{b_1}{\Coarse_n\sigma)}  A_{\rm super},
\eeq
and
$\Coarse_n(\sigma)$ is obtained by replacing any $A_{\varepsilon_1}\ldots A_{\varepsilon_n}$
with $A_{\rm super}$.
\end{definition}
In other words, the coarse graining of a set of promises, in which
only the promisees are aggregated, leads to the same set of promises
(because promises originate from the promisers), made to the smaller
set of aggregate agents, with only loss of detail and possibly
redefined scope. This is now a question of definition, as the information
about specific promisees and scope membership is lost as agents are
aggregated.

\subsection{Intra-agent language uniformity}

The assumption of the previous sections has been that every agent
shares a single uniform language. As super-agency forms by aggregation
of parts, there is nothing to constrain the language inside the
super-agent boundary.  Hence, we must expect linguistic diversity
inside a super-agent, limiting the interactions on the interior.

\begin{definition}[Language of a super-agent]
For the purpose of all exterior promises, the effective language of a super-agent $S$
must be considered the union of all sub-languages:
\beq
\beta_{\rm super} = \bigcup_i \; \beta_i,
\eeq
where $i = 1, \ldots, \dim S$. The promised language will only be understood with a 
reduced probability however, since the coverage by sub-agents is non-uniform.
\end{definition}
This information about linguistic diversity is also lost under coarse-graining,
hence it must become part of the coarse-graining directory.

\subsection{Irreducible promises at scale $M$, and collective behaviour}\label{irreducible}

In addition to the exterior promises, which emanate from composite
super-agencies, as the remnants of microscopic promises belonging to
component sub-agencies, we may also observe that completely new promises are
possible at each scale, which do not belong to any specific agent
inside the surface.

\begin{definition}[Irreducible super-agent promises]
Let $M$ be an agency scale, and $A_s$ be a super-agent formed
by an aggregation of agents. 
A promise with body $b_s$ made by $A_s$
\beq
\pi_M: A_s \promise{b_s} \Unspec\\
A_s = \{ A_i, \ldots \}
\eeq
may be called irreducible iff there is no set of sub-agents $A_i \in A_s$,
for which
\beq
b_s \subset \bigcup_a  b_a, ~~~~~a= 1, \ldots {\rm all\,promises\,to\,\Unspec}
\eeq
where $b_a$ are existing bodies promised to $\Unspec$ by $A_i$, and $A_i \not= A_s$.
\end{definition}
In other words, if there exists a combination of promises made by one
or more sub-agents (and we assume that the sub-agent is not
the same as the super-agent), then that is semantically equivalent to the full promise made
by the composite agent, when one could say that the super-agent promise
could be reduced to the promise of one of its components. As long as no
single agent, working alone, can make such a promise, it makes sense
to talk about the collective super-agency making a new promise
that is not explicit in the capabilities of its sub-agencies.
Thus, irreducible promises at scale $M$ take into account emergent
effects, and collective effects of agent interactions.

\begin{figure}[ht]
\begin{center}
\includegraphics[width=8cm]{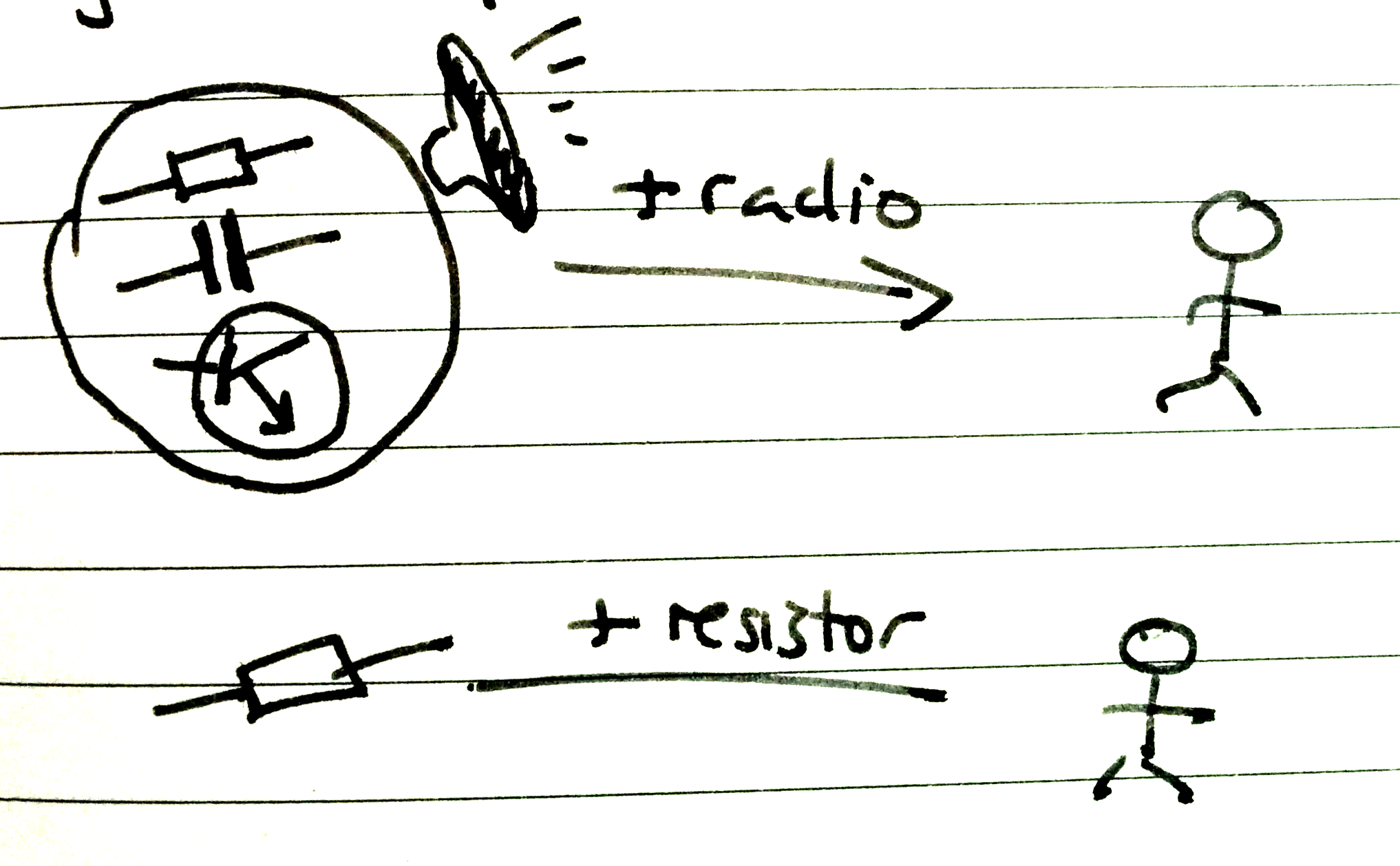}
\caption{\small A radio is an example of a composite `super'-agent, with
interior agencies and promises of circuit connectivity, and exterior
promises to play music. The radio and listener define one agency scale,
but the components and a repair engineer could form another. Scales of
interaction are not necessarily subject to uniform coarse-graining in 
a world of semantics.\label{radio}}
\end{center}
\end{figure}

\begin{example}
  A radio is a composite agent that makes the exterior promise of
  playing radio signals on a loudspeaker (see fig \ref{radio}).
  Interior promises are made by many component agents like resistors,
  capacitors, transistors, etc. The radio has a semantic surface which
  interacts with people as promisees\footnote{In fact, in the case of a radio,
one could argue that it is the outer casing which makes the promise
of being a radio, and that the other components are tenants of
the outer casing agent. I'll return to the issue of tenancy in the latter
part of these notes.}. Does this mean that the
  components inside the radio have to interact with the components
  inside a person (cells, organs etc)?  Not really. Implicitly, this
  might be partially true by the distributive rule, but clearly that is
  not a requirement. Any agency can interact, semantically, with any
  other agency at any agency scale, provided there is a physical
  channel for the communication to work.
\end{example}
Irreducibility is thus a result of collective phenomena in the
underlying semantics and dynamics.  The latter answers classic
objections against the naive reductionism of systems. Yes, we can
decompose systems into a sum of parts, but one must not throw away the
promises when doing do, else it will not be possible to reconstruct
both semantics and dynamics. This is a consequence of boundary
information or systemic topology. 

The form or collective identity of a super-agent can be enough to
signal a promise, by association. This is an emergent promise in the sense
of \cite{promisebook}, i.e. one inferred by a potential promisee rather than
given explicitly by an agency.
Nevertheless, some scaled objects do make promises: tables, chairs

\begin{law}[Reduction law]
  When reconstructing a system from components from finer-grained
  scale $M'$ to a scale $M$, all super-agencies at scales below $M$,
  and their component promises must be retained in order to
  reconstruct the system as the sum of its parts. This is achieved if
  each super-agent promises its coarse-graining directory, providing
  dynamic transparency.
\end{law}
Thus increase of detail downwards may not be at the expense of loss of
upward irreducible information. If we try to view a compound agent as
a collection of parts, with component promises (on the interior of the
super-agent), the component promises revealed do not replace the high
level exterior promise: they are simply prerequisites for it.  The
only reason to disregard irreducible promises belonging to a coarse
grain is because one is focusing on the internals of an agent in
isolation (what one calls a closed system in physics).

If a super-agent has no agency of its own, how can it make a promise?
Clearly, a collection of agents has agency through its members, and
these may or may not have the ability to keep promises related to
scale $M$. So where do promises come from in a super-agent?
\begin{itemize}
\item $+$ promises: the collective appearance of a super-agent, at a
  certain scale, must provide the information to signal a promise.
  The declaration of the promise may or may not come from a single
  agent.  The keeping of an irreducible promise does not come from a
  single agent, by definition.  New irreducible promises are dependent on the
  individual agents, only through their cooperative behaviours.

\begin{example}
  A troop unit promises to surround a house (+ promise). The troop
  leader can make the promise on behalf of the group, but no single
  agent can keep this promise, but collectively they can. In this
  case, the promise would often be given by a team commander, with a
  centralized source of intent, and subordinate agents. However, a
  team can also arrive at this promise by cooperative consensus.
\end{example}

\item $-$ promises: a promise to accept another promise, made by an external agent,
might only apply to certain sub-agents with the appropriate capabilities.
It must be provided as an exterior promise based on the 
by cooperative agreement amongst the agents. 
\begin{example}
  The Very Large Array of radio telescopes in Mexico has 27 receivers.
  Each receiver (- promise) is coordinated with the others so that
  they act as a single super-agent.  No single agent can see what the
  full array can see working together, up to diffraction. Hence the
  combined array can make promises that individual agents can't.
\end{example}

\end{itemize}
A super-agent's {\em agency} is thus conditional on the existence of
its sub-agencies, even though its promises are not locatable in any
single agent.  In both cases above, the promise made by a super-agent
could not be made by a single agency (this is what we mean by a
`host'), or by distributed consensus. Irreducible promises are thus
conditional, not only on the uniform cooperation of the sub-agents
about a single promise, but on their making all the promises that
indirectly lead to the irreducible property.

\begin{lemma}[Irreducible promises are conditional promises, for all scales greater than $M_0$]
  Irreducibility is not expressible directly as a sum of component
  promises of the same type as the irreducible promise, but it is
  second-order expressible in terms of sub-agent promises, by building
  on the existence of these contributing promises.
\end{lemma}
With no promises to build on, a collective agent cannot make any kind
of promise, since nothing can be communicated, and it has no
independent agency. Crudely, a super-agent is indeed the sum of its
parts, as far as agency and promise-keeping are concerned; but, it is
not merely a direct sum as new promises are possible through
cooperation.

If the scope of exterior promises extends to the interior of a
super-agent boundary, that scope becomes ambiguous under
coarse-graining. External observers can no longer see which agents are
make or receive the interior promises, nor is there any way to refer
to them independently after coarse-graining. Observers can only assume
that the scope of a promise includes all sub-agents, but this might
not be the case, and might result in erroneous expectations.
Indeed, there is no reason why the sub-agents would even all
use the same body language. We explore this more in section
\ref{resolve}.

\subsection{Scaling of promise impact, and generalized force}

As we scale agency to deal with larger entities, it seems unreasonable
to expect the impact of a single promise from a sub-agent to have the
same impact on the super-agent as on the microscopic parts before
coarse-graining.  When might we be able to disregard a promise? If
system is stable, i.e. small influence leads to small effect.  Without
going into too many details, I'll sketch how this can be dealt with.

\begin{example}
In dynamics one has the notion of change of momentum (force) and
pressure (force per unit area).  It seems natural to try to construct
such a notion of force density or pressure for promises too.
However, we would note that even though momentum transfer alters in
magntiude, the semantics of momentum transfer do not change, i.e.
Newton's law of momentum conservation does not change.
\end{example}

An effective force (see section \ref{disting}) is one
possible measure of impact. It's attraction is that it works for
semantics and dynamics, though it is clearly modelled after physical
dynamics.  If we tried to scale this relation, we would scale the
promises first, but then we also need to scale the assessment
function. I'll leave that exercise for another time, or as an exercise
to the reader.

The coupling of one scale to another has some coupling strength which
depends on or describes the transmission effect of information between agencies.
In physics we have the law of conservation of momentum as the currency
of influence, and energy as a `stored wealth'. For semantics, we need an impact
if intent where intent is preserved, but not necessarily outcome. This is more
analogous to the conversation of charge in physics.

There is thus a reasonable algorithmic procedure for progressively
disregarding the impacts of certain promises relative to the
scale of certain agencies, as we coarse-grain a system.

The characterization of weaknesses, cracks, and defects, in spatial
promise structures, is an obvious follow-up question to this notion of
semantic impact. If two super-agencies or subspaces come into contact,
could there be catastrophic outcomes by which one region might not be
able to withstand the influence of the other? This introduces concepts
like {\em dynamical stability} and {\em material failure}. Here,
instability transduces the smallest dynamical effect into the
largest semantic importance.

\section{Agent hierarchies}

A full range of agent dynamics should be able to mimic all the
processes of the natural world, as well as artificial behaviours based
on computation. Promise theory's goal is then to explicate the
semantics of these processes. As in the scaling and renormalization of
physics, this leads to hierarchical ideas. However, the implications
for semantics go further than those dynamical ideas, as function is often
tied specifically to a fixed scale.

In the foregoing sections, I've shown explicitly how the sub-agencies
of one (super-) agent can be contained within its boundary, and even
promised to others as a resource, by emission\footnote{Agents can be exchanged, by
  emission from one agent and absorption by another, as the actual
  information of the promise body. This is how one models the exchange
  of forces in quantum theory, via gauge bosons like the photon, or
  gluons, etc.}.  Agency thus exists and interacts in coarse, bounded
`packages', much like Milner's notion of bigraphs (see paper I), and all the time
on top of a substrate of basic adjacency promises that we call
fundamental spacetime.

\subsection{Resolving interior details of super-agent structure during coupling}\label{resolve}

Every time an observer zooms out by coarse-graining, the detail wiped
out by the formation of a grain can be captured as a map called a {\em
  directory} or {\em index} (see discussion in section \ref{pcg}).
Preserving this map can help an external agent to resolve and interact
with the the sub-agencies inside a super-agent's boundary\footnote{In
  information technology, these maps are called variously directory services,
  name services, indirection tables, or data indices.}.

This is paradoxical: in order for an agent to exchange promises with a
super-agent at scale $M$ (which has no physical boundary or form other
than its constituent parts), an external agent perceiving the
effective promise at scale $M$ would surely have to be directed to an available
sub-agent to provide the directory is available to the external agent.

\begin{example}
A bank super-agent might promise to give you cash, but the promise
still has to be given meaning and carried out by an actual bank teller.
The bank itself has no interface to bind to without its sub-agents.
\end{example}
As a fictitious boundary, a super-agent could simply be imagined by the
assessments of an observer, with no coordinated intent of its own.
However, a promise made to or by a super-agent has to be a promise made
to or by its components somehow.  If a promise binding is only conceptual,
this might not be an issue, but if an actual transfer of information is
implied, there must always be a real source and a real receiver. 

\begin{definition}[Transparency]
An agent may be called transparent if it promises an index or directory of all its 
internal sub-agents and their promises.
\end{definition}

In the remainder of this section, we address how such a coupling of
agency scales, given what we know about irreducible promises (see
section \ref{irreducible}), from the dual perspectives of dynamics and
semantics.

\subsubsection{Distribution or dispatch of promises at super-agent boundaries}\label{distributive}

So, what happens at boundaries when a promise is made to a
super-agency?  What happens to the information?
Let's try to answer the question dynamically first,
since everything is dependent on what is dynamically possible.

When we make a new promise at scale $M$ to a super-agent, we need to
understand what this means for the component sub-agencies at a
finer-grained scale within it.  Two possibilities present themselves:
\begin{itemize}
\item {\em Distribution/flooding (broadcast)}: Promise bindings made to a super-agent
  are broadcast or diffused throughout the sub-agents that comprise it, spanning
  multiple agent locations, like the behaviour of a gas or fluid
  {\em flooding} into contact with an interface. 

\item {\em Direction/dispatch (switched)}: Promises are routed to a
  subset of sub-agents, or representative binding sites, in a solid
  state, making an exterior use-promise on the surface is responsible
  for accepting the promise. The routing can be direct from the
  promise to the interior sub-agent (if the super-agent exposes its
  directory), or it can be made via a proxy routing agent inside the
  super-agent (if it exposes only a gateway).
\end{itemize}
Why and how these possibilities should happen at all merits some
further discussion. The details almost certainly depend on the scale
and context. The generality of the questions (and the occurrence of
examples in the natural and technological worlds) is what makes them
most intriguing.

\begin{example}
  Consider an example of an extended super-agent $\{a,b,c,d\}$ bound by some
  cooperative promises, which we neglect to mention here.  These may
  occupy a space of similar extent $\{A,B,C,D\}$, as in figure
  \ref{pathtenancy}. 
  This scenario is a realization of many possible scenarios, e.g. a
  journey in many legs (plane, train, network routing), in which the
  promise of multi-tenant sharing of several sequential host resources
  forms a journey in which several hosts have to cooperate as a
  super-agency (an inter-network cloud) (see figure
  \ref{pathtenancy}). A traveller, (a tenant of the journey) has to be
  authorized for passage by each stage of the journey. This requires
  promises to authenticate credentials to be distributed throughout
  the path, and the collaboration of the hosting agencies in trusting
  the credentials.

\begin{figure}[ht]
\begin{center}
\includegraphics[width=12cm]{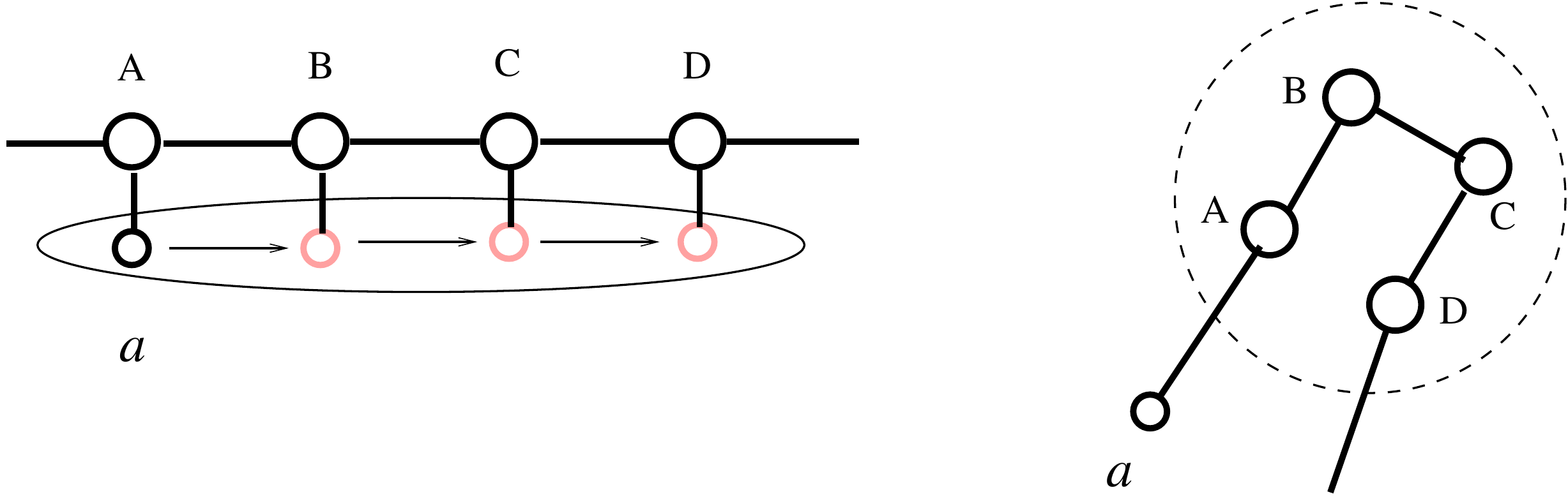}
\caption{\small Path tenancy. The circled agents form a single super-agent that
occupies the corresponding space in the line above. The circles region is
simply a super-agent with exterior promises at its end-points.
An agent binding to each site in transit can only do so at the scale
of the sub-agents but the tenancy binding can be made
as a distributive promise to the super-agent.\label{pathtenancy}}
\end{center}
\end{figure}
\end{example}
\begin{example}
  Directing or dispatching promises, through a specialized agent, is
  like using a reception desk, service portal, in an office or hotel.
  Routing of information requires the underlying adjacency
  infrastructure to be able to direct messages to particular
  addresses.
\end{example}
We may now state the two methods formally, for
clarity:

\begin{definition}[Distributive promise, at scale $M$ (flooding)]
A promise made to a super-agent $A_s$
\beq
A \promise{+b} A_s
\eeq
is assumed made to all agents within $A_s$:
\beq
A \promise{+b} A_i, ~~~~\forall A_i \in A_s.
\eeq
The agents $A_i$ voluntarily accept the promise, if they are suitable recipients,
hence selecting by brute force rather than intentional labelling.
\end{definition}
\begin{example}
In information technology, flooding is used to make a `bus architecture'. Ethernet and
wireless transmission are examples. 
\end{example}
\begin{definition}[Directed promise, at scale $M$ (dispatch)]
  A promise of type $\tau$, made to the super-agent, is assumed directed to a named subset
  of (one or more) members, on behalf of the entire super-agent.
\beq
A \promise{+b} A_i, ~~~~ A_i \subset A_s.
\eeq
The subset $A_i$ voluntarily accept the promise, if they are suitable recipients,
which we may assume is likely, given the intentional direction.
\end{definition}
\begin{example}
  In information technology, dispatch to a directed address is used in
  queue managers, like load balancers, or memory and storage devices,
  to route data to a labelled destination.
\end{example}
Stating these methods does not imply that they are possible in all cases.
To understand whether diffusion of promise information is realizable
we need to understand the small scale adjacency structure of spacetime,
and its effect on promise scope.

\subsubsection{Adjacency between external agents and super-agents}\label{extadj}

Can promises, made by an exterior agent, reach all the internal
sub-agencies in a super-agent, then be comprehended and accepted?  A
promise made to a super-agent has to be transmitted along the network
of underlying adjacencies. 

\begin{figure}[ht]
\begin{center}
\includegraphics[width=9cm]{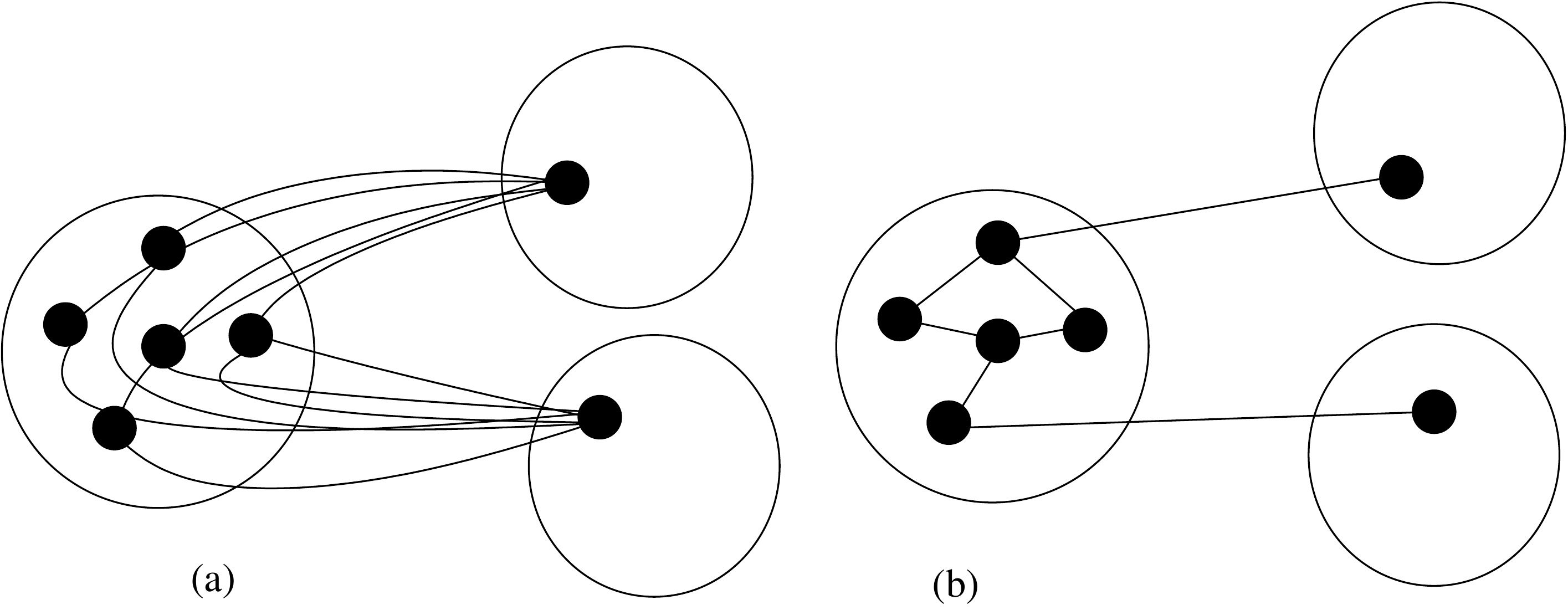}
\caption{\small Regardless of whether promise diffusion has the
  semantics of flooding or directed dispatch, one is limited by the
  actual adjacencies that mediate communication in the space.  In (a)
  agents are directly adjacent or `patched' to all promisers of a particular type,
  and hence the promise diffuses naturally. In (b) adjacency is only
  to specific `front desk' agents, who must then arrange adjacency virtually by
  proxy.\label{direct}}
\end{center}
\end{figure}

Both dispatch and distribution approaches to dissemination and binding
assume that promises can be made directly between the sub-agents of
neighboring super-agents.  The communication needed to make and keep
such promises depends greatly on the network substrate of adjacency
made at the lowest spacetime level.  So the question becomes one about
how spacetime adjacency is wired (see figure \ref{direct}).

\begin{example}
To visualize promises made at coarse-grained scale, imagine
a water authority that promises electricity to a town. Both these agencies
are super-agents composed of many sub-agents. Where (which agent) does the promise
come from, and who receives it? What adjacency allows the promise
to be transmitted?

A generic promise made in the name of the company, depending on its
legal department might make the promise. Every resident in the town is
a potential recipient, as long as they can receive the information
directly or indirectly, i.e. as long as they are in scope.  The
adjacency might be by postal communication and by water pipe.
\end{example}

In figure \ref{direct}a, the external agent promising to a super-agent
is directly adjacent to every sub-agent inside it. In figure
\ref{direct}b, the external agent only connects to a binding site.
How these adjacencies come about, in practice, depends on the phase of
the agents.  There are two possibilities:
\begin{itemize}
\item Agents in a disordered gaseous state (no long range order), agents have no prior knowledge about
one another without random walk meetings, and binding to one another. Thus discovery
is a kind of Monte Carlo search\footnote{This is somewhat like the way the immune system binds
to cell sites.}, and communication is like broadcasting or flooding with messenger agents.
\item Agents in an ordered phase, can assign fixed coordinate
  locations which can be indexed and used to access agents by design.
  This requires a mapping in the index between promises and locations,
  and adjacency between the index and each agent inside the
  super-agent boundary. Thus an index or directory service must act as
  a switch, routing promises to intended destination.

This, in turn, can be done in two ways:
\begin{itemize}
\item by exchange of agent contents at the boundary, and exterior lookup
\item by encapsulation of agent contents and routing with interior lookup
\end{itemize}
\end{itemize}

\begin{figure}[ht]
\begin{center}
\includegraphics[width=9cm]{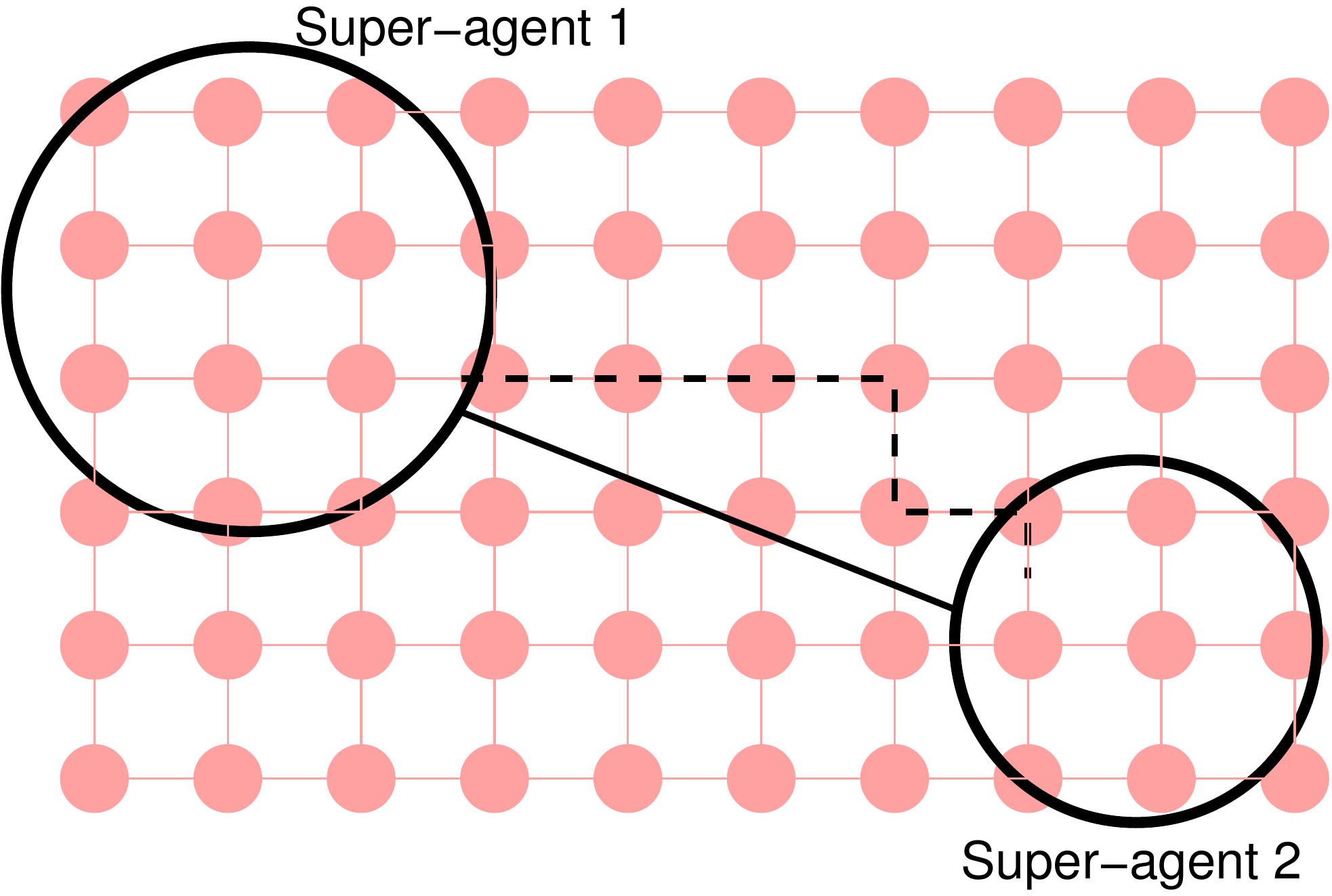}
\caption{\small A schematic `virtual' coarse-grained view of promising, on top of the real
adjacencies between sub-agents. If adjacency is mediated through intermediate
  adjacencies, the entire structure has to support the dispatch and/or
  flooding of messages to keep promises through the substrate of true
  adjacencies (whether gas or solid).\label{adjsub}}
\end{center}
\end{figure}

\subsubsection{Semantics of promise scope for super-agency}

The semantics of scope need to be clarified during scaling, since
scope represents the boundary of information about a promise's intent,
and the ability to distribute a promise depends on the underlying
adjacencies of spacetime. If two agents are not adjacent, they might not
be able to occupy the same scope.

\begin{lemma}[The scope of a promise to a super-agent]
Consider a promise made to a super-agent $S_M$ at scale $M$: 
\beq
\pi: A \scopepromise{b}{\sigma_M} S_M.\label{boundary}
\eeq
Without the coarse-graining directory $\pi_{\rm directory}(S_M)$ for
super-agent $S_M$, the scope of $\pi$ is only defined to the boundary
of $S_M$. It is not possible to say which sub-agents of $S_M$ are in
scope of the promise. If access to the directory is promised
to the promiser:
\beq
S_M \promise{+\rm directory} A\\
\sigma_M \rightarrow \sigma_M + \sigma_{\rm directory} \label{avscope}
\eeq
If agent $S_M$ promises access to its coarse-graining directory, 
an external agent $A$ can infer the scope of a promise in terms
of the sub-agents of $S_M$.
\end{lemma}
The promise could be visible to all the sub-agents $A_i \in S_M$,
which are adjacent to $A$, as well as the extra scope $\sigma$. However,
a promiser can say which agents are reachable by a promise message
if and only if the directory $\pi_{\rm directory}(S_M)$ is available.

The implication of this is that promises do not scale automatically by
replacing an agent with a super-agent: the scope of a promise made to
a super-agent is not necessarily distributive, because of the loss of
information in coarse-graining.

\subsubsection{Scale transduction: coupling to a super-agent boundary (gateways and advertisements)}\label{transduce}

If all the specific actionable agencies are concealed by a super-agent
boundary, how can any promise message reach them from outside?  The
sub-agents are the agencies that must ultimately act to keep any
promises of the super-agent.  This becomes a question about information
transmission. Loss of transparency, i.e. loss of information both about
adjacency and agent type hinders promise resolution.

As seen in section \ref{extadj}, the underlying adjacency of spacetime
remains even under the coarse-graining of super-agency, though it
might not be visible inside a boundary.  A promise could thus reach
the sub-agents by flooding, or by direct dispatch, following the
adjacencies within the boundary, provided sufficient adjacenct ecists.
Dispatch can be directed by the promiser (if transparency is granted
to an external agent by access to the coarse-graining directory) or by
an intermediary agency (acting as a relay gateway) within the
super-agent. It is assumed that there are no changes to agent semantics
simply by aggregation:

\begin{assumption}[Promisee autonomy is preserved in super-agency]
Once a promise reaches a sub-agent, it is up to the sub-agent to
accept the promise or not, and behave accordingly, unless it has
voluntarily promised to subordinate itself to another agent. 
Even if the promise is transmitted as an imposition, autonomy can never be
violated.
\end{assumption}
Thus the scaling of cooperation remains does not change the rules of
autonomy; the `voluntary cooperation' assumption of autonomous agents
persists.  It is up to sub-agents to use any
promise made to them. However, this does assume that they are in the
scope of such an external promise.  Hence to ensure the coupling of an
external agent to a super-agent, transparency has to be restored or relayed.

\begin{figure}[ht]
\begin{center}
\includegraphics[width=10cm]{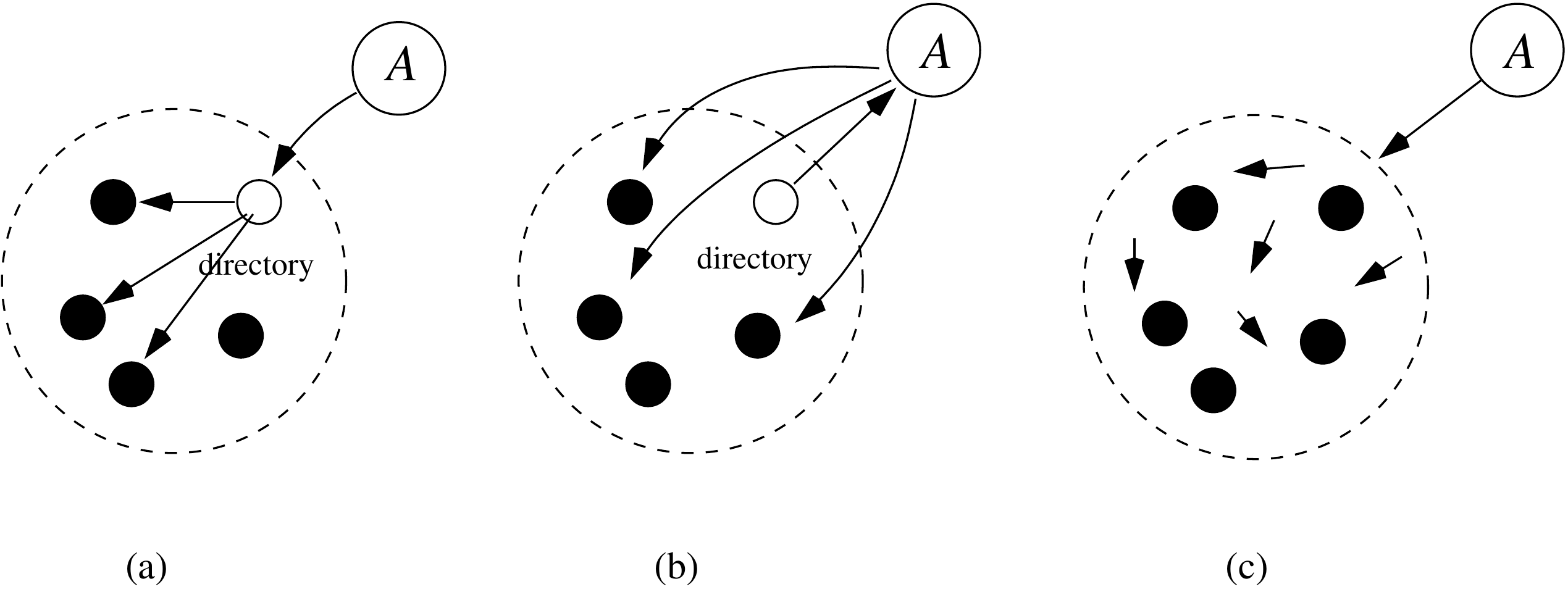}
\caption{\small Scale transduction for incoming promises to a super-agent: (a)
the coarse-graining directory is internal to the super-agent and the
external agent makes its promise to a gateway `receptor'; (b)
the coarse graining directory is exposed and makes the super-agent transparent
to the external promiser so that it can promise directly to the sub-agents;
(c) there is no directory, and internal information is lost. Promises
are flooded to all sub-agents, and may or may not be picked up. The efficiency
of flooding depends on the solid or gaseous state of the super-agent.\label{directory}}
\end{center}
\end{figure}
Under coarse-graining, a super-agent is seemingly replaced by only its
boundary, possibly with a promise to access its index/directory of interior information.

Promises from external agents outside the boundary can only refer to the
super-agent as promisee or body-tensor coordinate.  How then would the
internal sub-agents know what to do with the promise?  How do they
find one another?  

In order for one scale to couple to another scale, we can thus
introduce the idea of a scale transducer: a combination of necessary
and sufficient conditions for promises made {\em to} a super-agent
boundary, to be resolved without mentioning any sub-agencies.

\begin{example}
  What part of a radio makes the exterior promise of being a radio,
  rather than a collection of electronic components? Whether the device
  is switched on or off, its function is not clear without a promise.
  The agency that explains this is usually the packaging of the radio,
  i.e. the casing.  It might further be packaged in a box, but that is
  not a part of the radio itself. In this case, the enveloping a
  casing becomes an agent with a promise that tracks the super-agent
  boundary\footnote{I've usually drawn agents schematically as atomic dots, 
but the shape of an agent is not defined. Nothing prevents it from appearing as a shell.}.
\end{example}

\begin{definition}[Scale transducer]
  Let $A_s$ be a super agent at scale $M = \{ A_s \}$, and let $\pi_s$
  be a promise made $A_s$, by any agency $A$ (see figure \ref{directory}).  Recalling that the
  scope of a super-agent includes all agents inside it, and coarse
  graining limits scope (see section \ref{cgpromisee}), we define a
  scale transducer by:
\begin{enumerate}

\item A number of promises addressed to the super-agent boundary, along exterior adjacencies.

\item One or more sub-agents that make exterior use-promises, within the super-agent, to accept the promise
made to the collective:
\beq
\exists A_i \in A_s : A_i \promise{U(\pi_s)} A.
\eeq

\item An index for directory $\pi_{\rm directory}$ is promised
  externally to agents outside the super-agent boundary. This provides
  them with information about how to address their promises.

\item If there is more than one agent adjacent to the super-agent,
  i.e. more than one possible agent that can make promises with the
  exterior, these agents can be located through the directory.

\item One or more sub-agents in $A_i$ play the role of gateway and dispatcher, to conditionally forward messages
  to interior agents with matching use-promises. This is a standard entry point for the agent, e.g. it is the 
agent adjacent to all exterior agencies, thus forming a skin or boundary between exterior and interior agencies.
\beq
\exists A_i, A_j \in A_s : A_i &\promise{U(\pi_s)}& A\\
A_i &\promise{+\pi_s| \pi_s}& A_j ~~~i \not= j
\eeq
In this case, the super-agent must advertise the location of one or
more gatekeepers, entry-points or directory agencies that promise to
relay information.
\begin{enumerate}

\item The sub-agents inside a super-agent boundary of a given type $\tau$
are advertised in the index/directory $\pi_{\rm directory}(\tau)$. They are symmetrical
with respect to {\em what} they promise (promise type $\tau$), but
they might not necessarily be equivalent in {\em how much} they will
promise (their promise bodies might differ in all details except the
type). This means that the directory must advertise any promises to
this effect also.

\item If a gateway is used as a proxy relay (see figure \ref{directory}a),
  it must additionally make promises that select promisees from the sub-agents
  inside $A_s$,
  e.g. policies might be distributed (by flooding) or directed by dispatch.

\item Gateway agents might also need to translate between the language
  $\beta_s$ assumed for the super-agent, and the language(s) of
  recipient sub-agents $\beta_i$, as part of transducing through the
  opaque boundary.

\end{enumerate}
\end{enumerate}
\end{definition}
We may note that messages from an external agent might be forcibly constrained to
a particular route by spacetime structure, i.e. by a limited adjacency (a
bottleneck, as in section \ref{extadj}). 
Also, a gatekeeper need not be a single agent, or even a localized cluster
in the role of gatekeeper.  It could, itself be a fully distributed
collective agency, embedded within a specialized set of sub-agents,
and coordinated by mutual cooperation.  e.g. like a cellular skin.
This idea of exterior agents binding to specialized `docking sites'
leads us naturally to consider the idea of {\em tenancy} in semantic
spacetime structures. Indeed, I'll return to this later in these notes.

From the list of requirements for transducing promises between
scales above that the following corollary applies:
\begin{lemma}[Dynamical requirements for coupling between external agent and super-agent]
  Any agent $A$ can probe scale information in a super-agent, at
  scales finer than its boundary, provided the super-agent promises its
  coarse-graining directory:
\beq
\pi_{\rm directory}(S_M): S_M &\promise{+\rm directory(S_M)}& A\\
\overline \pi_{\rm directory}(S_M): A  &\promise{-\rm directory(S_M)}& S_M
\eeq
\end{lemma}
Alternatively, we can say that a coarse-grained agent $S_M$ may be
made transparent by promising its coarse-graining directory.

\begin{example}
To couple to a single interface in an electronic device, 
there has to be an exterior promise to bind to (e.g. USB), and
an address of the agent inside.
\end{example}
\begin{example}
  In order to affect a nucleus within an atom (e.g. NMR), a field
  basically floods its promise to all sub-agencies blindly, hoping to
  excite the resonance (receptor use-promise).
\end{example}

\subsubsection{Addendum on scaling of scale transduction itself: queue dispatch}

Since scale transduction is itself a dynamical process, dependent on
underlying spacetime, its efficiency is also subject to scaling
issues.
Consider first the coupling issue from a semantic perspective, of
interfacing instead.  When a promise is made to a super-agent boundary
(as in equation (\ref{boundary})), the promise information has to go
to an agent that is listening.  A super-agent is not such a real
agent, it is only an abstraction. This suggests that, in the case of
super-agents, the boundary itself might be represented by an explicit
agency that can perform routing and forwarding of messages between the
super-agent's fictitious boundary and its sub-agencies.

While this direct dispatched routing of promises makes sense
semantically, dynamically, the idea seems contrary to the notion of
scaling: to replace a scaled mass of agents by a single gatekeeper or
router creates an obvious bottleneck and fragile dependence. This is
the cost of coarse-graining, especially in a discrete spacetime, and it suggests
that the grain size should never become too large, else a super-agent
becomes hindered by interfacing issues.

\begin{example}
  In queueing theory, a dispatcher is an agent that processes a queue
  of incoming messages and routes them to a service
  agent\cite{kleinrock1,gross1}.  Load balancers introduced into
  networks as `middle boxes' are single-agent dispatchers to multiple
  sub-agents within an super-agent of servers. These middle boxes
  break the equivalence of the system under re-scaling, by forcing all
  adjacency through a single route.
\end{example}
\begin{example}
  A reception desk at a company accepts promises and information on
  behalf of the collective organization of sub-agents. An agent works
  at this desk to receive messages, and dispatches, routes or forwards these
  messages to other relevant agents using an internal directory
for responsibility.
\end{example}

An alternative is for the super-agent boundary to promise flooding
contact and allow the surface agents to coordinate internally, as
redundant gatekeepers, each deciding how to resolve what happens if
multiple agents receive the same message. If promises are not, by
their nature, exclusive to a single gateway, then oordinating
exclusivity adds $N^2$ complexity of promise coordination.
\begin{example}
Using the coarse-graining directory to give transparency, any external
agent could perform its own dispatching / load sharing without loss
of scalability. For example, a directory service used by software could
replace `middle box' load balancers in software, with the help of a software
interface used to select a specific sub-agent server within the super-agency of all servers.
\end{example}
Let's summarize the implications of loss of scope from the past previous section.
\begin{itemize}
\item Routing of messages to internal agents through a single gateway breaks scale invariance.
\item A scale transducer may be introduced, for mediating interactions directly by granting transparency,
using a directory.
\item Directory promises allow external agents to look up lists of sub-agents encapsulated within
enabling scale transduction, i.e. a kind of microscope for crossing the semantic scale boundary.
\end{itemize}

\subsection{Addressability in solid structures, and the tenancy connection}\label{addresses}

To see how we can route messages back to the specific sub-agents, we
need to understand addressing.  The ability to give every agent a
predictable address, and then be able to have messages forwarded to it
uniquely, using that address, depends on a number of promises being
kept. To illustrate this, let's construct a semi-lattice by iterating a simple
asymmetric message pattern.

\begin{figure}[ht]
\begin{center}
\includegraphics[width=6cm]{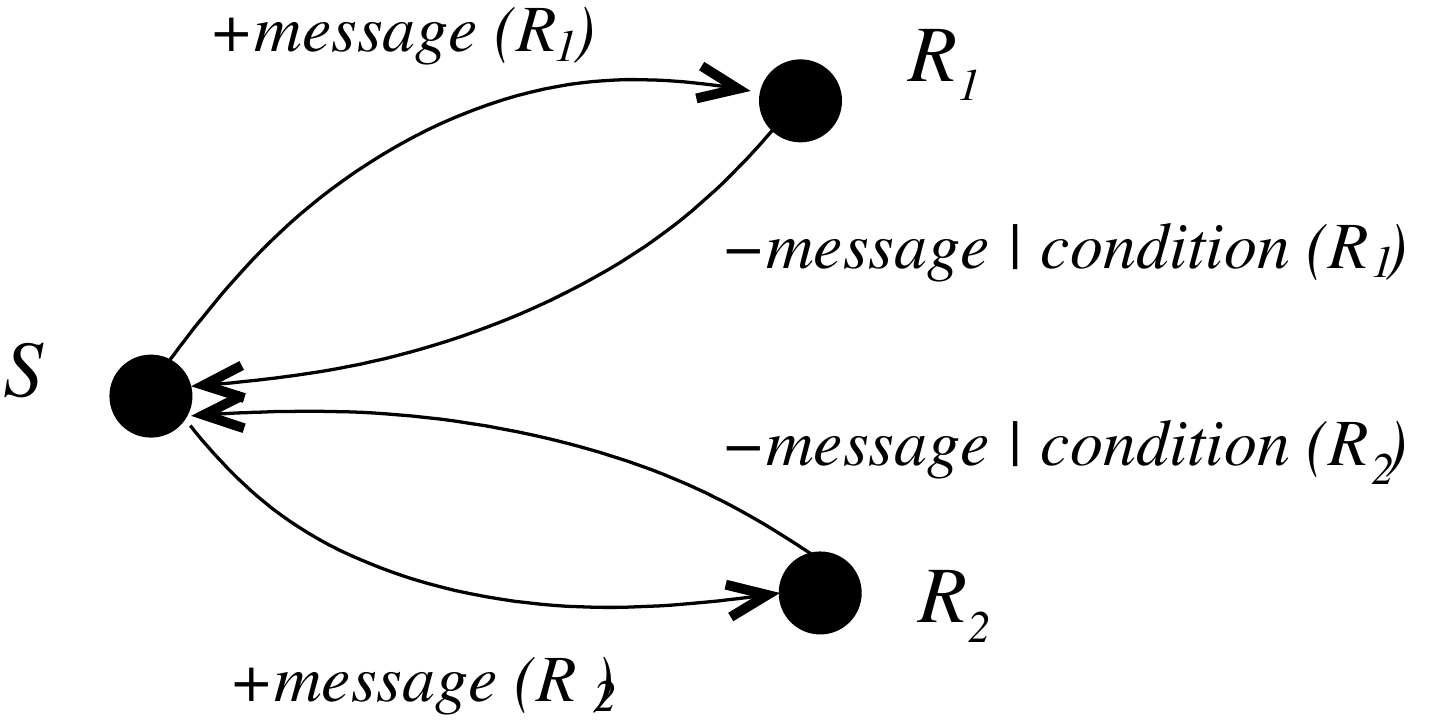}
\caption{\small Two promise bindings, leading from a source agent $S$
to two different recipient agents $R_1$ and $R_2$. This forms the basis of
a routing structure for address decomposition, where part of an address
can lead to selection or rejection of a particular route.\label{forward1}}
\end{center}
\end{figure}

Let $S_i$ and $R_i$ be two types of roles for a set of agents $A_i$, and
consider bindings between two kinds of promise:
\begin{itemize}
\item A vector promise to dispatch messages to a recipient agent $R$: 
\beq
S \promise{+\rm \text{dispatch message to R}} R
\eeq
\item A use-promise to accept messages from a source $S$, only if its address is compatible
with the agent's conditional expression for forwarding:
\beq
R \promise{-\rm message | \text{addressed to me}} S
\eeq
\end{itemize}
Building on these two promises, we may construct uni-directional
adjacency-like binding for use as a template to build larger
structures.

In the figure \ref{forward1}, we see a node with two such promise
bindings sporting different conditionals in different directions.
Notice how the choice to forward a message from $S$ to $R$ is a
voluntary act by $S$.  It can send in different directions to
purposely separate messages, or it can send along different paths for
traffic management or load balancing.  Note also that the difference
between a flooding promise (sending messages to all recipients) is
simply a scalar version of the dispatch promise, in which we take away
a target from the promise body, i.e.  without exclusivity to
promisee/body vector. The result is the same, but the efficiency
is compromised; efficient routing is assisted by long-range cooperation, and
ultimately by long-range order.

The receiver $R$, has the last word in accepting a message. So no
message will arrive at the wrong location no matter whether it was
forwarded by targeted dispatch or broadcasted to all agents.  Agents
have to promise their unique identities, both so that they may be
recognized by neighbours, and so that they can recognize messages
directed to them.

The consequence of creating an ordered tree from these promises is
to create a dumb filter, which routes messages along a unique path
depending on their address.

\begin{example}
  Coin sorting machines create unique pathways the sort and select
  different sized coins, allowing the to roll only one way through a maze
of pathways. This is the basic principle of a semantic sorting process.
The pathways for a treelike structure, and the end points of the tree
all have a unique address. By placing a coin of a particular kind into the
process at the root, it is like placing a message with an address (the type
of coin), and having it sorted until it reaches its destination. Coins
with the same address will end up at the same location.
\end{example}

Armed with this tool for spatial sorting, we may iterate these promise
patterns to generate semi-lattices of greater size. Having a
coordinate system within a super-agent boundary, for example, would
allow agents to be located in a targeted manner, assuming only that
they are connected.  To iterate the pattern, we simple make each
receiver into a source for the next iteration, and so on.

\begin{example}
  Figure \ref{forward} shows to iterated patterns formed from
  branching source-receiver iteration. The first (a) is a simple tree
  structure.  Every leaf node of the tree has a unique address, and
  can be reached from the root by a unique path. This is the property
  of trees (and spanning trees).

\begin{figure}[ht]
\begin{center}
\includegraphics[width=11cm]{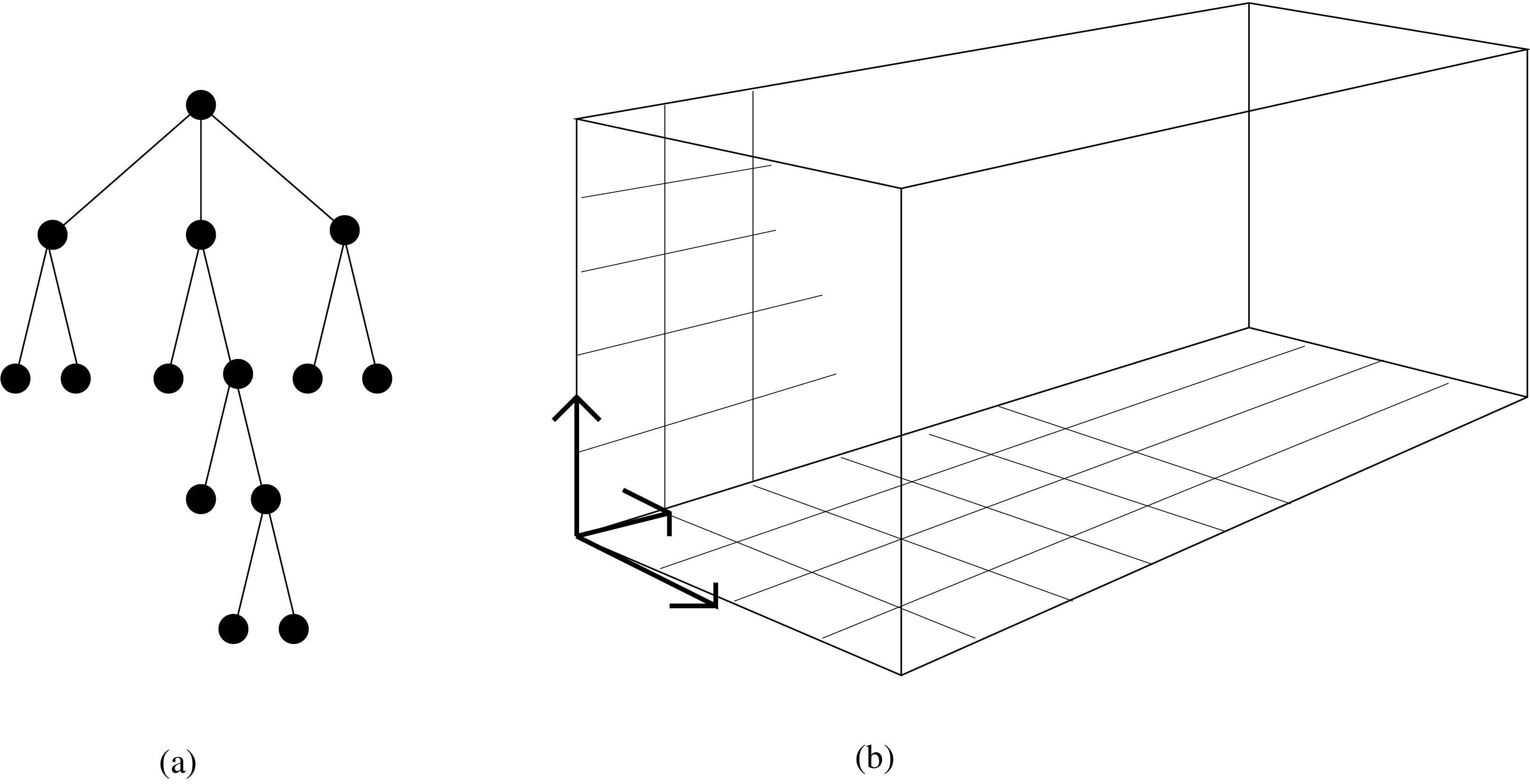}
\caption{\small Iterating a pattern of promise bindings over a
  spanning tree allows unique labels to be associated with
  destinations in a graph. The symmetry can be a minimal tree as in
  (a), or, by adding redundant links it can be turned into a
  semi-lattice (b). Arrows are not shown for simplicity; however, both
structures are uni-directional.\label{forward}}
\end{center}
\end{figure}

The second case (b) is also a tree formed from three-way branchings $S
\promise{+b(R_1, R_2, R_3)} \{R_1, R_2, R_3\}$, iterated homogeneously
and isotropically. The agents then fall into a three dimensional,
Cartesian arrangement, which we may call the Cartesian semi-lattice.
By filling in some redundant promises from each point, one can arrange
multiple routes from any node to any other, but still only in one
direction (radially outwards from the origin). 

By making the promises in each of the three dimensions sort forwarding
of messages according to a different component in a vector tuple,
forwarding can be encoded as a purely local operation\footnote{This is
  essentially how routers and switches forward datagrams in
  information infrastructure.}, e.g.  forward only if the tuple value
is greater than the current tuple address of the agent for the current
lattice location.
Furthermore, by completing the reverse
direction, as a mirror image, with opposite semantics, addresses can also
be navigated in the opposite direction, completing the lattice.
\end{example}

\subsection{Conditions for a uniform coordinate covering of an ensemble of agents}

What are the conditions for being able to address agents using
contiguous coordinates without loss of locatability? This is a slightly
different question to the one about naming and promise body continuity,
because it requires us to preserve the partial ordering of agents in a
lattice. It is helpful to explain addressability by introducing two concepts
that cover the semantic and dynamic aspects of location:

\begin{definition}[Semantic addressing]
An agent is said to have a semantic address if it is labelled only by a tuple of names
that do not form part of an ordered pattern. Semantic addresses contain no
relevant kinematic or dynamical information about an agent's location in a space.
\end{definition}
Semantic addresses act only as sign-posts, and a directory is needed to 
map which adjacency will eventually lead to the named agents. This is the approach
used in Internet routing.

\begin{example}
  Internet addresses (aka IP addresses) are semantic addresses,
  despite being composed of numbers, because they have no requirements
  of spatial order. Any agent can assign itself an address with any
  number, and these numbers do no not imply information about where
  agents can be found.  In order to locate IP addresses, a directory
  called a routing table is needed, which maps the random numbers of
  the addresses to physical adjacencies of the cabling. This is the
  function of a router or switch.  The advertisement of these local
  directories is performed by services, which are called routing
  protocols (BGP, OSPF, RIP, etc).
\end{example}

\begin{definition}[Numeric (metric) addressing]
An agent is said to have a numeric or metric address if it is labelled by a tuple of values 
in which each value map to a unique integer. Numeric addresses represent kinematic or
dynamic information about a space.
\end{definition}
Agents with numeric addresses are partially ordered in a
multi-dimensional lattice. The addresses form a coordinate system in the usual
sense of mathematics.

In lieu of setting up a proof, I'll hypothesize this informally for now:
\begin{lemma}[Hypothesis: Conditions for a uniform coordinate covering of an ensemble of agents]
placement of fixed and ordered address labels on an ensemble of agents,
in a voluntary cooperation structure

\begin{itemize}
\item Fixed locations.
\item Addresses ordered by location.
\item Promises to sort and relay messages to destination.
\item Long range order in address promises, i.e. cooperation in
  behaving uniform sorting/routing to addresses.
\item Overlapping regions of $\beta$-language relevant to address
\end{itemize}
\end{lemma}
This does not refer to any particular topology, so it can be solved by
multiple adjacency patterns.  A suitable address-sorting process
can be satisfied in a number of different ways, reflecting the
encoding of the address in relation to the structure of spacetime.
Network structures are typically tree-like, and IP addresses are
prefix-based with distributed routing tables based on tree branching
assumptions. Toroidal structure and Cartesian lattices using tuples
are based on a pre-ordered layout, as in a warehouse, for instance.

\begin{example}
  In networking, regions of a network (like the Internet) can become
  cut off from the rest by a loss of routing information. Because the
  Internet is a gas, with semantic rather than numerical (metric) addressing,
  each super-agent boundary must contain a routing table that points
  to the next signpost to the destination.  To prevent this from
  getting out of hand, a `default route' is normally used as a
  wildcard (go this way to find any unspecified agent $\Unspec$),
  allowing regions to compress information about how to reach
  non-local agencies by handing off to centralized routing hubs.
\end{example}

There is an interesting suggestion here, that semantic naming tends to
favour the formation of a hierarchy in order to scale.  Such a
hierarchy is unnecessary for metric naming. This warrants further
study.

\subsection{Efficiency of addressing in a semantic space}

Consider a network of agents, with unique names and which are all
interconnected by a sufficient number of adjacencies to allow
full percolation. Suppose these agents promise to cooperate in relaying
messages to one another, by passing the message along one of their
adjacencies until it reaches the unique name (i.e. address).  How much
information has to be available to each agent in order to know how to
forward messages to every other agent?

The maximum size of directory information may be computed as a sum
over every location, which keeps a table of every other named agent in
the space, paired with the `next hop' neighbouring agent that brings
the message closer to its destination. This applies for every agent in
the space.

\begin{itemize}
\item If every agent is independent, then every agent needs a list of
all $N-1$ agents, with
\beq
(\text{Agent name, direction of agent}) 
\eeq
and a direction in which to forward. In total memory required for this information is of order $N(N-1)$
in the number of agents.
If there are few agents, this is easy. If there are many, the search
cost rises linearly, and the distribution of information by
flooding brings high cost.

\begin{example}
This is exactly like routing in the Internet, imagining there
are no network CIDR summarization prefixes, which would correspond to super-agent boundaries.
\end{example}

\item If one can replace atomic agents with super-agents, which can handle their
own internal forwarding, then the amount of information one needs to exchange
is less. 
\beq
(\text{Super-agent of every node, direction of super-agent}) 
\eeq
Aggregation of clusters leads to a cost of order $(N-1)\log N$ memory.
Scaling of addressese now depends on the ability to delegate responsibility to
agents `further down the line' by using super-agent
container names to route messages, as in the coin sorting machine
or lattice. This leads naturally to hierarchical naming and routing.

\begin{example}
Postal addresses refer first to town, then street, then building, and so on.
By referring to larger container boundaries first, one can delegate the
detail of finding the final destination to agencies within the boundary
of the super-agency, e.g. the town. This assumed encapsulation comes at a price, however.
It introduces inter-dependency into the end-to-end communication.

Today, postal addressing also now uses metric post codes, which are non-hierarchical.
Given modern computational resources, a simple brute force approach can be used
to look up these codes from directory information.
\end{example}
\begin{example}
  This is like IP routing with CIDR prefixes.  Internet (IP) addresses
  were originally designed to reduce routing cost by aggregated along
  certain prefixes, originally of fixed length (called class A, B, C
  networks).  By grouping addresses under a smaller number of prefix
  patterns, and assuming that all such addresses were contained in the
  same super-agent boundary, routing tables could be kept small.
  Later, as these limited prefixes became consumed, they were
  subdivided into more, causing routing table growth.
\end{example}

\item If there is a regular lattice, e.g. $(x,y,z)$, with long range order, and tuple addresses,
then the amount of information is now of order 1.
It is like asking which way is `up'? Irregularities
(like holes) can be routed locally at no extra cost.
\begin{example}
For a Cartesian lattice, one knows left or right, forwards or backwards for
each address because of the ordering of the integers.
\end{example}
\end{itemize}

Delegation, or deferred evaluation, is an attractive idea for scaling
linearizable searches, however we must note that, an agent cannot ask
another agent for help without already being able to know how to reach
it; so, with no basic pattern to compress by, there is no way of
centralizing this routing information in the manner of a
coarse-graining directory.

Consider the following worst-case scenario in which every agent has a
random name, i.e. a spacetime addressed by random numbers. Then, every
location has to have a complete map of every other location, with zero
possible compression.  The need to flood all that information to all
parts of spacetime adds a significant cost to promise keeping, and
might exceed the capabilities of any or all agents.

\begin{example}
Instead of handing out metric addresses to visiting mobile devices (as
a parking lot, or hotel, would do to its visitors), the internet hands
out local semantic addresses (by DHCP), and tries to map them into its
routing infrastructure. This makes sense for ephemeral gas-phase devices,
but is quite inefficient for the repurposing of solid phase agents, like
virtual machine slots, or process containers.
\end{example}

In practice, we see, from the stages of address scaling above, that
the information is compressible only if each agent can replace a
collection of addresses with a single promise. Hence to coarse-grain
addresses into a hierarchy of containers, without loss of information,
we need to restore the information lost using a directory at each
super-agent boundary\footnote{This corresponds roughly to the design of the Border Gateway
  routing tables, for IP addressing, as known from the Border Gateway
  Protocol BGP.}. If we want to keep directory information small, we
need {\em long range order} in the structural addressing promises (and
presumably the adjacencies too) to enable logarithmic aggregate
summarizability. Asymmetric tree structures can be adequate, but
bi-directional lattices, like a Cartesian lattice, are better
still\footnote{One can see the historical reasons why the Internet was
  not designed as a Cartesian lattice, but modern datacentre fabrics
  still have the opportunity to repair this choice.}.

\subsection{Summary of agency properties in semantic spaces}

The scaling behaviour described thus far allows us to `inflate' (or
scale-up) any functional arrangement of promises by substituting an
arbitrary agent with a super-agent composed of sub-agents making similar
promises. Then, one may define the exterior promises in
such as way as to integrate the sub-agent members seamlessly to agents on the outside
of the super-agent boundary.  
There is a progression: 
\beq \rm agent \rightarrow \text{super-agent}
\rightarrow role \rightarrow subspace 
\eeq 
In other words, as an algorrithm to scale given a single agent, we replace it with a
black-box super-agent.  Then we proceed to fill it with multiple
sub-agents that are connected to the outside agents by exterior
promises. These similar sub-agents are symmetrical with respect to the
outside, so they form a role by association.  Eventually, as we scale
each of the original agencies and connect them to scale the promises,
what remains is a set of non-overlapping sub-spaces, one for each
agent, embedded in a larger semantic spacetime\footnote{This is
  essentially the method of top-down decomposition used in procedural
  programming, but with constraints that allow parallelized scaling of
  execution. One could imagine programming and representation
  languages that support this kind of model in many different walks of
  life. This is probably how we should be teaching programming, and
  service management, instead of the linear imperative models of
  today. In paper III, we'll see why the linear storyline approach has
  its own naturality.}.

\begin{example}
Name-spaces, walled/gated communities, zones of privilege, service providers, etc are examples
of agencies which scale from a single agent to collections bounded by some kind
of contact surface.
\end{example}

Semantic spacetime (agents)
have a number of scalable properties:
\begin{itemize}
\item They have discrete languages of intentions, easily translatable, in order for promises to be effectively communicated.
\item They can be observed and interpreted at a multitude of scales, at the behest of an observer.
\item They can effectively cluster their own promises into coarse-grains, through cooperation.
\item The number of agencies can grow or shrink, i.e. spacetime itself grows or shrinks, as new points are added or retired.
\item There are simple rules for transforming from one agency scale to another, analogous to renormalization transformations.
\item Promises behave like tensors in general, with directionality.
\item Causal influence is passed by vector promises, and principally through use-promises, by the principle of autonomy.
\item Biological organisms offer a useful measuring stick for spacetimes with strong semantics.
\end{itemize}

So far I've focused on preserving symmetries and semantics, while
piecing together the underlying connectivity of space.  The asymmetry
in these promises for routing to fixed addresses has a general
utility, and it can be associated with the idea of {\em tenancy},
which is the subject of the latter part of this paper.  Functionally, this
asymmetry is the most important tool for making anything happen in
space or time, and is worth exploring in more depth.

\section{Occupancy and tenancy of space}

Let's now turn to a different topic: how to fill the space we've built
up.  A semantic space is richer in structure than its underlying
connective graph so it contains information that goes beyond pure
adjacency. In particular, as we add autonomous observers with their
own agency, we quickly arrive at the need for agents to extend their
realm of autonomous control through {\em occupancy} and {\em
  ownership} of resources. The question of occupancy and tenancy are
thus about how we draw the boundaries of agency on a background of
spatial adjacency So far we've focused on symmetry and scale in
discussing agency, however strong functional semantics are a result of
asymmetry, hence we must now pursue the effects of broken symmetry.

\subsection{Definitions of occupancy and tenancy}\label{tendef}

Tenancy goes beyond simple aggregate membership in a cluster.  A
tenant is understood to be an agent that `occupies' or utilizes a
resource or service, provided by a host, often in a temporary manner,
and for mutual benefit (symbiosis).  Tenants have separate identities.
When we think of tenancy in every day affairs, we do not usually
imagine a tenant as merging with its host, and becoming a part of it
(though merger and acquisition is certainly a process one can discuss,
as absorption).  Tenancy is rather an association between two separate
agency roles (host and tenant), each of which retains its autonomy.

To relate this to our spacetime discussion, consider the following
question (which, at first glance, might seem purely facetious): {\em
  does a suit occupy space when no one is wearing it, or does space
  occupy the suit?} The space inside a suit is simply empty before
someone climbs into it. Try replacing `suit' with `car' and `wear'
with `sit inside'. 

This peculiar question is closely related to the considerations
surrounding the kinds of motion described in paper I, section 5.12.
Since we are modelling space as a resource, this is not only a
meaningful question, it is essential to understand what kind of volume
a suit occupies.  Does the presence of suit matter replace space,
occupy it, attach to it, or overlap with it? These have different
semantics.  

Recall, from paper I, that motion of the second and third kinds
distinguish between the idea that space and its occupants are either:
ii) a visitation by a separate entity, or iii) a change in the state
of the same entity. In other words, does space get filled by matter or
does matter transform the nature of space?

To address some of these issues, we need to formulate defintions using promises,
building up the distinctions in a rational way.
Let's begin with occupancy. Its semantics are difference from mere presence,
as there is an assumption of valency.
Within the scope of promise theory, we can define the following:
\begin{definition}[Occupancy]
An asymmetric association of one agent (the occupier) with
another representing a host (the location) at agency scale $M$,
in which the valence of a promise make by the host is reduced by
its binding to the occupier. The resource $R$ may be any scalar, vector or
tensor type:
\beq
\host_M &\promise{+R\#n}& \Unspec\\
\occupier_M &\promise{-R\#1}& \host_M
\eeq
In other words, a host makes a finite promise $+R$ to a number of agents in scope,
and each occupier reduces the valency by making a use-promise $-R$
\beq
\valence(R,\host) =\valence(R,\host,\occupier) + 1
\eeq
\end{definition}

\begin{example}
Our understanding of the semantics of occupancy has many possible interpretations. Here are some examples:
\begin{itemize}
\item Occupation of a territory without necessarily being there. e.g. a table reservation.
\item Occupation of space.
\item Occupy a car, a suit, a dress.
\item Occupy a time slot in a calendar.
\item Filling a space with something.
\item In physics, bosons can occupy the same space, like voices in a
  song, but Fermions have exclusion, like the bodies in the choir
  themselves they occupy space.
\end{itemize}
\end{example}
From here, we may state a basic template for tenancy for application to a variety of special cases:
\begin{definition}[Tenancy]
  Tenancy refers to the conditional occupancy of a location, by an agent, together
  with the provision of one or more services by the host, which may be
  considered a function $f(R)$ of resource $R$. These services are provided conditionally on a promise of $C$
from the tenant: 
\beq
  \host_M &\promise{+R\#n | C}& \Unspec\nonumber\\
  \host_M &\promise{-C}& \tenant_M\nonumber\\
  \tenant_M &\promise{+C}& \host_M\nonumber\\
  \tenant_M &\promise{-R\#1}& \host_M\nonumber\\
  \host_M &\promise{+f(C,R) | -R}&  \tenant\label{deftenancy}
\eeq
This is the basic template for tenancy, which may be extended by additional promises.
\end{definition}

\begin{example}
A landlord promises a rentable space for a single occupant $+R\#1$, conditionally on the signing
a contract of terms (i.e. the promise to abide by terms and conditions) $+C$.
\beq
L \promise{+R\#1, f(C,R) \, |\, C} \Unspec
\eeq
A tenant quenches this exclusive resource, by signing up and promising the terms:
\beq
T \promise{+C, -R} L.
\eeq
The terms and conditions contain a composite promise body, detailing the services
$f(C,R)$ offered as part of the promise:
\beq
C =\rm \{ +payment, termination\, date, \ldots \}
\eeq
and 
\beq
f(C,R) = \rm \{ +power, +heating, \ldots \}
\eeq
\end{example}
Tenancy is a service-like relationship between a host and a tenant.
This may be contrasted with the notion of residency at a location,
which is related to definition of boundaries within an observer's
realm.  Tenancy is a also relative concept (relative to promise
semantics).

\subsection{Laws of tenancy semantics}

It is basic to promise theory that we distinguish between a promise
made by an agent, and the agency itself.  Hence, we begin by noting that:
\begin{assumption}
  Promises are neither occupants nor tenants of the promisers or
  promisees, since they have no independent agency.
\begin{itemize}
\item Tenancy and occupancy requires two agencies to become associated.
\item Agents can be promised, but promises are not agents 
(they do not possess independent agency).
\end{itemize}
\end{assumption}
In a sense, a promise emanating from an agent seems to be attached at
the location represented by the agent. However, we do not call this
tenancy. A promise is a property of an agent, but it has no
independent agency, thus it cannot be a tenant. 

\begin{example}
An agent $A$ can be the subject of a promise, e.g.
\beq
A_1 \promise{+A} A_2,
\eeq
but it is not the promise itself, which belongs to $A_1$.
\end{example}
The semantics of the promises in (\ref{deftenancy}) select an inherent directionality
for the provision and use of a resource.
\begin{lemma}[Tenancy flows in the direction of the resource being used]
Tenancy flows towards the host, i.e. towards to source of 
the hosted resource.
\end{lemma}

\begin{figure}[ht]
\begin{center}
\includegraphics[width=10cm]{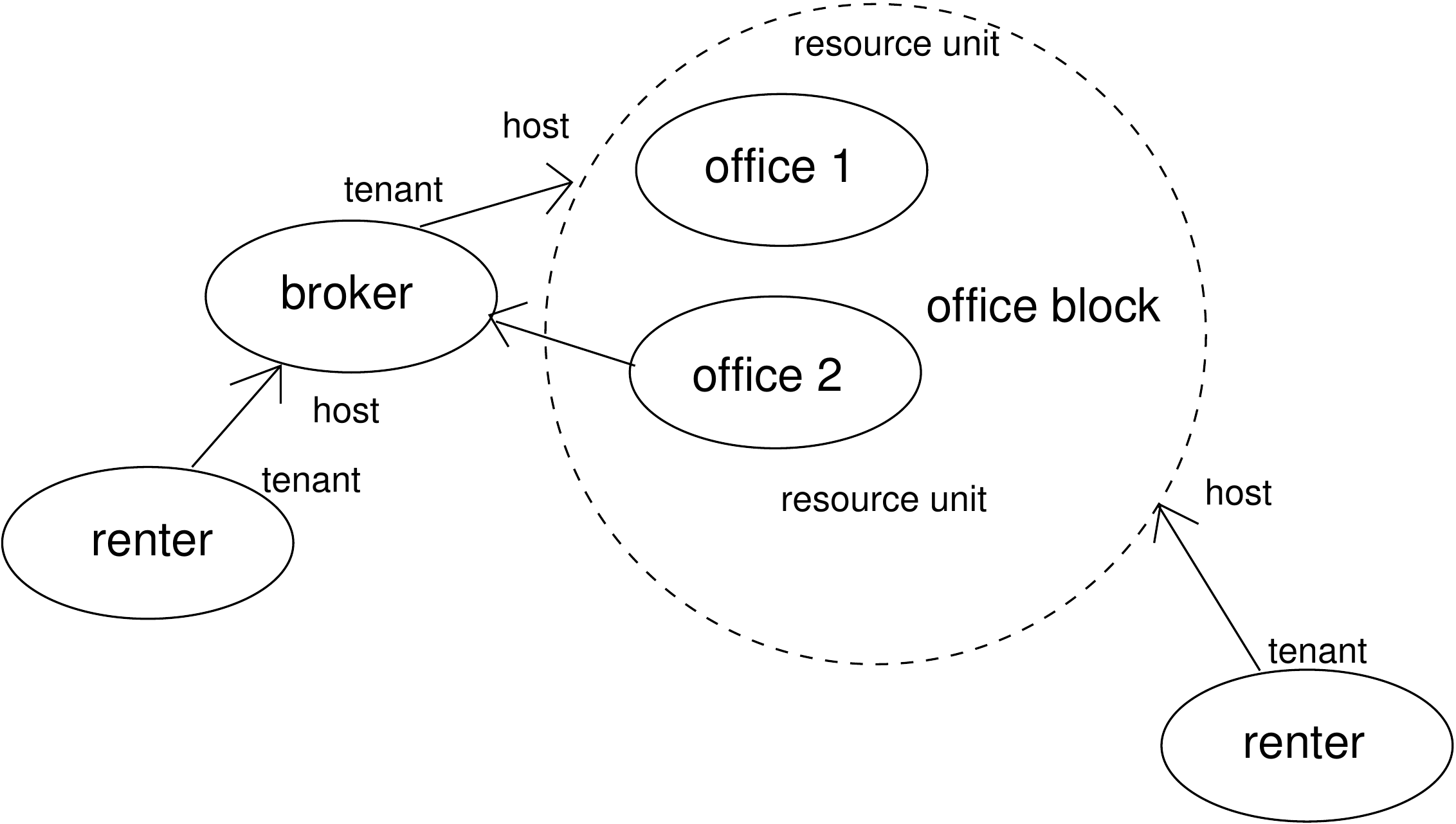}
\caption{\small Identifying tenants and hosts correctly requires us to follow
the tenancy law carefully. In each case, the arrows point towards to host resource sought by the tenant.\label{tenancylaw}\\(1) A renter may be a tenant of either an office block (providing multiple offices to multiple renters),\\ (2) A renter may be a tenant of a broker (providing multiple client offices to multiple renters),\\ (3) An office block or single office may be a tenant of a broker (offering multiple renters as a resource) to multiple office blocks or office.}
\end{center}
\end{figure}

It is important to bear in mind the semantics when looking at host and tenant.
Consider the following case in figure \ref{tenancylaw}. From the perspective
of a renter going directly to a hosting apartment block, the tenant

\begin{definition}[The host:tenant binding is 1:N]
  A host can have any number tenants, at any one time, keeping full
  promises, up to and including the valency of the host resource
  promise.
\end{definition}
There is an exclusivity between a tenant and a resource, which is a question of definition.
Tenancy with a super-agent scales like any other promise (see section \ref{distributive}). When we speak of a tenancy,
it refers to a single relationship, even though an agent might be engaged in multiple
similar tenancies.
\begin{example}
A rider on a horse is a tenant of the horse. A rider cannot ride a herd of horses,
at the same time.
Moreover, the rider and horse
are not joined by encapsulation, forming the embodiment of a super-agent.
A driver in a car however, is a tenant of the car, and is encapsulated
by it. The car is a tenant of the driver's direction.
Hence, while the two have independent agency, they seem to form an encapsulated super-agent.
\end{example}

\begin{example}
  In the OSI network model, the layers from L1-L7 form a tower of
  dependence, in which network resources (at the bottom) are shared
  out between different applications and users which are tenants of
  the basic service. These layers farther up the stack depend on the
  lower layers, hence the arrow of tenancy points down to the L1 
physical layer. L2 is a tenant of L1, L3 is a tenant of L2 and so on.

Network encryption is a tenant of L3, and computing applications are
tenants of the encrypted stream.

When these layers are implemented as encapsulations, the tenancy increases
into the core of the encapsulation, and the host is the outside part.
This seems to be the opposite of the way we are taught to think about
networking, from a software engineering perspective.
\end{example}

\begin{lemma}[Causation is partially ordered by pre-requisite dependency]
Promises and intentions may be partially ordered by conditional dependencies,
from the conditional promise law. This leads to a hierarchy of directional
intent, for fixed semantics.
\end{lemma}

We can distinguish tenancy from simple scaled agency by this partial
ordering of tenants to hosts in the direction of a named resource.
However, in most cases, the law of complementarity of promises allows
us to transform one tenancy into the reverse relationship interpreted
as a different promise.  In either case, the orientability of tenancy
gives agents topological `hair' which can be combed in a certain direction, as a vector field.

\subsection{Forms of tenancy}

Let's look at some familiar exemplars to see how this general pattern
is realized in different scenarios.

\begin{itemize}
\item {\bf Club membership, or passenger with ticket} 

  The issue of club membership is one where an agent associates itself
  as one of a group of typed agents: a vector promise directed to a
  specific host. The host offers the tenant a membership, and the tenant accepts the membership lease.
\beq
C    &\rightarrow& \rm  membership\, fee\\
R    &\rightarrow& \rm  membership\, credentials\\
f(C,R) &\rightarrow& \rm benefits\, and\, services
\eeq
Membership in a club is a label, i.e. a property of an agent. However, in the case
that a separate agency validates this label as evidence of an association, we can
view the members as guests of the hosting club. The condition $C$ is typically
some kind of subscription, the membership itself is promised with a badge
or access credentials, and the additional services that accompany membership
require showing of the credentials.

If a club is exclusive, then the promise of $+R$ has finite valency, else it
has infinite or unlimited valency.

\item {\bf Employment} 
An immediate corollary of membership is employment at an organization.
\beq
C    &\rightarrow& \rm  work\, performed\\
R    &\rightarrow& \rm  employee\, status/badge\\
f(C,R) &\rightarrow& \rm benefits\, and\, wages
\eeq
In this case, an employee is a tenant of the hosting company
that pays for membership with his/her daily work. Tenancy is fulfilled
by access or credentials (the company badge), and benefits include wages, lunch,
travel costs, etc.
Tenancy is always symbiotic, by nevertheless asymmetrical. The relative values of $C$ and $f(C,R)$ are
in the eyes of the beholders. When trading promises, what is valuable to
one party is usually not valuable to the other, else they would not be
motivated to trade.

\item {\bf Privileged access (territorial access)} 

A further corollary is the use of credentials to gain access to territories, e.g.
foreign visas, password entry, identity cards, etc.

\beq
+C    &\rightarrow& \rm  identity\, credentials\\
-C    &\rightarrow& \rm  authentication/access\, control\\
R    &\rightarrow& \rm  access\,passport/visa\\
f(R) &\rightarrow& \rm territorial\,access/resources
\eeq

\item {\bf Shared exclusive resource usage (multi-tenancy)} 

  Now consider the case where we add a finite valency to a limited
  resource, as well as a condition of fair sharing. A fair sharing
  promise, up to a maximum valency of $n$, becomes an additional
  constraint on the host, of the form:
\beq
+R_j\#n \; \Big|\; \sum_i^n R_i \le R,\label{Rconstraint}
\eeq
for each qualifying tenant $\tenant_j$, paying its tenancy cost $C_j$.

See figure \ref{tenancy}
\begin{figure}[ht]
\begin{center}
\includegraphics[width=4cm]{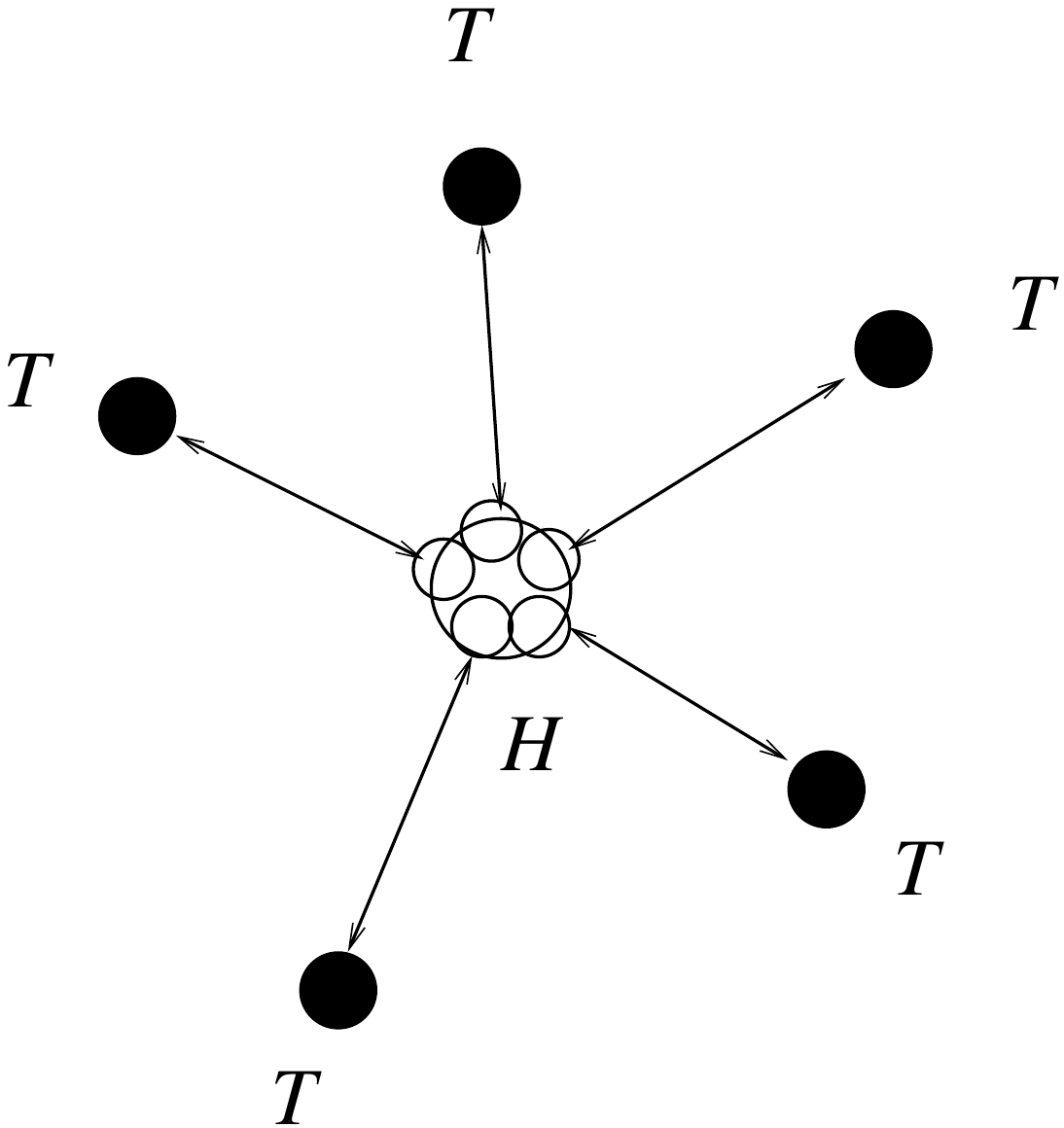}
\caption{\small Tenancy of a singular resource by multiple agents. This is the same
as membership, and containment. If multiple tenants occupy the same space,
then the host effectively promises independent constituencies, so
it has internal structure.\label{tenancy}}
\end{center}
\end{figure}

So the total promise set becomes:
\beq
&\host& \promise{+R_j\#n \Big|\; C_j\,,\; \left(\sum_i^n R_i \le R\right)} \tenant_j\\
&\tenant_j& \promise{-R_j\#m} \host\\
&\tenant_j& \promise{+C_j} \host\\
&\host& \promise{-C_j} \tenant_j
\eeq
where $m < n$ by necessity due to the valencies. Services $f(C,R)$, based upon
$R$ might be subject to additional constraints, but they are also naturally
limited by the constraint (\ref{Rconstraint}).

\item {\bf Representation by proxy} (spokesperson)

  In some cases a hosting agency's purpose is to be a proxy or
  representative for a client. This is the case for modelling
  agencies, writers' agents, sales representatives, public facing
  spokespersons, and even accountancy firms. Examples include
 `Intel inside', goods on a shelf in the shop that represent their brands.

In this case, the value added service to signing up is the representation of
the tenant itself:
\beq
+f(C,R) \rightarrow \rm \tenant\,representation
\eeq
Representation or brokering for the tenant does not necessarily
imply constraints on the tenant's autonomy (this depends on other promises).
This is not exchange of the tenant, like sending a letter, or transporting a passenger.
Notice, furthermore, that nothing promised here can prevent the tenant or host
from acting as separate entities in other ways.

\item {\bf Catalysis} (special semantic environments)

In a chemical process, some tenants need the help of a tailored environment
to make a transition to a new state. A host plays the role of catalyst

A pit-stop for tyre change, or a port/dock for loading and offloading, or repair
of transport vessels.

In the human realm, start-up labs and incubators are catalysts for companies
and biological processes. The womb is a host for infant morphogenesis.

In each case f(C,R)
\end{itemize}

\begin{example}
Users are tenants of multi-user software, logging into walled communities with login credentials.
Processes are tenants of operating systems.
Operating systems are tenants of computer hardware.
Computers are tenants of networks and datacentres.
\end{example}

\subsection{Tenancy and conditional promises}

It should already be apparent from the definition of tenancy, in
section \ref{tendef}, that there is a likeness between the
pre-condition for tenancy (denoted $C$) and the resource relationship
(denoted $R$).
From the conditional promise law\cite{promisebook}, a conditional binding
to provide service $S$ takes the form
\beq
A_T \promise{+b} A_1,
\left.\begin{array}{c}
A_1 \promise{S|b} A_2 \\
A_1 \promise{-b} A_2
\end{array}
\right\rbrace
\simeq A_1 \promise{S} A_2
\eeq
Notice how the exchange of the condition has the same structure as 
the tenancy relationship. This is because both are examples
of a generic client-server relationship, based on vector promises.

\begin{figure}[ht]
\begin{center}
\includegraphics[width=7cm]{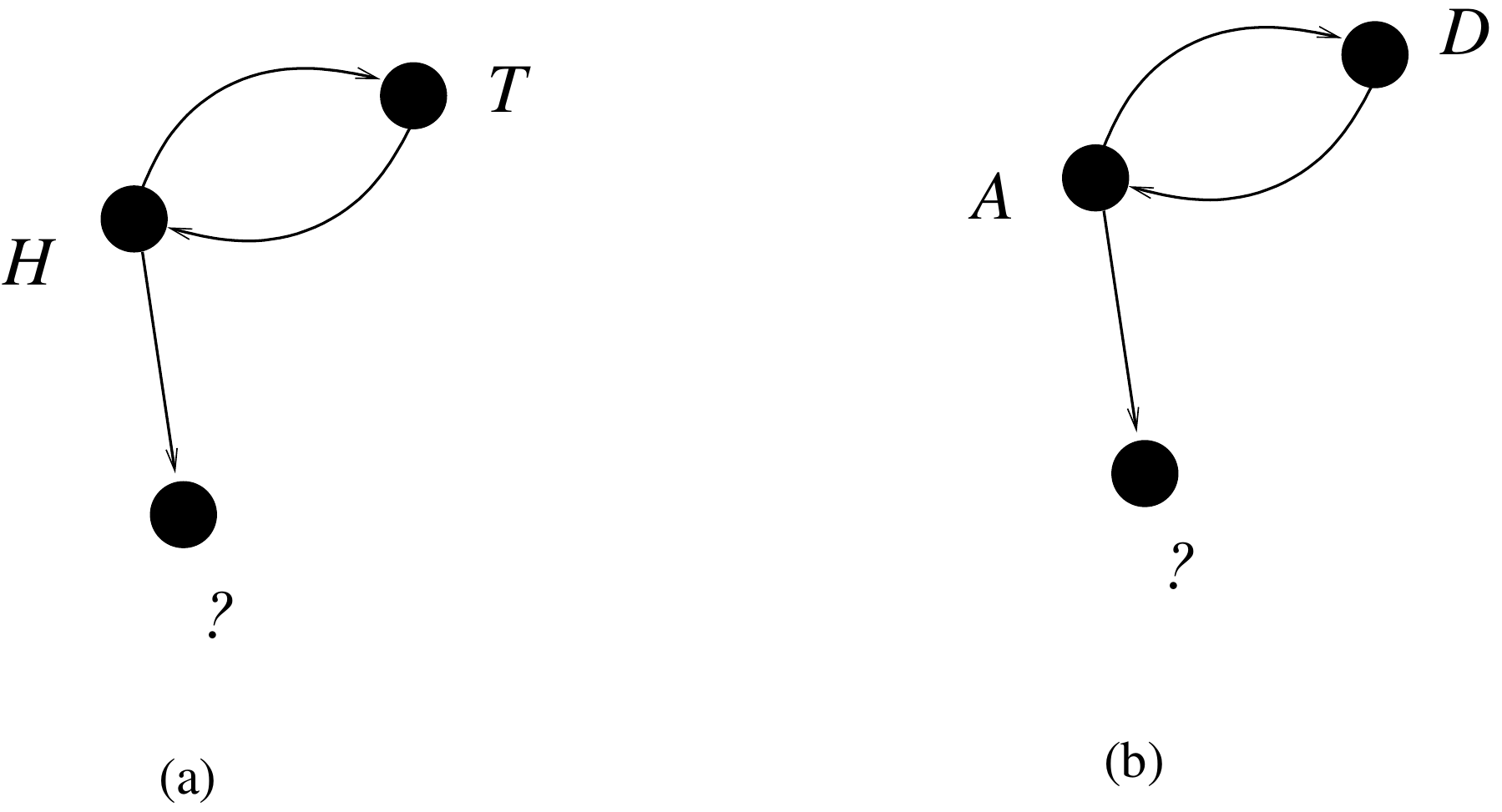}
\caption{\small The likeness between tenancy and a conditional promise involving a third
party.\label{tenancondition}}
\end{center}
\end{figure}

This can be formalized this further to show that a tenancy is really a conditional 
promise (see figure \ref{tenancondition}).

\beq
  H   \promise{+R}         \Unspec  &vs& A \promise{-c} D\\
  T \promise{-R}         H  &vs& D \promise{+c}\\
  H   \promise{+f(C,R) | -R} T  &vs& A \promise{+b|c} \Unspec
\eeq
Thus the tenant is the assumed recipient of functional promises derived from the tenancy
relationship, whereas in a general conditional promise this is unspecified.

Note, we shouldn't worry too much that the sign of the $+R$ maps to a
$-c$, as the complementarity rule (see \cite{promisebook}, section 6.2.2)
allows us to re-interpret the signs. For example, $+R$ could represent the active
garbage collection of resources, while $-R$ represents quenching with resources.
Similarly, $+R$ could represent employment, while $-R$ is work done to fulfill
the employment moniker. In both these cases, the $+$ promise takes on the character
of a receipt of service, often associated with $-$ promises.

The tenancy relationship is just an extended version of the basic client-server
relationship, with the special focus on identity\footnote{The description of services
in \cite{promisebook} treats clients and essentially faceless, generic entities.}.

\subsection{Remote tenancy}\label{tenadj}

\begin{figure}[ht]
\begin{center}
\includegraphics[width=7cm]{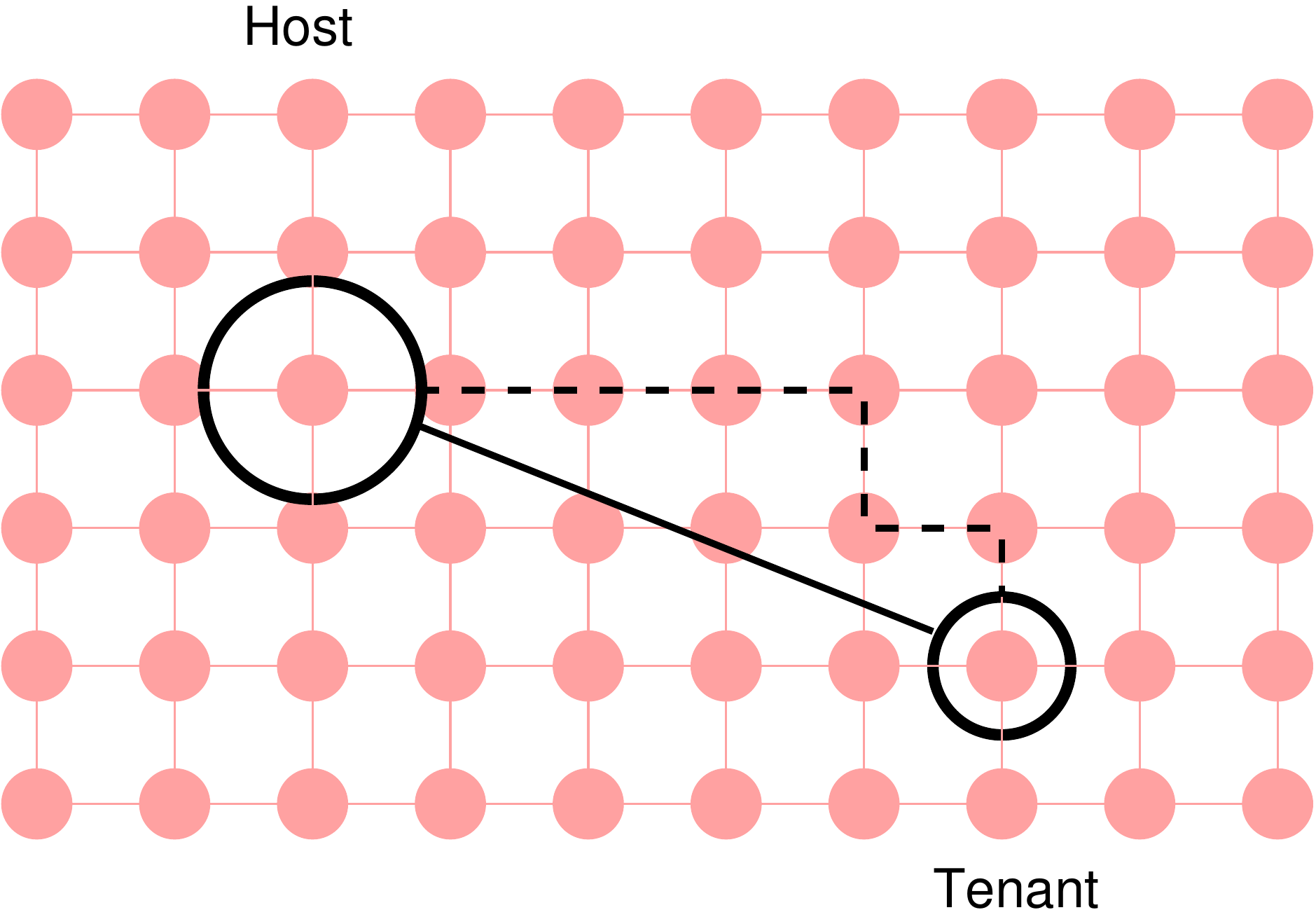}
\caption{\small Promising tenancy virtually, over a substrate of truly adjacent intermediaries,
often requires the distributive rule
adds to the complications at the finer-grained scale, and scaling away these details
requires implicit trust. \label{tensub1}}
\end{center}
\end{figure}

If we consider the case in which tenancy is not between agents that are actually
adjacent to one another, then the promises are delivered by proxy, in the sense
of a delivery chain (see \cite{promisebook}, section 11.3, and figure \ref{tensub1}).

When carried out via proxy, every adjacent node in a connective path through the
adjacencies of the carrying spacetime becomes a
possible point of failure or loss of integrity, and the cost of promising 
explicit integrity increases as the square of the number of agents along the
path taken via adjacent agents.

\subsection{Asymmetric tenancy}\label{asymm}

The semantics of tenancy are always asymmetrical, by definition.
Adjacency is usually symmetric and mutual, at least when locations are
equally weighted.  However, if a superior location is next to an
inferior location, according to some weighted importance ranking, then
the symmetry is broken, e.g. pilot fish surrounding a whale, shops
surrounding a mall. Unifying locations like malls, hubs, planets are
natural host-roles to shops, spokes, and satellites, regardless of
their relative size, because they connect agencies into an accessible
nexus. Their `size' may be thought of in terms of their network
centrality\cite{graphpaper}, for example, which gives them semantic importance.

\begin{lemma}[Adjacency is a form of tenancy, or tenancy is `rich adjacency']
  By symmetrizing over the host-tenant promises, and directing unspecified
promisees to mutual neighbours, we reproduce
  adjacency. 

\beq
  H   \promise{+R}         (\Unspec=T)  &,&   T   \promise{+R}         (\Unspec = H)\\
  T \promise{-R}         H  &,&  H \promise{-R}         T\\
  H   \promise{+f(C,R) | -R} T  &,&   T   \promise{+f(C,R) | -R} H 
\eeq
which reduces to
\beq
H &\promise{\pm R, f(C,R)}& T\\
T &\promise{\pm R, f(C,R)}& H
\eeq
Thus $R$ plays the role of adjacency, and identifying $R \rightarrow \adj$ and $f(C,R) \rightarrow \emptyset$, we see
that adjacency is equivalent to mutual tenancy in its weakest form.
\end{lemma}

\subsection{Scaling of occupancy and tenancy}\label{scaletenancy}

The ability to use space and time in a functional and operational way
is the key to building organisms and organized processes.  When we
speak of scaling these semantic forms, we implicitly expect to
preserve symmetries, asymmetries, and functional relationships, while
inflating the overall size of a semantic space by introducing more
agents. Coarse-graining should then allow us to see the functional equivalence of the
larger and the smaller system.

The asymmetry inherent in the ideas of occupancy and tenancy suggests
that we are not generally going to see scale-free phenomena. What
characterizes tenancy and occupancy is the retention of a
differentiated cooperative relationship between agencies. Specific
agents are bound together with intentional directionality.  This
contrasts with the idea of absorbing new agencies into a singular
agency.

\begin{example}
In a business partnership, or symbiosis, businesses or organisms
retain their separate identities and work together for mutually
beneficial returns. In a merger or acquisition, one company or organism
subsumes the other, hoping to control it without worrying about explicit cooperation.
\end{example}

The homogeneity of host-tenant semantics often play a role in the
coordinated, functional usage of space. Long range order helps us to
utilize space in a regular way.  Without it, many aspects of space and
time are simply opportunistic.

\begin{example}
  In a parking lot, the spaces need to be homogeneous in size else you might
  not be able to park your car in just any space. The same applies to
  the width or refrigerators, washing machines and kitchen
  appliances, block and sector sizes on disks.
\end{example}
We need to account for both strong and weak couplings, homogeneity and
inhomogeneity, to understand the wealth of possibilities in the world
around us.

\subsubsection{Extending tenancy with structural memory}

Homogenization is a forgetting process, while inhomogeneous
differentiation encodes a memory into spacetime. In a semantic
spacetime, memory is encoded through promises, and structure might
refer to any one of the aggregation, residency, occupancy and tenancy candidates.

\begin{figure}[ht]
\begin{center}
\includegraphics[width=7cm]{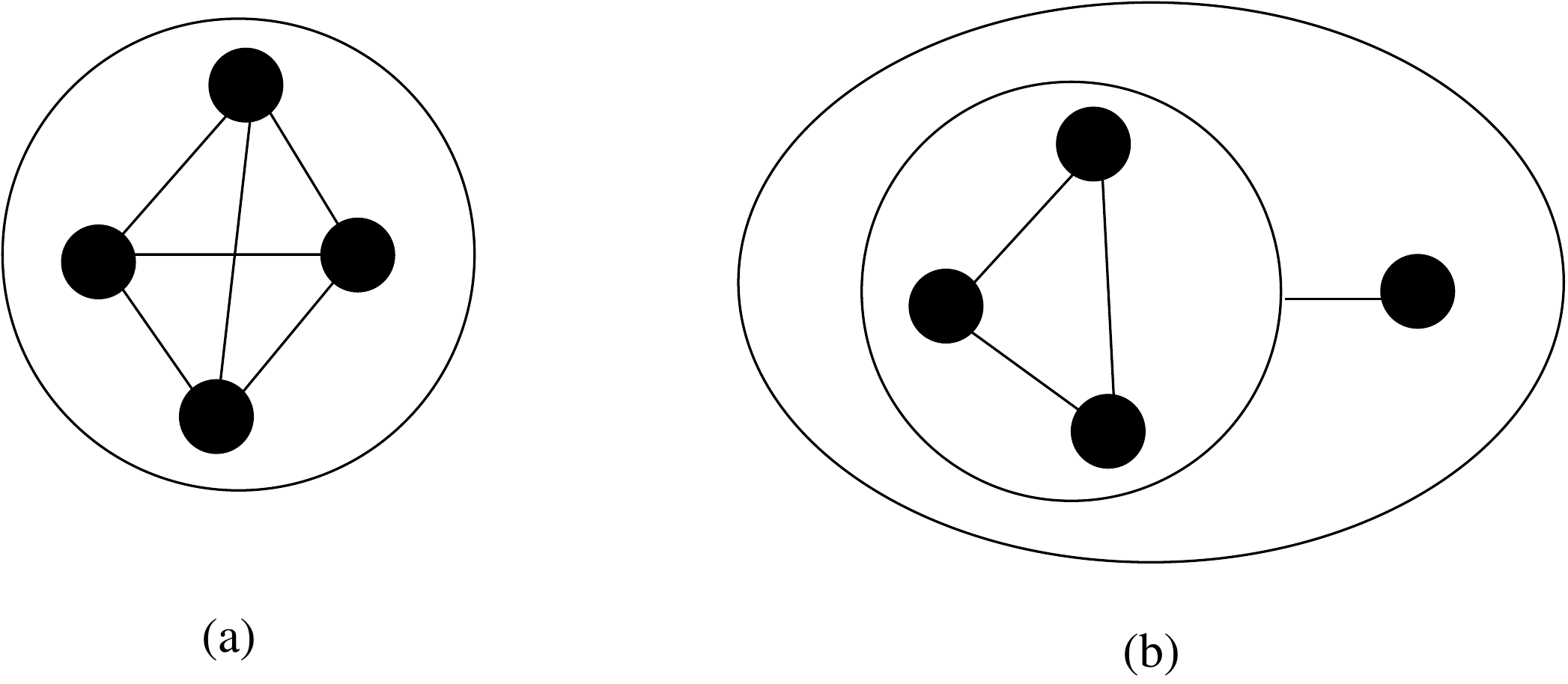}
\caption{\small The scaling of membership can retain a memory of its process. 
The introduction of an agency scale can introduce artificial asymmetry.
In (b), a promise binding is made to the super-agent, but this has yet to be
realized by a physical agency within this virtual/logical boundary.\label{agentscaling}}
\end{center}
\end{figure}

Let's contrast the ideas of scaling by absorption and tenancy more carefully.
Consider the two scenarios in figure \ref{agentscaling}, which
contrasts a symmetrical form of cooperation (a) with an asymmetric
tenancy configuration (b). The solid circles represent agency scales,
forming various levels of super-agency.

In the first case (a), the cooperative arrangement is completely
symmetric, and no information about the order in which the agents came
together is retained in the structure. If we assume that the tenancy
binding is based on a promise with some body $X$, then we may characterize
this arrangement by:
\beq
A_i &\promise{\pm X}& A_j, ~~~~~\forall i,j = 1,2,3,4. \\
A_i &\promise{\pm C(X)}& A_j
\eeq
We see exchange promises for $\pm X$, and symmetrizing coordination promises
$C(X)$.
In the second case, an intermediate step is apparent which singles out
a distinction between the set $A_s = \{A_1,A_2,A_3\}$ and $A_4$:
\beq
A_i &\promise{\pm X}& A_j, ~~~~~\forall i,j = 1,2,3. \\
A_i &\promise{\pm C(X)}& A_j~~~~~\forall i,j = 1,2,3.\label{coop}\\
A_s &\promise{\pm X}& A_4
\eeq
In this case, we see a memory of the structure which sees $A_4$ joining an already
established agency. As yet, the agent $A_4$ is not completely symmetrical
with the other three. The asymmetry of intent remains, although an observer watching
how these promises are kept might not be able to tell the difference between
scenarios (a) and (b). This depends on how we interpret the promise between $A_4$ and 
the super-agent $A_s$. There are two possibilities:
\begin{itemize}
\item $A_4$ binds distributively, assuming that each of the agencies $A_1, A_2, A_3 \in A_s$
makes individual promises so as to behave symmetrically (by virtue of (\ref{coop})),
allowing:
\beq
A_4 &\promise{\pm X}& A_j ~~~~~\forall i,j = 1,2,3
\eeq
The addition of these promises completes the symmetry that turns
scenario (b) into (a), eliminating the memory of their inequivalence
to an observer, and absorbing $A_4$ effectively into a cooperative
entity. Regardless of how a partial observer might draw its
super-agency boundaries, it is now able to identify a symmetrical {\em
  role by cooperation} (see \cite{promisebook}) between all four
agents. This exercise gives us a clue about what absorption means, in a formal
sense.

\item $A_4$ binds only to a single representative of the collective $A_s$.
This remains asymmetric, and we would consider $A_4$ to be
a non-resident occupant or tenant of $A_s$. The memory of the inequivalence is coded into
the intentional behaviour, by promises.

\end{itemize}
These figures illustrate how the promise configurations document the
history by which an arrangement of agents was constructed, and allow
an entity formed by multi-layered cooperation to retain a memory
of past states.

What we see, in figure \ref{agentscaling}, is that adding promises
can effectively remove asymmetry between host and tenant, meaning that
tenancy can be eliminated by `acquiring' an agency\footnote{This
  happens frequently in merger and acquisition of companies of
  course.}.  However, it is important to also consider scaling in
which the tenant never becomes a part of the host, i.e. we maintain
the strict asymmetry, as this implies no loss of autonomy between the
tenants.

\subsubsection{Scaling of the tenancy law}

Now, let's see how the extended scaling of internal structure in
either host or tenant affects the promises made in a tenancy
relationship. This scaling applies to all promises between
coarse-grains, not just tenancy promises.

As the number of sub-agents in internal structure grows, it 
becomes natural to consider them both as embedded subspaces of the
surrounding semantic space. Such a subspace may be either
a solid lattice, with long range order, a
gaseous state, or forest-like (molecular).

Consider the full tenancy relationship:
\beq
A_{\rm super} &\promise{+X\#n \, |\, C}& A_{\rm tenant}\\
A_{\rm super} &\promise{\pm C}& A_{\rm tenant}\\
A_{\rm super} &\promise{+f(C,X)\, |\, -X}& A_{\rm tenant}, \Unspec\\
A_{\rm tenant} &\promise{-X\#m}& A_{\rm super} 
\eeq 
Suppose we try to add agents in the manner of a scale perturbation.
We preserve the implied roles (the tenant $A_{\rm tenant}$, and host $A_{\rm
  super}$), and the structure of the binding between them.  What
features would change, if we now attempted to scale this relationship by
adding new internal structure to either host or tenant?
With the further addition of $A_{\rm pert}$, assuming this adds to the valency, this becomes:
\beq
A_{\rm super} + A_{\rm pert} &\promise{+X\#(n+1) \, |\, C}& A_{\rm tenant}\\
A_{\rm super} +A_{\rm pert} &\promise{\pm C}& A_{\rm tenant}\\
A_{\rm pert} &\promise{\pm C}& A_{\rm super}\\
A_{\rm super}+A_{\rm pert} &\promise{+f(C,X)\, |\, -X}& A_{\rm tenant}, \Unspec\\
A_{\rm tenant} &\promise{-X\#m}& A_{\rm super} + A_{\rm pert}
\eeq 
Hence, the symmetry and internal structure of the super-agent has a
material effect on the binding properties in a tenancy arrangement.
This is a formal scaling, but it does not really explain what happens
to communicate intent across a coarse-grain, as discussed in earlier.

The exterior promises of a host need to scale to provide valency $n$
binding sites.
The first issue to satisfy is the valency binding constraints, we seem to need $m \le n$ in the tenancy
promises. There are several ways this might be solved:
\begin{itemize}
\item One sub-agent is allocated to one tenant.
\item Sub-agents host multiple tenants each.
\item Several sub-agents working together service a single tenant.
\end{itemize}

Similarly, we have to answer the question: how does a tenant know with
whom it should interact? Will tenant and host sub-agent be able to
locate one another (see figure \ref{direct})?  Referring to figure
\ref{direct}, and the previous discussion of promising at super-agent
boundaries in section \ref{distributive}, we ask: should external tenants
bind (a) directly to independent sub-agents, or should they (b) go
through brokers and interfaces?

\begin{example}
  A tenant in figure \ref{agentscaling2} might represent a person or
  a population looking for an unoccupied apartment through a fronting
  organization; or a car looking for a parking space in a collection
  of parking lots, a cubicle in an office space, and so on. While the
  super-agencies like apartment blocks and parking lots formally
  promise space (valency), the potential tenant needs to locate the
  empty slots in order to bind to them.
\begin{figure}[ht]
\begin{center}
\includegraphics[width=7cm]{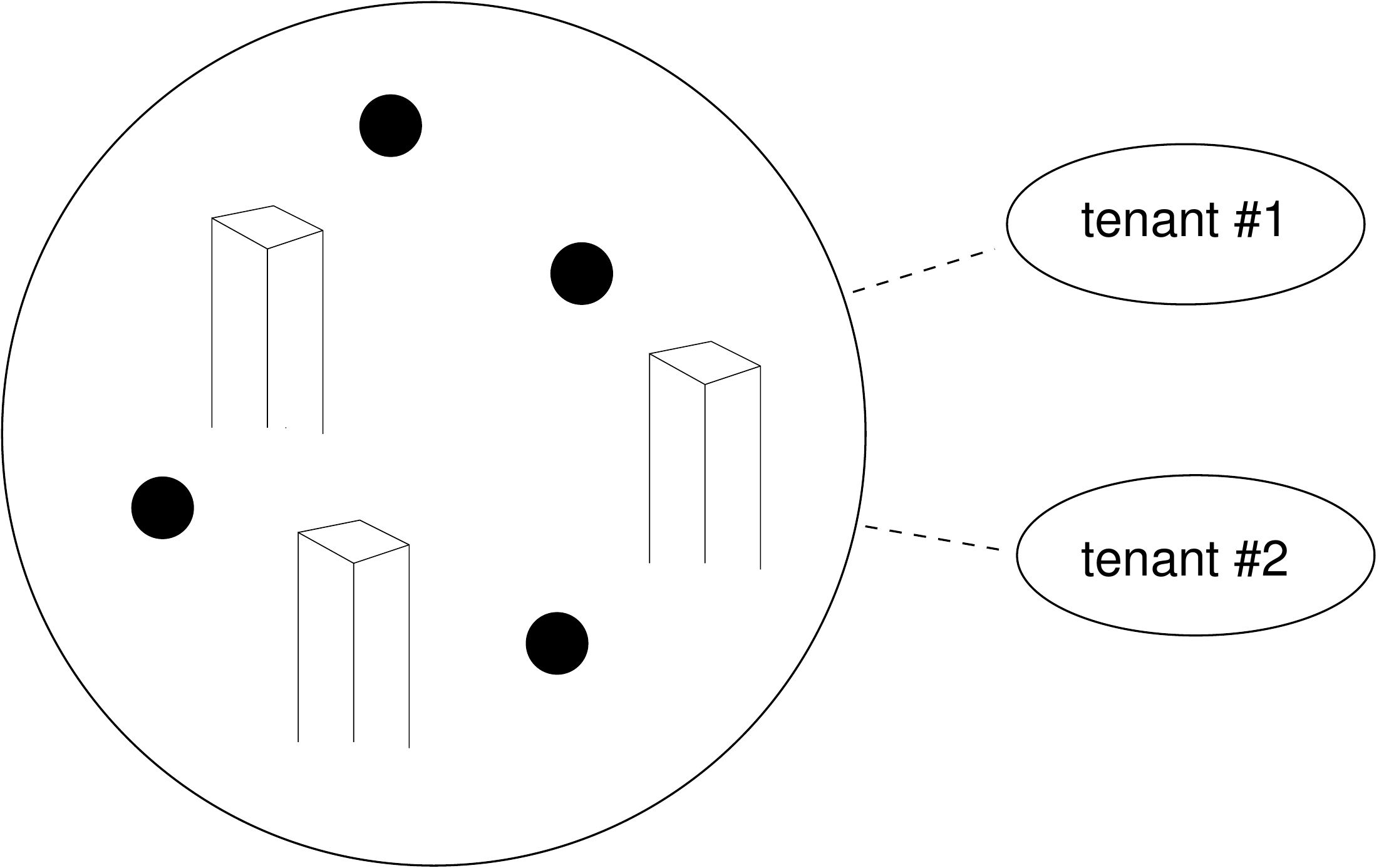}
\caption{\small Tenants, \`en masse, can contact
a front organization like a website to search for housing, but
still need to connect with the agency's internal components (individual landlords)
that can provide actual rather than logical service. e.g. a person needs to 
make direct contact with the apartment block that offers apartments to rent,
not merely the super-agency of `real estate' which has no physical reality.
This if one tenant is asking for more service than a single host
can offer, an issue becomes whether the service can be delivered in practice,
in spite of the apparent size of the super-agency.\label{agentscaling2}}
\end{center}
\end{figure}
\end{example}

\subsubsection{Distributive scaling of tenancy relationships}

In order for an outside agent to form an adjacency or a tenancy
binding to sub-agent, its has to know of the other's
existence\footnote{This is actually a hierarchy problem. At the lowest
  fundamental level of agency, there is no intermediate solution to
  this: either agents are adjacent or they aren't. We have to assume
  that they can sense and discover one another somehow (see section
  \ref{distributive}).}.
In all cases, tenant and host need to be able to locate one another,
or be introduced, in order to communicate, both form a promise binding
and to keep the promise. There are only two possibilities here:
\begin{itemize}
\item Host and tenant are directly adjacent (see figure \ref{direct}a).

\item Host and tenant can communicate with the aid of intermediaries (see figure \ref{direct}b).
This requires cooperation between the sub-agents.
\end{itemize}
The super-agent coarse-graining directory plays the key role here in 
making these details transparent.
\begin{example}
  Viewing virtual or switched private networks as tenants of a series
  of hosts that cooperate as super-agency: this requires tenancy at
  each independent host, and coordination between them.  In addition,
  the agencies are connected by adjacency promises between the chained
  carrier hosts (see figure \ref{sharednet}), or virtual adjacencies
  via intermediaries or proxies (see \cite{promisebook}, section
  11.3). The scaled tenancy relationship allows tenant spaces to
  appear contiguous, even though they might be distributed.
  Structures like overlays and tunnels act as virtual adjacencies,
  which rely on a substrate which we coarse-grain away.

\begin{figure}[ht]
\begin{center}
\includegraphics[width=10cm]{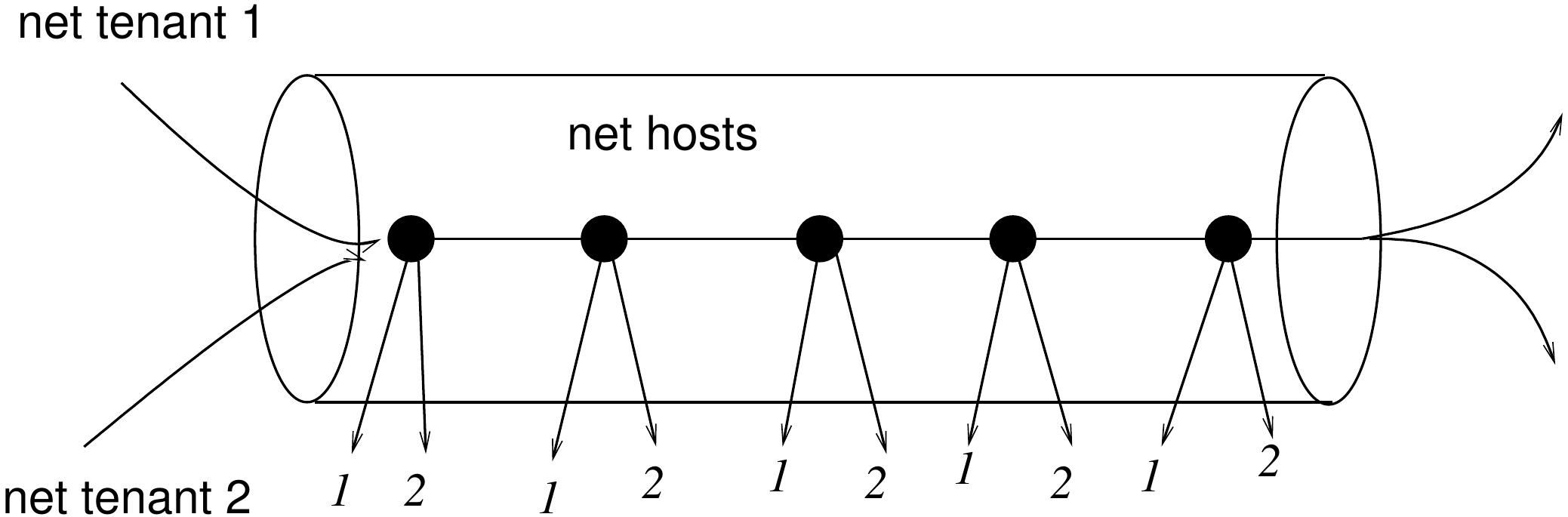}
\caption{\small Shared network requires the distribution of the entire tenancy binding.
The two hosts cooperate on handing over their resources to make a collaborative
channel, while maintaining the integrity of the segmentation of tenants.\label{sharednet}}
\end{center}
\end{figure}
\end{example}

To scale tenancy on the host side, there are some basic requirements:
\begin{lemma}[Distributed tenancy law]
In order to distribute a tenancy to a collection of autonomous hosts sub-agents, 
there must exist:
\begin{itemize}
\item A cooperative agreement between all sub-agents acting as hosts to share its local resources
according to the collective multi-tenancy agreement.
\item An index for adjacency coordinates of the host sub-agent, if not directly
adjacent.
\end{itemize}
\end{lemma}

\begin{example}
A customer of a bank can visit any branch to access their account, by mutual
cooperation between the branches, else the customer has to deal directly with
their home branch. (Remarkably, this is still an issue even in 2015, in spite of
the prowess of information technology.)
\end{example}
\begin{example}
  Binding sites on cells, immune cell responses, MHC etc, allows a chance
  encounter for antigen to locate a cellular tenancy. 
\end{example}

\subsubsection{The significance of functional asymmetry}

\begin{figure}[ht]
\begin{center}
\includegraphics[width=12cm]{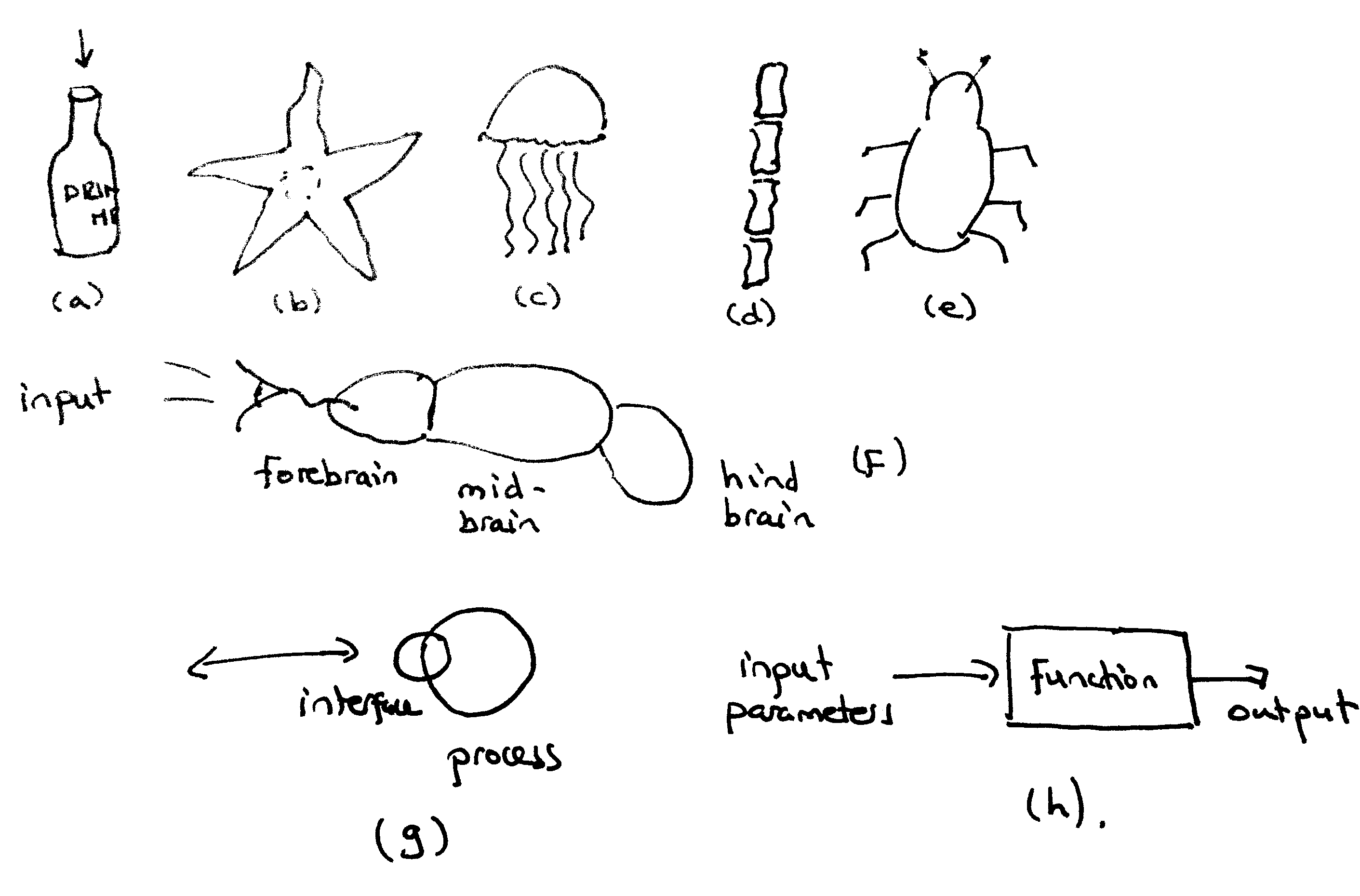}
\caption{\small Functional behaviour is more associated with symmetry
  breaking, than residual symmetry.  Maximal symmetry or disorder may
  be regarded as the default state of structure, with minimal
  cooperation. Cooperation is a force for order, and symmetry
  breaking.\label{asymmetry}}
\end{center}
\end{figure}

The economics of tenancy (regardless of the currency used, e.g.
energy, money, prestige, etc) brings tenants and hosts together,
leading to an asymmetric relationship (see figure \ref{asymmetry}).
Functionality is associated with broken symmetry: the default state of
maximal symmetry has very little going for it to attach semantics to.
As seen in earlier sections, pure scaling of agency, in the absence of
constraints, does nothing to break symmetries, and hence leads to a
maximal radial (spherical) symmetry, like a cell; however, boundary
conditions from the environment break symmetry allowing functional
behaviour.  There are many examples of this in the world of biological
organisms.

\begin{example}
Reproductive organisms begin as a single egg, which grows and differentiates. Initially
it grows symmetrically, dividing into a ball of cells, then boundary conditions from the environment
during morphogenesis create preferred directions with chemical signals. This leads to dorsal-ventral
asymmetry, etc. Selective apoptosis leads to further local asymmetries, with functional consequences, such a fingers.
Cephalization (formation of a head) is associated with the appearance of brains in the nervous systems
of organisms, and the brain is at the front where the organism meets and senses its environment.
\end{example}
\begin{example}
The shift from a spherical symmetry to axial and bilateral symmetries has happened prolifically
in biological evolution. Axial symmetry is typically associated with orientation of an organism
with respect to a flow, e.g. a jelly fish. In computational terms organisms orient along their input-output axes.
\end{example}

\begin{example}
  A bottle (box, container) is usually a round axially symmetric
  structure, but this is irrelevant to its function.  It does not
  matter whether the axial structure is symmetric or asymmetric,
  round, square, oblong, ellipsoid, etc.  The functional shape of a
  bottle only depends on one end being open to be able to receive the
  substance it will contain. This asymmetry makes it bind with
  specificity with functional consequences.
\end{example}

Figure \ref{asymmetry} shows organisms that can be modelled as tenancy
relationships with various levels of symmetry and asymmetry. The more
decentralized organisms are, the more symmetrical they tend to be, and
the less `smart', i.e.'  with functionality that is more
uncoordinated. In biology, brains are associated with `cephalization',
or the evolution of asymmetry to deal with an axis of input-output.
Segmentation along that axis (e.g. figure \ref{asymmetry}d,e,f), or
the formation of cells or functional compartments can form through
non-linear transmission of boundary information in the manner of
B\'enard cells, from the presence of a preferred length scale, and
aids in semantic differentiation. The segments then often work
together in the manner of a host-tenant relationship, binding through
specific functional promises. Thus scaling of tenancy (a strong
coupling regime) looks very different to the scaling of simple
aggregate residency, such as how herds, swarms, and societies scale
(weak-coupling regime).

\begin{example}
Mitochondria and cells co-exist, with mitochondria residents of their super-agent cells.
Herds and flocks scale in the manner of residency, but remain loosely coupled
without direction intentional `encoded' functional cooperation.
\end{example}

The scaling of intentionality tends to promote asymmetry, through specialization.
This places much greater emphasis on strongly coupled, and hence fragile, semantics.
It is no longer sufficient to have safety in numbers, as in a herd or tissue
design. Functional uniqueness, like the components in a radio, bring fragility\footnote{It seems likely 
that survival instincts and intelligence would evolve to be strongest in those species that cannot find
obvious safety in numbers.}.

\subsection{Tenancy as a state of order}

The structural memory of tenancy leads to asymmetry, but the long
range effects of that asymmetry depend on the phase in which spacetime
elements find themselves. The phase becomes a part of the strong or weak
coupling contraints.

\subsubsection{Tenancy in gaseous state}

If agents that are seeking to bind in some kind of tenancy
relationship are in a gaseous state, it is easier for them to
rearrange and attach directly to a host binding site.  Conversely,
this makes their meeting more haphazard, since no agent is in a known
location.
Agents are free to move, so it is harder to trust their identity. 
On the other hand, they can move around and form adjacencies directly
without the need to hand off to intermediaries.

Any spacetime location that is not fixed is in a gaseous state, as a free agent.
In technology, this applies to humans, mobile phones, vehicles, satellites, etc
There should be sufficient similarity between agents that they can
cooperate? Else they will remain inert noble gases.

\begin{example}
The Internet, for example, appears superficially solid, with all of
its cables and boxes, but its lack of a robust adjacency relationships
makes it just a slow liquid (like the amorphous solids or pitch or
glass). Moreover, the lack of a fixed spatial coordinate system means
that it has a cellular structure on a large scale: it has small
regions of brittle coordinatization, loosely floating in a
structureless soup. Names and addresses have no global significance
within the scope of a symmetrical pattern, and hence have no long
range predictive value.
\end{example}

In contrast to simple absorption, any cooperative tenancy
configuration must bring additional stability to the combined system
of agents in order to persist: a symbiosis which confers a positive
advantage to being a solo agent. This suggests an instability leading
to the condensation of cooperation, first into amorphous liquids and
swarms, and then into more rigid crystalline structures.
The binding of a tenant and a host into an $H-T$ molecule is locally
solid, but might be globally a gas. One could imagine a phase
transition in which a bi-partite crystal was formed when
scaled, with the structure of an alloy, formed from donor and a
recipient.

As a general principle, flexibility and agility require fluidity; but,
in order to stabilize semantics, time or change need to be eliminated.
The more time plays a role, the less significant the information is.
This points us towards solid structures.

\subsubsection{Tenancy in a solid state}

The solid state is familiar to us through regular spacetime structures
including hotels, parking lots, warehouses, hard-disks, and computer
memory, to cite a few examples.  In a solid spacetime, agents are not
free to alter their adjencies over `time' (see paper I, section 5.12),
so we trust their identities and relative promises more easily.

At the lowest atomic levels of agency, tenancy in a solid state must
lead to asymmetry of space itself, as we'll see in addressable
structures.  When agents are locked into a solid structure, tenancy
often leads to a large scale asymmetry, such as that seen in
biological organisms. This seems to have implications for being able
to pinpoint functional locations. Hence the loss of maximal, random
spatial symmetry (or the rise of so-called long range order)
is associated with functional behaviours, and hence functional promises.
There are numerous ways in which functional asymmetry can manifest and
materialize:
\begin{itemize}
\item Tenants and hosts are sufficient in number to be able to symmetrize 
on a larger scale, say in $n$-dimensions, and give the appearance of minimal loss
of symmetry, e.g. a metallic cubic crystal, such as steel, with tenant impurities.

\item Back to back, tessellating structures mixing various
  orientations are a possible solution to long range order however.

\item Hierarchical structures which fill partial spaces, with self-similarity (fractals).
\end{itemize}

At coarser scales, where `virtual' agencies can form tenancy bindings
on top of an underlying solid adjacency substrate.  The fixed
locations cannot easily change and be replaced, so they are easy for
other agents to trust local neighbours, however being locked into a
fixed number of neighbours now mean that long range promise
relationships have to be prosecuted through intermediaries (the
so-called end-to-end problem discussed in \cite{promisebook}).  The
presence of intermediaries means that information integrity is now in
peril, and trust in non-local relationships is not automatic.

The tighter coupling of agents and reliance on intermediaries to
transmit information leads to the possibility of topological defects
in the structure, such as crack propagation, and sudden catastrophes.
The homogeneity of a solid crystal is important, as mentioned earlier.
Re-usability of space suggests quantization of resources in a `first
normal form'.  The relational data normalization rule of first normal
form is about the re-usability of
space\cite{date1,burgess04analytical}.

\begin{example}
In a parking lot, the spaces need to be the same size else you might
not be able to park you car in just any space. The same applies to the
width or refrigerators, washing machines and kitchen appliances,
block and sector sizes on disks.
\end{example}

\section{Multi-tenancy, and co-existent `worlds'}

Multi-tenancy in information systems is one of the key challenges of
operational effectiveness.  As the custodian of the shared resource,
the host has to be able to keep promises on an individual basis, while
still being constrained by the behaviours of all its tenants. This
means that there will be contention at the host. Moreover, the host
is often the seat of mediation between the tenants and the outside
world, acting as a gate-keeper, and sometimes as a barrier or firewall
between them.

How the host isolates these resources from one another is one of the
key issues in a functional space. One tenant should not be able to
bring down another through its misbehaviour, either directly or
through the host as proxy.  This is the thinking behind insulated
private rooms, isolated electrical circuits, multi-user systems and
even virtual machines in computing. In these scenarios, the special
semantic role of the host makes it vulnerable: a `single point of failure'.

\subsection{Defining multi-tenancy}

What are the promises that make a spacetime multi-tenant? See the example figure \ref{sharednet}.
\begin{definition}[Multi-tenancy at scale $M$]
Consider a agency scale $M = \{ H, T_1, T_2, \ldots, T_m \}$ 
for some $m > 1$. An agent $H$ is said to exhibit multi-tenancy
if a number of autonomous agencies $\{ T_1, T_2, \ldots T_m\}$, called the tenants (making no a
  priori promises to one another) independently form a tenancy binding
  to $H$, in the role of host, for a share of a single resource $R$ that $H$ promises.
\beq
  H &\promise{+R\#n | C}& \Unspec\nonumber\\
  H &\promise{-C}& T_i, \nonumber\\
  T_i &\promise{+C}& H\nonumber\\
  T_i &\promise{-R\#1}& H\nonumber\\
  H &\promise{+f(C,R) | -R}& T_i\nonumber\\
&m \le n&\nonumber\\
&\forall i = 1\ldots m&\label{multenancy}
\eeq

\end{definition}
The issues for multi-tenancy include all those for general tenancy,
and also segregation, mutual isolation, addressability of tenants,
scaling of naming, sharing of resources between tenants, and the
possibility of tenants sharing amongst themselves, without involving
the host.

\subsection{Branching processes: subroutines, worlds, and asymmetry with hierarchy}

A brief digression on understanding the dynamics of aggregation of
tenants around a host, introduces the notion of a {\em branching
  process} (and its inverse, the merge, confluence or aggregation
process).  Branching is what happens as the possible states or
locations of a system fan out and increase in number from $n$ to $n'$,
where $n'>n$, selecting a preferred direction, either over space or in
time.

\begin{figure}[ht]
\begin{center}
\includegraphics[width=8.5cm]{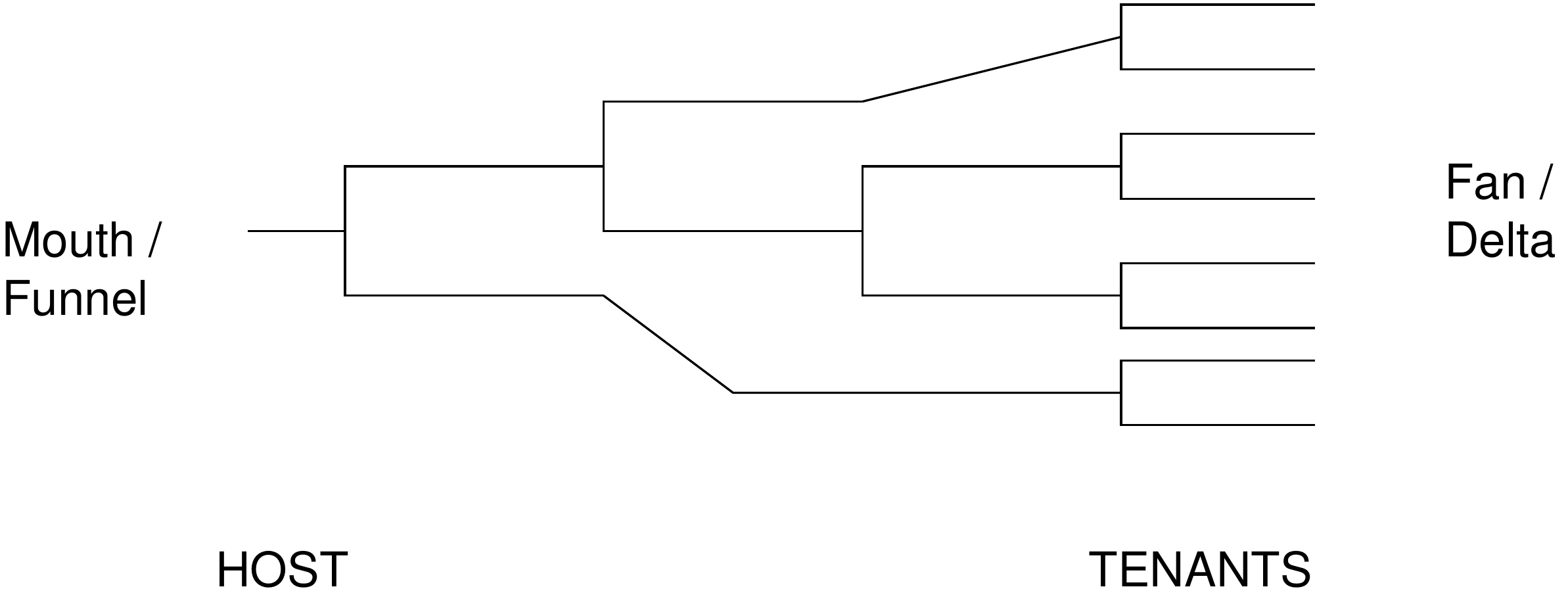}
\caption{\small A schematic branching process, showing asymmetry of process.\label{branch}}
\end{center}
\end{figure}

A branching process is an evolutionary sequence of changes in which
the level segregation or multiple agency increases either in space (over
distance) or time (spacelike hypersurfaces) (see figure \ref{branch}).
The branch points are associated with instabilities of the system both
dynamically (bifurcations) and semantically (if-then-else reasoning).

Branching has an arrow of directionality from the root, along each
local vector to nearest neighbour, which marks a gradient from fewer
to more states. Thus as the process unravels (in time or space), the
number of outcomes increases, favouring an increase in system entropy,
and decreasing the likelihood of arriving at any chosen one outcome.
The structure is not necessarily linear in total: it
could be radial (like a Cayley tree, or snowflake), but there is a
coordinate direction in which the number of locations is increasing,
quasi-exponentially.

\begin{itemize}
\item {\em Branching} is associated with fine-graining, sharing from one to many,
reasoning about different possibilities, and it is also about
increasing specialization.  Branching is a $+$ promise of the host.

\item {\em Merging}, the inverse of branching, is associated
with generalization, coarse-graining, averaging over
possibilities, and shared or common resources. Merging
of aggregating is a $-$ promise of the tenant.
\end{itemize}
As a companion to branching, we also have a notion of {\em hierarchy},
which is sometimes confused with it. Similar to branching, hierarchy comes about by ranking
of states. Thus, if one assigns greater importance to having more
or less states, one could call a branching process a hierarchy, as the
 branching asymmetry implies an order from head to tail.
\begin{definition}[Promise hierarchy]
  An iterated process, over a series of promises $\pi_i$, in which the promises
  $\pi_i$ exhibits asymmetric semantics, with respect to the promiser
  and promisee. Hence the process generates a chain or sequence of ranked or
  ordered elements (agents), in the manner of a semi-lattice.
\end{definition}
\begin{example}
  For example, a promise to use (dependency, requirement, etc) is
  asymmetric.  Promises of mutual adjacency, on the other hand, have
  symmetry, hence there is no preferred direction for semantics. A branching process
  over mutual (symmetric) adjacency does not form a hierarchy, only a
  tree or forest structure.
\end{example}

\begin{definition}[Branching hierarchy]
A branching process in which there is directionality of both dynamical
branching and semantic intent.
\end{definition}
\begin{example}
Dependency trees, and taxonomies form branching classification hierarchies.
\end{example}

Because the branching of a single host into multiple tenants has the same asymmetry,
clearly branching is a key process in interpreting multi-tenancy. It is
the inverse of the accretion of multiple tenants to the single host.
The converse of the tenancy law is that tenancy is a branching process
of {\em host resource usage}, branching from one host into multiple
tenants, and thenceforth.
Semantically, branching may come about in a number of ways:
\begin{itemize}
\item Through a breakdown of cooperation between existing agents, leading to a
  {\em fine-graining}\footnote{The vernacular `dis-aggregation' is common.} or fragmentation of roles.
\item As a spawning of new agents, increasing in number.
\end{itemize}
Dynamically, the branching can be either 
\begin{itemize}
\item Due to a change in the receiver, e.g. a (-) use-promise no longer aggregates agents.
\item Due to a change at the source, i.e. a (+) promise now differentiates agents.
\end{itemize}
Branching processes lead to proliferation of `parallel worlds'
(disconnected sub-spaces), and possibly a change of dynamical scaling.
Keeping track of these worlds becomes a divergent problem of {\em knowledge
  management}, without a counter-process.  If there is exponential growth of agency, this may be
{\em intractable}, in the sense of computational complexity.

\begin{figure}[ht]
\begin{center}
\includegraphics[width=9cm]{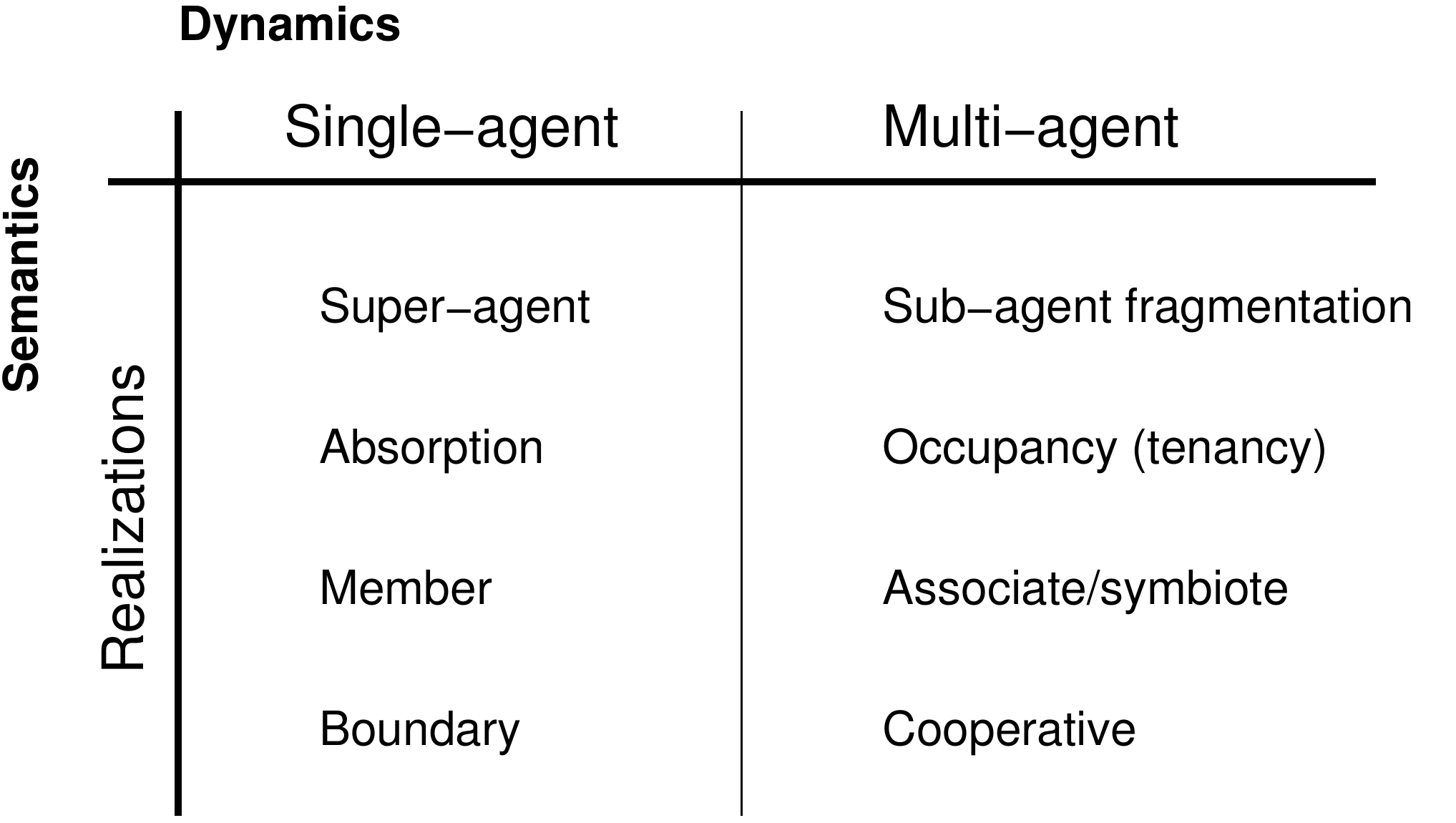}
\caption{\small Dynamical and semantic aspects of multi-tenancy eventually
merge through the semantics of scaling. This illustrates the importance of
scale in the total semantic content of an observer's realm of assessment.\label{multiten}}
\end{center}
\end{figure}

Branching processes may be examined through the lenses of semantics and dynamics:
\begin{itemize}
\item {\em Semantic}: Disambiguation, tenancy, security, etc.
\item {\em Dynamic}: Isolation, bifurcation, cell division, renormalization, etc 
\end{itemize}
Figure \ref{multiten} contrasts a few of the scale issues mentioned thus far.
Notice how, at what might be perceived in one way, at one scale, could be perceived
another way from the perspective of a different scale.

\subsection{Tenant world segregation, and resource multiplexing, in privately hosted spaces}

A host shares its resources into slots, which can be occupied by a number of tenants
in a number of ways:
\begin{itemize}
\item Segmentation or partitioning of adjacencies in an underlying graph, or spacetime.
\item Labelling of regions by promises, marking gradients (stigmergic typing of agents).
\item Growing and spawning new agents.
\end{itemize}
\begin{example}
Division multiplexing is a common strategy in hosting:
\begin{itemize}
\item A parking lot (host) divides up its space into slots that can be used by cars (tenants).
\item A hard disk (host) divides its surface platter into sectors and blocks, for user-data (tenants).
\item A time-sharing computer operating kernel (host) divides up its computational resources into
processes that can be used by jobs (tenants).
\item Memory address spaces (host) divide up addresses into pages, for occupation by process data (tenants). 
\end{itemize}
\end{example}
Isolation of a subspace at the host can come about by two means:
\begin{itemize}
\item The absence of adjacency between the tenants.
\item The absence of reachability: i.e. an agent cannot be reached because it has:
\begin{itemize}
\item No adjacency path/route, with or without the help of intermediaries (vector promises).
\item No name or (unique) identity, or address, to locate it by (scalar promises).
\end{itemize}
\end{itemize}
Thus, both scalar promises (e.g. names) and vector promises (e.g. adjacency) play roles in
the segmentation of shared spaces.

\subsection{Mouth formation at host boundary, and gatekeeping credentials for tenants access}

As remarked in paper I, the picture of spacetime described in Milner's
bigraphs represents a containment view of the world; such a
representation is not an efficient way of addressing objects for ease
of locating them. On the other hand, it is a useful way to describe
the semantics of interfaces.

One of the values of the concept of multi-tenancy is the ability to have the host act
as a broker for mediating contact with the tenants. We see this in many
everyday scenarios:
\begin{example}
Interface scenarios:
\begin{itemize}
\item A security checkpoint at the mouth of a host building, or secure area.
\item Passport control at the airport interface of a host country.
\item Gated access via host to locked tenant storage units.
\item The eyes and mouth of a host organism mediate contact to the tenant organs: brain, stomach, etc.
\end{itemize}
\end{example}
A tenancy leads to a form of super-agency seeded on (and mediated by)
a host.  The host acts as a kind of moderator or proxy for certain
communications with the outside world, though it might not be the only
source of adjacency, if the scale of the tenancy relation is based on
promises that are made over a fabric of lower-level adjacency.

\subsection{Tenancy formation and privacy as an additional promise}

\begin{figure}[ht]
\begin{center}
\includegraphics[width=11cm]{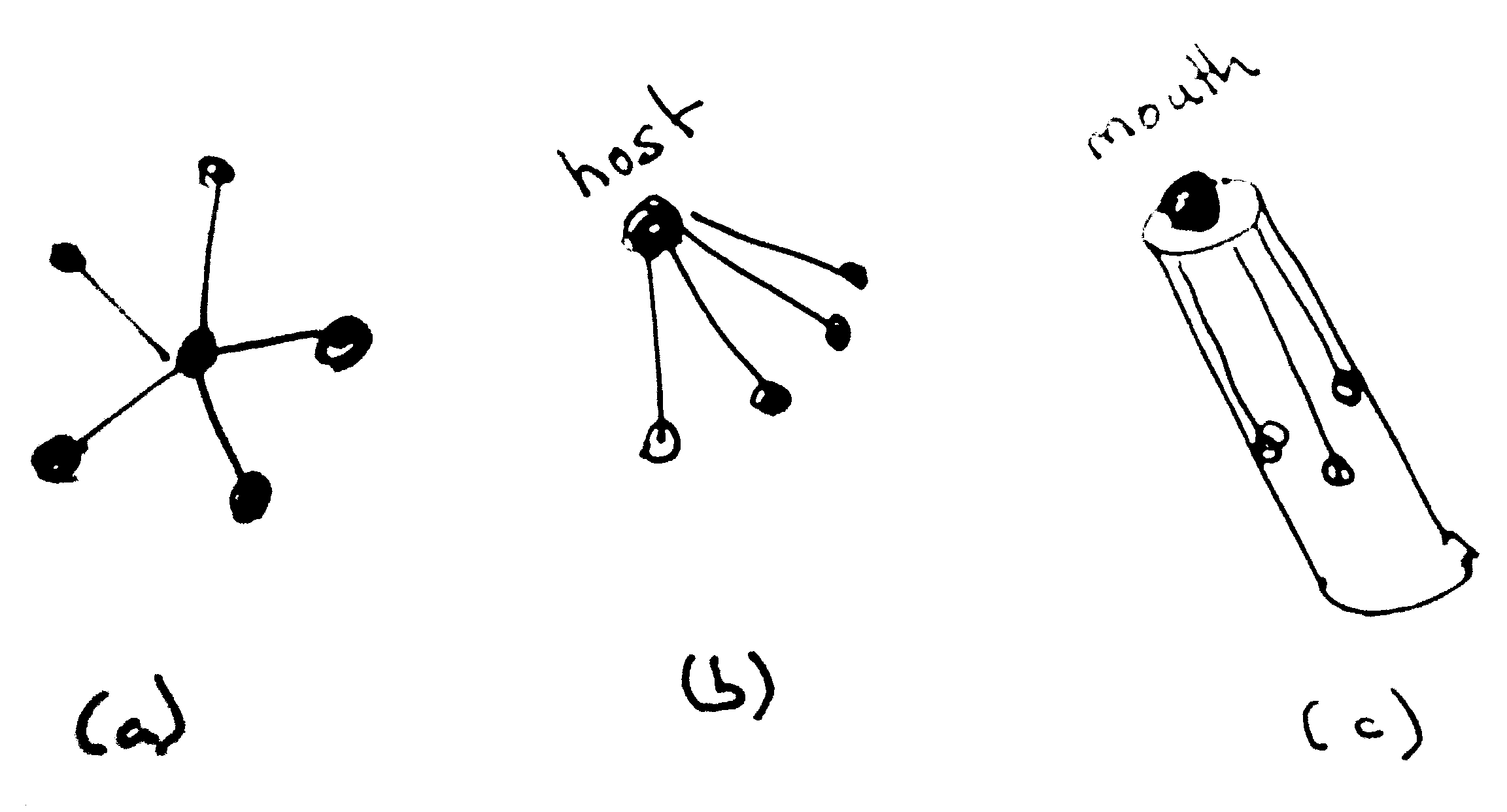}
\caption{\small How the preferred host role leads to an axial symmetry.
From spherical symmetry (a), a external flow selected a direction (b), which
eventually organizes along an axis. Segmentation along the axis then
marks out different levels of hierarchy in multi-stage tenancy.\label{folding}}
\end{center}
\end{figure}

There are two questions concerning tenant segmentation and containment in a hosted
super-agent:
\begin{enumerate}
\item {\em Who gets to decide whether a tenant or member can join a hosting collective?}
This might simply be part of the evolution of the design, or it might be a decision made by the host,
or indeed all of the tenants in concert.
\item {\em How are tenants kept disjoint from one another, and how is access to the tenants moderated?}
The connectivity involved in a tenancy with a host favours a radial symmetry
between host and tenants. This can be folded (see figure \ref{folding}) leading to the functional
asymmetry.
\end{enumerate}

The default assumption is that there is no cooperation  between tenants:
\begin{assumption}[Default null adjacency promise for tenant segregation]
The default adjacency state, at scale $M$, between tenants is no adjacency.
This is most easily accomplished in a spacetime gas phase. If tenancy
is built on a connected lattice in the first instance, then this
isolation might require additional promises to block adjacency.
However, the latter is a losing strategy: the amount of information
needed to `lock down' every agency is too large. You need to compress
the pattern into a list of exceptions.
\end{assumption}
Independent semantics of tenants might well go beyond this simple
dynamical observation however. Segregation of assets might be viewed
as being an important requirement of tenancy.
Naturally, this promise can only be kept by a single agent, hence
only host-mediated resources can be segregated as a promise to tenants.

\begin{figure}[ht]
\begin{center}
\includegraphics[width=11cm]{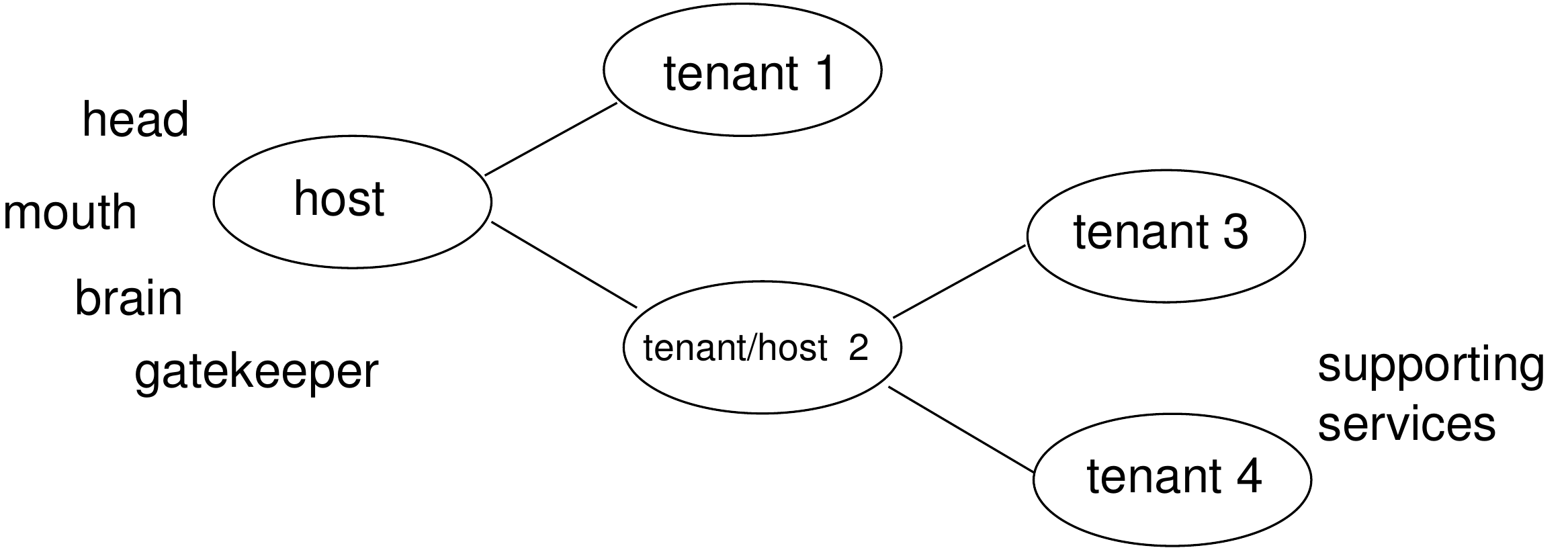}
\caption{\small Stacking tenancy branchings leads to an oriented access hierarchy. \label{brns}}
\end{center}
\end{figure}

If resources are mediated by the host, as is natural for this reason,
it can also act as a moderator, gatekeeper, or security monitor.  The
point of a gatekeeper is to limit adjacency to a narrow bottleneck, or
checkpoint (see figure \ref{brns}).  While this does forces a fixed
scale limitation on the choke point, it simplifies the semantics of
authentication.  The host has to be able to verify the identity of
tenants to keep its promises.  Having a gatekeeper interface at the
host is a simple way to do this.  This helps to turn the branching
process of the tenancy into a {\em hierarchy}, in which the host is
the gatekeeper to upper levels, mediating contact to the hosts below.

\begin{example}
  Many of the classic security blunders have been due to relying on
  the lack of addressability, in the belief that an item that cannot
  be named would not be accessed. This is a form of `security through
  obscurity'.  Systems that base isolation on prevention are much
  harder to police than

  As an exercise to the reader, consider what promises lead to `secure
  semantics'. How are keys or addresses for accessing tenants assessed
  by the host, and by users? What agency plays the role of gatekeeper,
  if not the host? Implement a `secure' system using conditional
  promises to segregate tenants.
\end{example}
The functional utility of the asymmetric tenancy structure thus seems to
lie the following observation:
\begin{law}[Hosting of input and output leads to axial symmetry]
  The functional arrangement of input/output mediated by the host
  leads to a natural head-tail asymmetry, in which the head is
  favoured in a hierarchy of longitudinal stages. This is known as
  cephalization.
\end{law}

\subsection{Spacetime sharing by tenants: serial time and parallel space}

When tenants subscribe to a resource in parallel, they share the
resources of the host.  This is called multiplexing in the resource
domain.  When a host-tenant network has net valency less than zero,
tenants can multiplexed in the time domain. Over-subscription of a
resource could lead to the need for time-sharing.  Time division
multiplexing is the way this is done in a serial queue\footnote{In
  data communications, so-called {\em frequency division multiplexing}
  is what corresponds to normal parallel resource sharing of $R$}.

\begin{example}
Time-division multiplexing (queue-processing) of an oversubscribed service:
\begin{itemize}
\item Time-sharing of apartments.
\item Car rental, or recycling of vehicles.
\item Aircraft/bus passenger seating is only fixed for the duration of a journey.
\end{itemize}
Finite-duration promises are the key to all queueing systems.
\end{example}

\subsection{Multi-stage multi-tenancy, and solid spacetime fabrics}

This section is a continuation of the iterated solid state tenancy
structures, used for locating agent addresses, described in section
\ref{addresses}. I return to the topic to emphasize that iterated
multi-tenancy leads to important scaling properties, both in the
information technology realm and in the biological realm.  Network
fabrics include toroidal networks, Clos networks, and Batcher-Banyan
networks\cite{banyan}. This example is about the popular 2x2 Clos
networks.

\begin{example}
  The example of a 2-valent Clos network is instructive because it is
  a dynamically simple case that is increasingly used in datacentres.
  Semantically, it is more complicated, because it combines
  multi-tenancy with multi-homing (multiple hosts) in order to create
  a cooperative self-organizing structure. To make addressing work as a
  repeated, tessellating pattern, each agent needs to be both a host
  and a tenant for different promises.

Today, Clos networks are intimately connected with a particular
implementation involving the Border Gateway Protocol (BGP)\cite{microsoftclos,juniperclos}. BGP is a
network route advertisement service that implements a set of promises
for promising route information between super-agencies known as
Autonomous Systems (ASN)\footnote{The similarity to promise
  terminology should not be a surprise: the two ideas are very closely
  related. BGP predates Promise Theory by many years, but through
  Promise Theory it gains a special clarity that cannot be seen when
  focusing on its irrelevant protocol.}.

Figure \ref{nclos1} shows the basic tenancy and dual homing. Two
adjacencies upwards, from each agent in the fabric (shown in bold),
connect each tenant in a lower tier to {\em two} redundant hosts in
the tier above (for resilience and load sharing).  Multiple
adjacencies downwards in a tier connect each agent, now in the
role of host, to its sub-tenants.
Thus there are two tenancies back to back promising resources:

The characteristic of the tree structure in a Clos network is that each
branch terminates at a definite `leaf location' with a definite and
unique address. This means that every agent in the pattern knows that
`down' means a specific location, and everywhere else is `up'. Hence, each
agent engages in two cooperative relationships, framed as tenancies:
\begin{itemize}
\item $R_\uparrow$: tenants forward messages that don't belong below me upwards for the host to aggregate and deal with. This could mean routing to one of the other parallel tenants, or it could mean sending the message out into the wider world beyond the host's boundary.
\item $R_\downarrow$: hosts forward messages that belong belong me downwards to the tenant they belong to.
They know which direction to send the message, because the tenancy requires this information to be
paid up as part of the condition $C$.
\end{itemize}
The tenancy boundaries thus lead to progressive layers of concentric
nested agency.  The agents (which are all network switches) play the
dual roles of host and tenant with respect to these different
services, in different layers. The tessellating pattern of all of
these woven into a fabric allows it to scale to number of independent
addresses that are greater than the fixed valency $v$ of any one
host\footnote{At the time of writing, the construction of a standard
  switch has valency of 48 possible tenants downwards, with fixed
  channel capacity, and a valency of two hosts upwards, each with
  greater capacity than the downward channels, for allow for
  aggregation.}.

When forwarding upwards and out, what was a host for the downward routing
is now a tenant of each of the lower layer agents that provide addressing data.
So the semantics of the roles are reversed depending on the process we consider.
\begin{figure}[ht]
\begin{center}
\includegraphics[width=10cm]{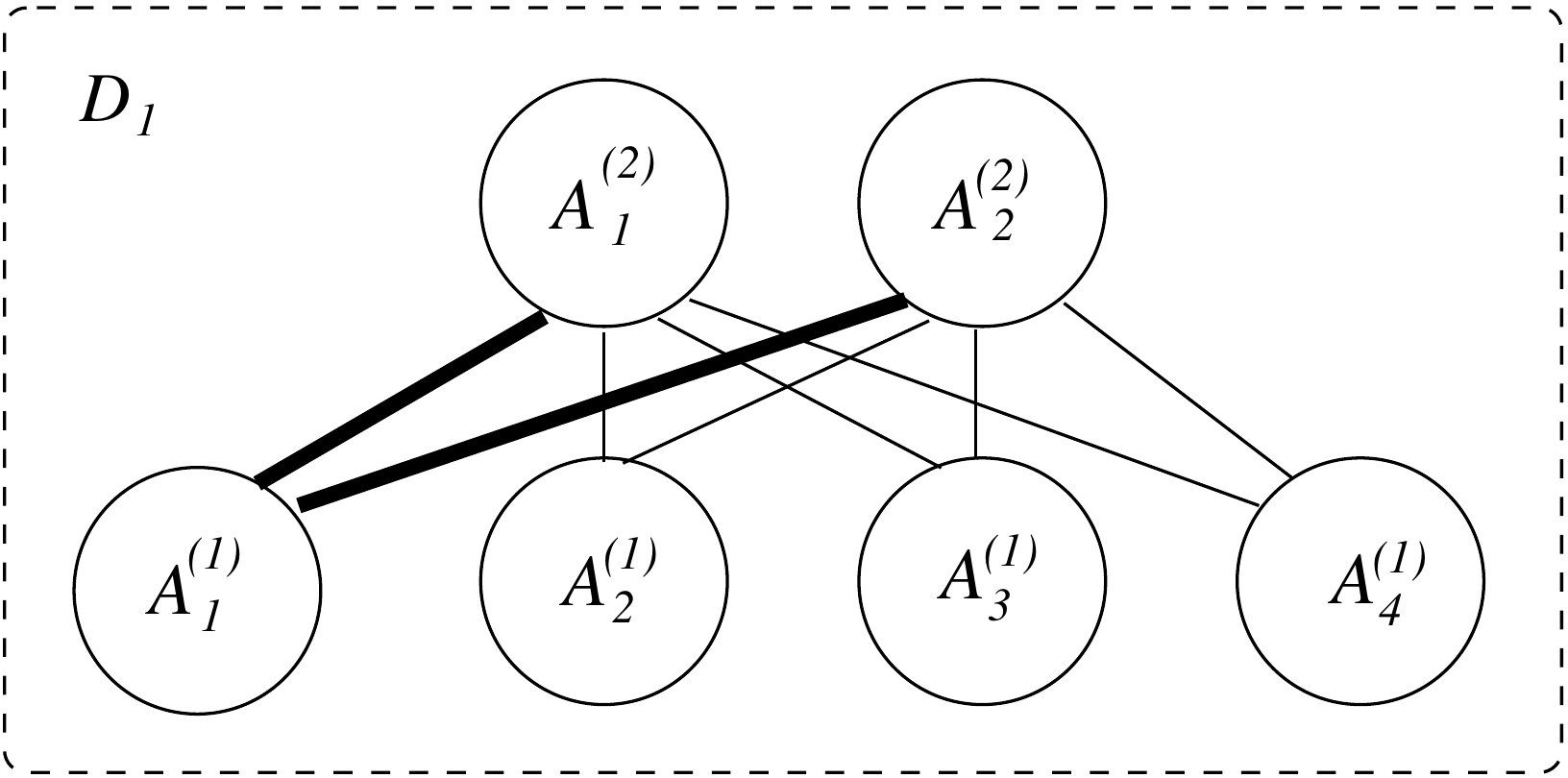}
\caption{\small Each agent promises to form tenancy agreements with two hosts
above it.\label{nclos1}}
\end{center}
\end{figure}

\begin{figure}[ht]
\begin{center}
\includegraphics[width=13cm]{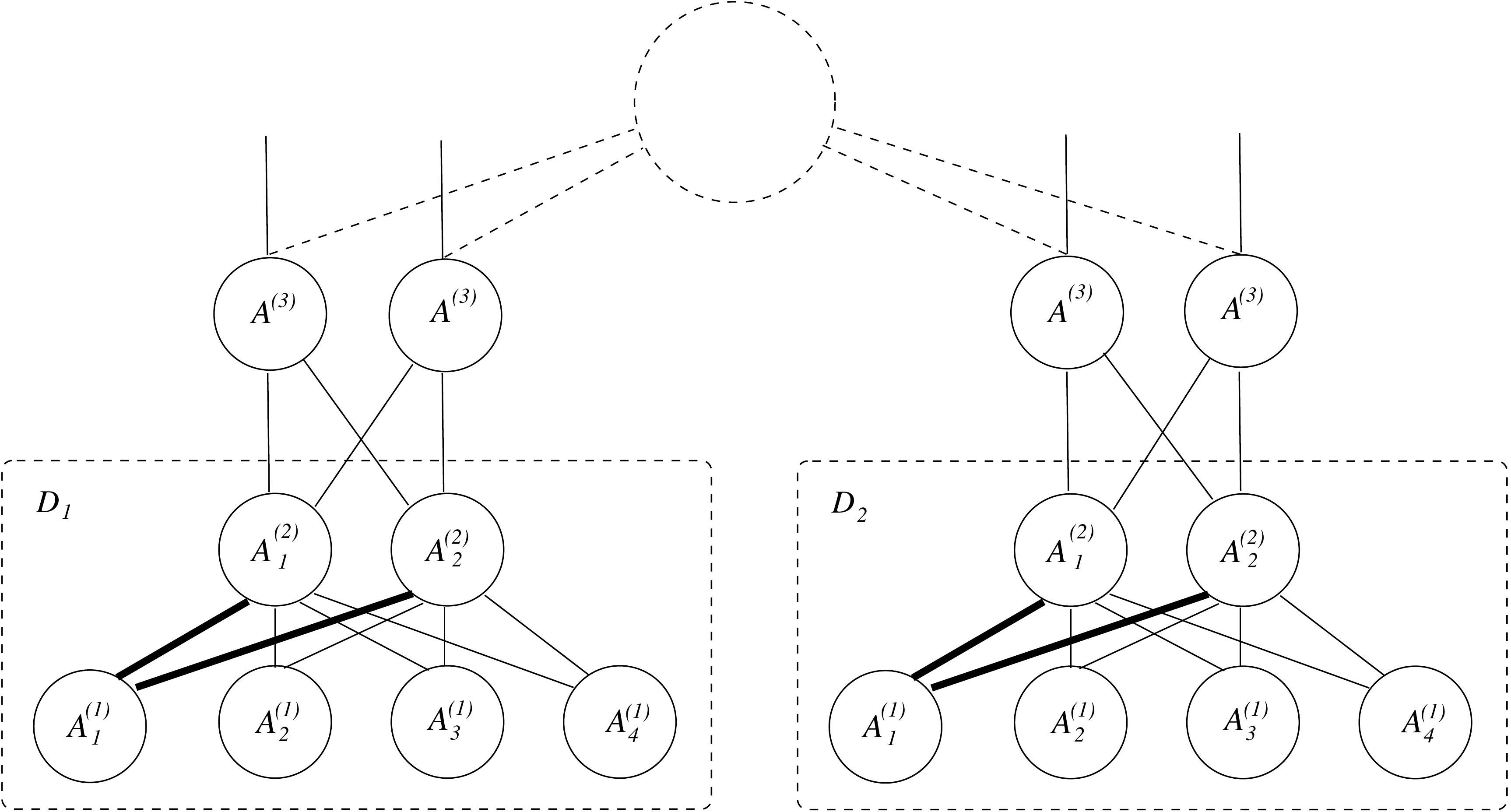}
\caption{\small The structure can be replicated across datacentre super-agents ($D_n$)
Notice that each tenant always has two adjacencies upwards, connecting it to redundant
hosts. Eventually, as a message goes up, it will reach the boundary of the Clos super-agent
and then the rules for regular (solid state) addressing must change.\label{nclos2}}
\end{center}
\end{figure}

The promises all rely on the correct positioning of the nodes to work.
In practice, the Border Gateway Protocol is typically used to
equilibrate that information and make sure the promises can be defined
relatively, with self-organizing consistency.

Keeping a simple notation for representing a pattern this complex is a challenge; however
we can illustrate the intent, along with the essential concepts of vector promises, and
tenancies. I denote the $i$th agent $A^{(n)}_i$ in tier $n$, either by $T^{(n)}_j$ or $H^{(n)_i}$,
depending on its role, where the indices $i$ simply run from 1 to 2 for the dual homed hosts,
and $j$ runs from 1 to the downward valency $v$.
Let's formalize the pattern using the multi-tenancy promises. We can label the tiers
by superfix $(n)$, and the tenants/hosts within a tier by subfix $i$.
Without taking into account the vertical edge conditions, we can write the tessellating 
mutual tenancies by the pattern:
\beq
T^{(n)} &\promise{+R_{\uparrow}^{(n)}(H^{(n+1)}_{2i-1})\#2|C_\downarrow}& H^{(n+1)}_{2i-1}~~~~\forall\,i = 1, 2\\
T^{(n)} &\promise{+R_{\uparrow}^{(n)}(H^{(n+1)}_{2i})\#2|C_\downarrow}& H^{(n+1)}_{2i}\\
H^{(n)} &\promise{R_{\downarrow}^{(n)}(T^{(n-1)}_{j})\#v|C_\uparrow}& T^{(n-1)}_{j}~~~~\forall\,j = 1\ldots v\\
\eeq
Downward forwarding assumes that an agent will only accept an address for forwarding
if it knows that there is a path to the address below it. That information could be
hardcoded into the pattern. In practice what happens is that BGP broadcasts this
information upwards from the bottom to the top, so that each agent tells the two agents
above it (its redundant upward forwarding hosts) which names and addresses it can forward to:
\beq
T_{(n)} &\promise{f_{\uparrow}^{(n)}(H^{(n+1)}_{1,2})\#2}& H^{(n+1)}_{1,2}.
\eeq
The summary of the parts may be written in terms of these vector promises:
\beq
R_{\uparrow}^{(n)}(A^{n+1}_{i}) &=& +\text{\rm forward up to $A^{n+1}_{i}$ if it represents the best path (route)}\nonumber\\
R_{\downarrow}^{(n)}(A^{n-1}_{i})\#v &=& +\text{\rm forward to best path (route)}~~~~\forall\,i = 1\ldots v\nonumber\\
C_{\uparrow} &=& \rm -R_{\downarrow}^{(n-1)}\; \text{\rm i.e. accept the upward forwarding from below as a trade (ACL)}\nonumber\\
C_{\downarrow} &=& \rm -R_{\uparrow}^{(n+1)}\;\text{\rm i.e. accept the downward forwarding from above as a trade (ACL)}\nonumber\\
f_\uparrow(C_\downarrow) &=& \text{\rm Inform about known addresses from below (BGP)}
\eeq
Each $v$-valent agent is a host to the $v$ agents below it, and a tenant of the agents above it.
The upward valency is 2.
As long as we are far away from the lower edge of this pattern, each host also knows that it has
two possible routes downward to its tenants also, because of the interwoven tenancy agreements.
However, at the bottom edge, there is only a single adjacency to the final address.

Throughout this patterns, the (-) use-promises play the role of access
control lists (ACL) for accepting data. I have suppressed most of this
to avoid overwhelming with detail; however, those details are
important the security and autonomy of the fabric.  Without individual
tenant control over its choices, the fabric becomes a homogeneous and
isotropic solid state space, with long range order. Certainly, this is
a good illustration of how such order is valuable, both semantically
and dynamically, but it doesn't really explain how it comes about in
practice between uncooperative agents.

When the Internet Protocol was designed, the routing of messages in this way
was not conceived in such dense and regular spaces. The Internet was assumed
to be a sparse network with clusters at the edges. Gradually, however, as the density
increased, and super-agent boundaries were drawn around organizations, a cooperative
agreement protocol (BGP) was introduced for mutual benefit. The sole effect of this
protocol is to propagate homogeneity along point-to-point adjacencies.
Thus, today, a Clos fabric is implemented as a BGP multi-tenant array, just like
a tenancy between two utterly independent organizations who want to cooperate
in order to forward messages within a shared address space.

Although we think of the Internet as having a single global address
space, there is no reason why this should be the case. What makes it
true in practice is the need to cooperate between `peers', i.e. between
private super-agents who can work symbiotically.
\begin{figure}[ht]
\begin{center}
\includegraphics[width=8cm]{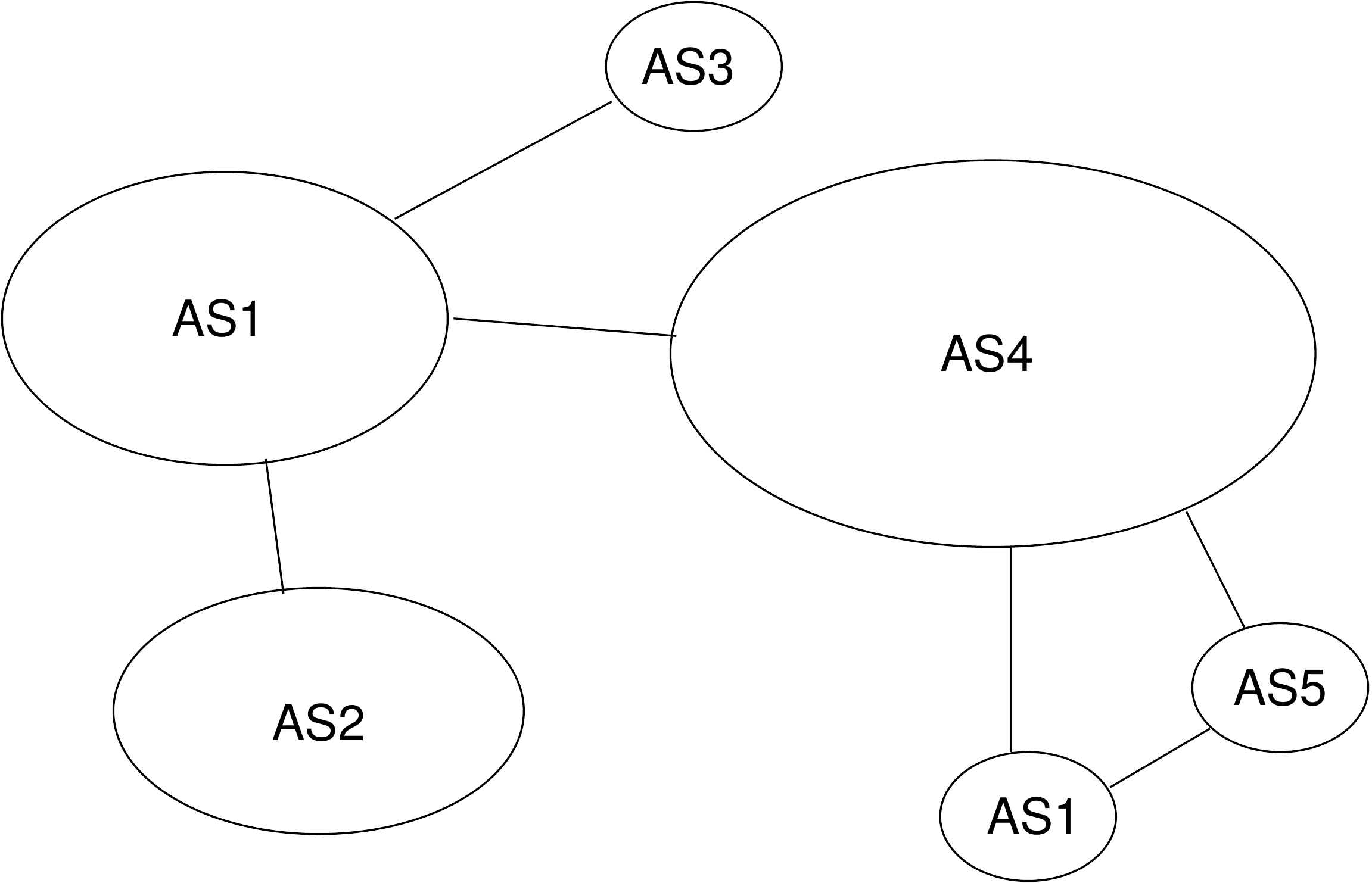}
\caption{\small The large scale structure of agent spacetime could
  easily have different coordinate patches, falling into independent
  namespaces (called BGP autonomous systems (AS)). Each super-agent
  (AS) can choose whether to cooperate with the community of
  addressing or go it alone. This has become common due to the poor
  scaling of the IPv4 address space, with Network Address Translation
  for partial splicing at the edges.\label{asn}}
\end{center}
\end{figure}

As discussed earlier, it is the breaking of symmetries that leads to
functional differentiation in a structure.  As a final comment on this
example of a Clos regular spacetime, we can note that there is very
little differentiation. So let's focus a moment on the natural
residual symmetry of the structure.  The functional asymmetry of the Clos network is a
compelling example of how multi-tenancy orients a structure, but it
has a less obvious cephalization. The structure has no obvious brain that
forms a master host for the entire structure, but the head is clearly at the top
or mouth of the outside world. The scaling is a
symmetrical interior scaling of the super-agent boundary: more of a
starfish than a cephalopod.

\begin{figure}[ht]
\begin{center}
\includegraphics[width=11cm]{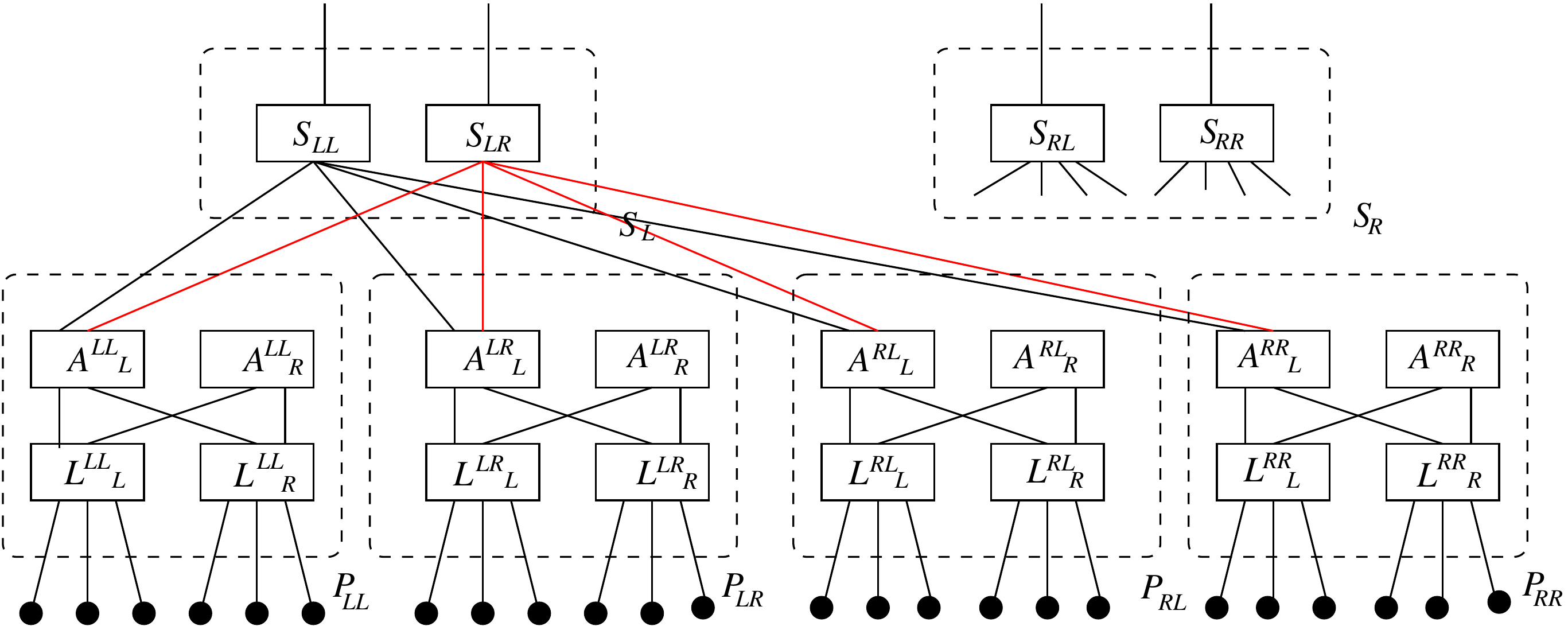}
\caption{\small A Clos network showing redundant multi-stage multi-tenancy.\label{clos}}
\end{center}
\end{figure}

Given such a level of long range order, and insignificant anisotropy
along the up-down left-right axes, it makes sense to study the
residual symmetry. On paper, we draw these networks as trees with
promise adjacencies that cross over one another (see figure
\ref{clos}), as if they were in a two dimensional Cartesian lattice.
However, we should not be fooled by the clumsiness of a paper drawing.
The fact that the adjacencies cross one another should be a sign that
this is wrong.  Tree-structures have a radial symmetry, and hence the
host-tenant decomposition lends itself naturally to polar coordinates,
centred on the host.

If we allow the structure to untangle itself, by going to three
dimensions (see figure \ref{clos2}), then its true structure begins to
make more sense. First the dual hosting can be symmetrized axially, going up and down instead
of just up. The then twisted pairs near the bottom of figure \ref{clos} can be untwisted to form
rings.
\begin{figure}[ht]
\begin{center}
\includegraphics[width=11cm]{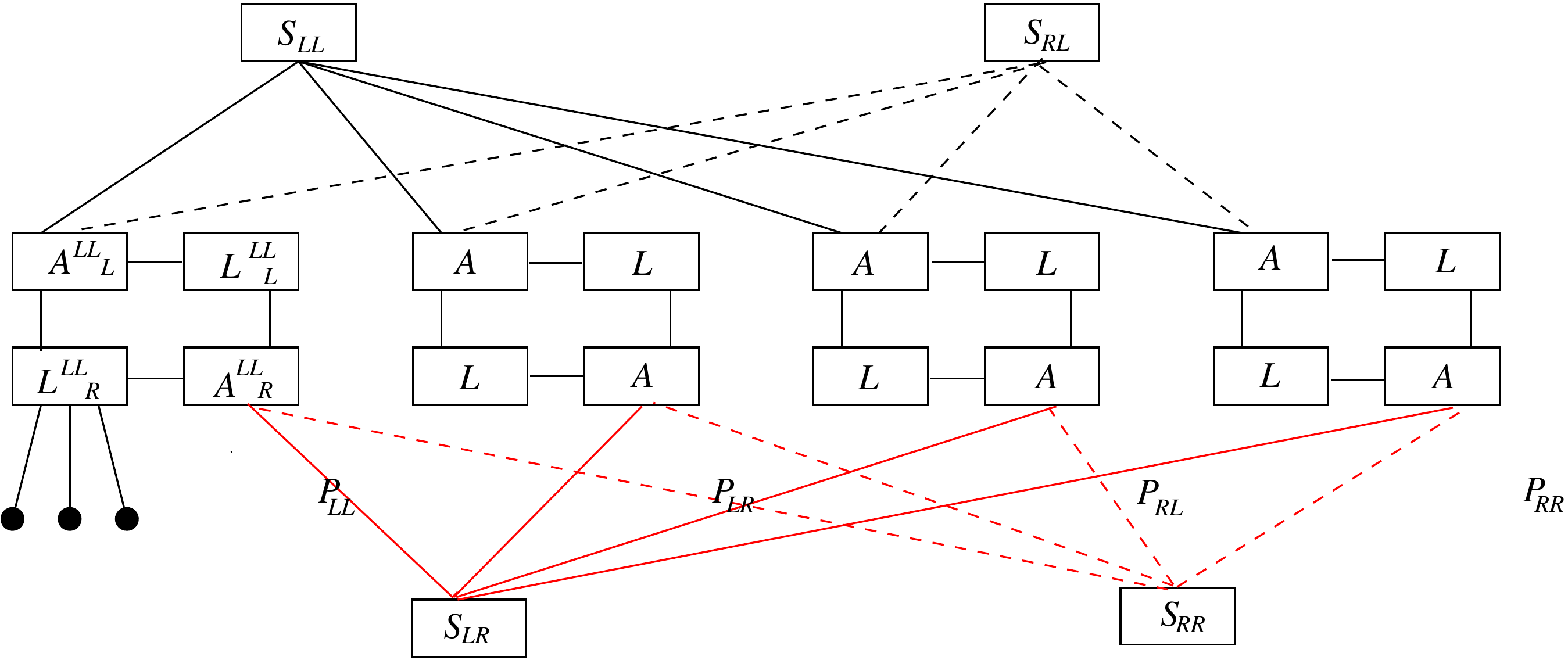}
\caption{\small A Clos network showing redundant multi-stage multi-tenancy.\label{clos2}}
\end{center}
\end{figure}
What one is left with is a hollow tube forming a toroidal geometry,
with symmetrical up-down mouths at the top and the bottom. Thus, we
have eliminated the quasi-cephalization of the structure and revealed
its natural form, which has no asymmetry. It can work top to bottom or
bottom to top interchangeably.
The final form is shown in figure \ref{clos4}.
\begin{figure}[ht]
\begin{center}
\includegraphics[width=8cm]{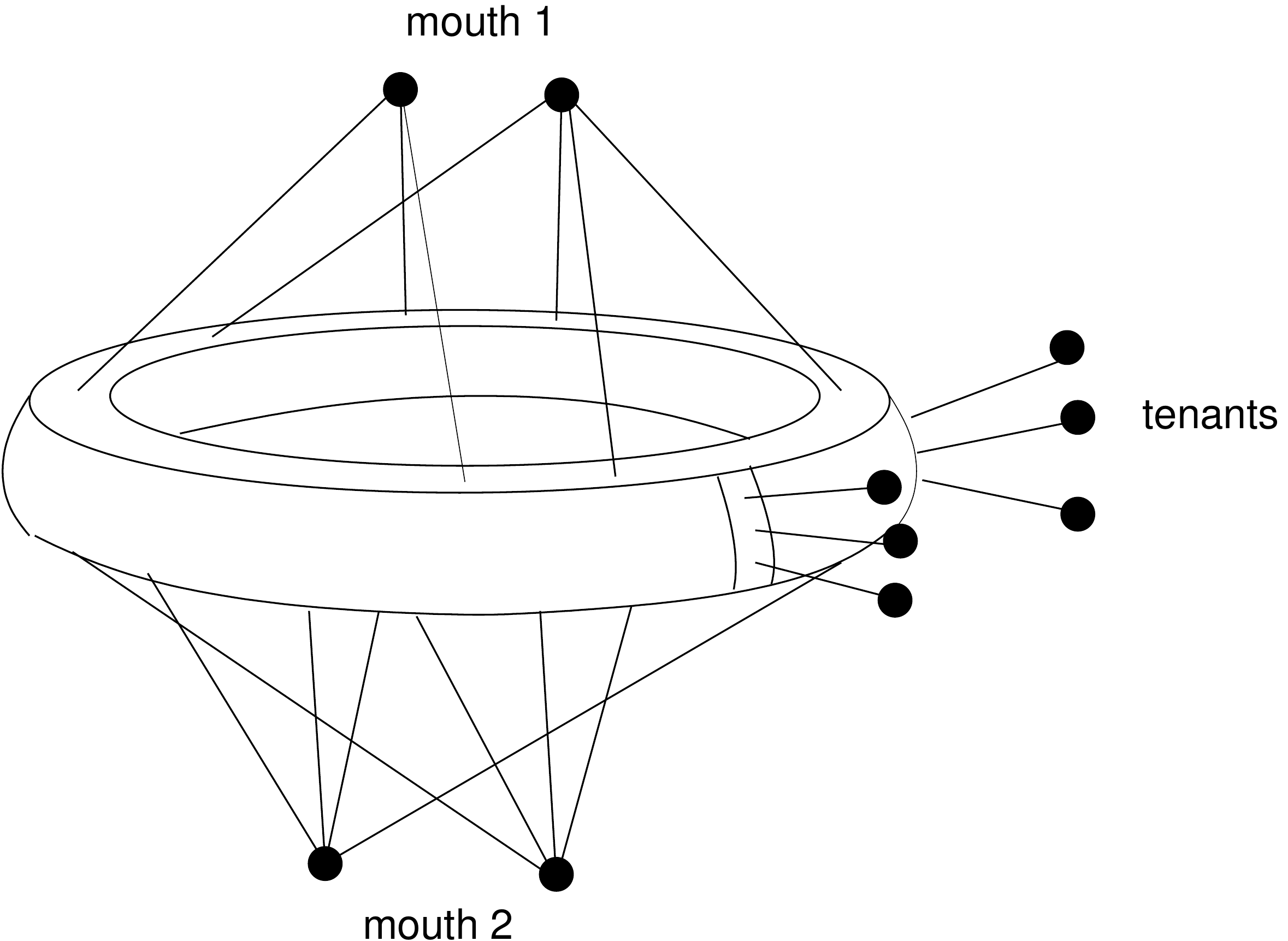}
\caption{\small Re-folding the radial symmetry into a non-oriented
  axial structure with no crossing paths. All connections are `line of
  sight'.\label{clos4}}
\end{center}
\end{figure}
This tendency for us to orient structures into hierarchies is a common
habit in human affairs. However, it often leads to scaling issues and
bottlenecking of promises.

It is interesting to think about how Software Defined Networking has
tried to recentralize network controls with brain-like controllers.
The annals of science would suggest this would be associated with a
natural asymmetry. But where is the asymmetry in a datacentre network
fabric? Where should the brain be located in order to fulfill its
roles as a rapid correlator of sensor input, and coordinator of
reaction?
\end{example}

\subsection{Collapsing private worlds (part 1: cross-cooperation)}

\begin{figure}[ht]
\begin{center}
\includegraphics[width=12cm]{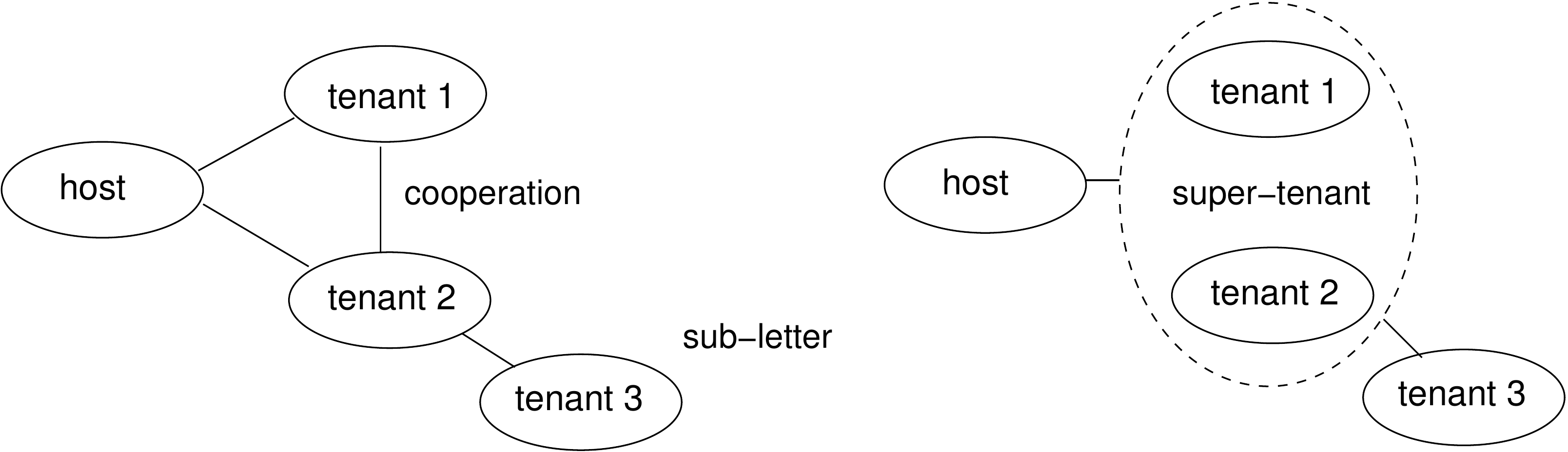}
\caption{\small Tenants might promise to cooperate and break their
  isolation. They are free to do this, unless the host agreement's
  conditions $C$ forbid it, in which case the host might cease to keep
its own promises.\label{ht2}}
\end{center}
\end{figure}

Tenants are assumed to be initially isolated from one another by
default. However, we need to ask at what scale is this isolation true?
The assumption might not be compatible with the underlying adjacencies
of spacetime, since the tenancy promises could easily be made on top
of an adjacency substrate.  
\begin{example}
Two students sitting an exam occupy desks next to one another, with
line of sight. They make no promises to communicate, but they are physically adjacent, not
isolated. Similarly, two rival companies share offices and computing
resources in a hosting unit. They make no promises to interact, they might even
promise not to interact, but they have an underlying adjacency making the
assumption ambiguous.
\end{example}
Segmentation, and non-interference rest on the default assumption
about tenants that they make no promises to one another, i.e.  that
their only notable promises are with the host only.
Sometimes separate tenancies may need to be combined. This process
can be carried out folllowing the scaling rules in earlier sections.
\begin{example}
Several military units under a common command are combined to 
carry out a mission, e.g. different NATO country members collaborate.
\end{example}
\begin{example}
The section of this document that you are reading now, and
the next, are separate tenants of the total host document. However,
they would better be written as a single section, and their isolation
can be worked around by making promises to cooperation (either through
the adjacency of the host, or otherwise).
\end{example}
If one tenant promises to sub-let part of its space to another, by
making a promise transverse to the longitudinal axis, this
is ok, as it does not affect the sharing promises of the host. If the
intermediary funnels resources from one tenant to the other, this
could undermine the host's intentions, but it can never cannot cause
the host to break its promise, thanks to the locality of promises (see figure \ref{ht2}).

However, the host might want to prevent this. It's channel for this is
in setting the conditions it is willing to accept from the tenants
(conditional $-C$).  Thus, at best, the host can try to protect its
own interests locally, and ask for the tenants to promise good
behaviour. It can punish bad behaviour (tit for tat) but it cannot
prevent it.

\begin{example}\label{radioexample}
Choice of agency semantics are a design decision in a functional world. For commodity items
like radios, television sets, watches, and so on, we design agencies to be easy to grasp by asking
the question: what semantics do we wish to expose to an observer?. Consider the illustration in
figure \ref{radio2}, showing two choices of agency boundaries for a common appliance.
The listener of the radio only needs to see the outer surface of the agency, not its internal
components. The battery is a component that needs to be changed frequently compared to the lifetime
of the device, so from the perspective it might seem to be a separable piece. However,
few users of a radio would want to have the battery alongside the radio as a separate entity.
\begin{figure}[ht]
\begin{center}
\includegraphics[width=9cm]{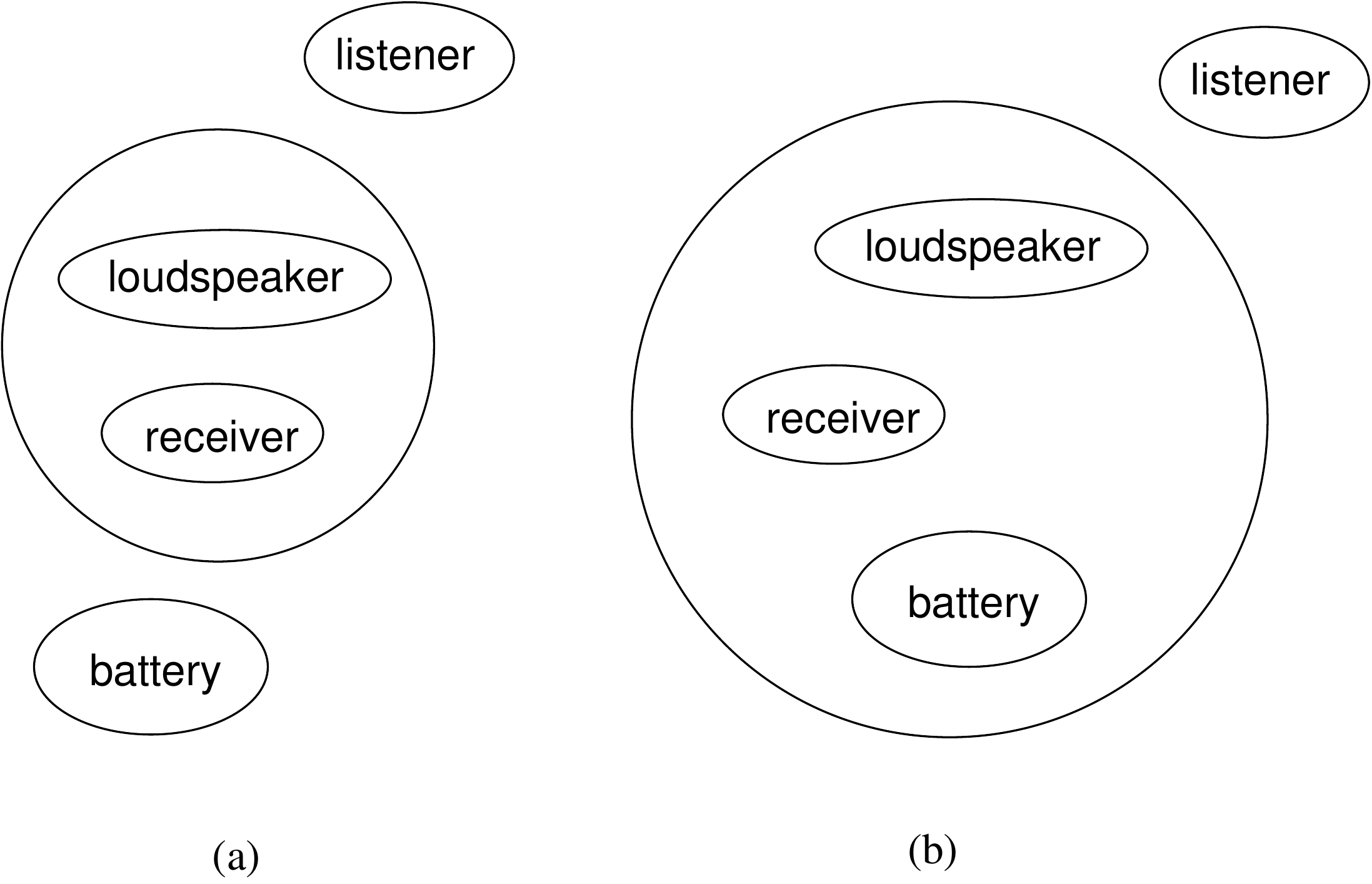}
\caption{\small Agency scales for a radio: (a) separate battery, (b) battery included.\label{radio2}}
\end{center}
\end{figure}
Thus, an agent in the role of radio listener (appliance user), a
single agency offers the preferred semantics of scale. However,
another agent in the role of radio-maintainer could prefer to see the
agent at a different scale, where replaceable components are exposed
as separable entities.
\end{example}

\subsection{Collapsing private worlds (part 2: failures of scalability)}

... is a cooperative promise.  It is not uncommon for mergers and
acquisitions of tenants to take place, in any realm of tenancy, so knowing that
this can be done without violating the isolation of any other tenants is important.

Following on from the subject of scaling agents in section
\ref{scaletenancy}, we can also consider how segregated tenants and
hosts behave when they promise to act as collective entities. Does
this change the tenancy relationship?
From the viewpoint of the host, the addition of a cooperative agreement
changes nothing.  Any resources are still provided as $+R_i$ and
$+R_j$ to tenants $i$ and $j$ respectively; thus sharing has to be
re-routed by the tenants to one another. This places them in the host
of mutual host for one-another, which might not be optimal in terms of
semantics (it is an unwanted complication of the design). The solution
for the long-term would be to renegotiate a new host-tenant relationship
for the new composite entity.

The asymmetric resource tenancy shown in figure \ref{ht2} is another
example of how a semantic space {\em remembers} the historical process
of its genesis, (recall figure \ref{agentscaling}). The resulting network is
a non-optimal promise bindings that are not simple scalings of the
intended relationships can come about by `hot-wiring' the promises to
work around the lack of scalability.  Such ad hoc cooperation can
solve a temporary need, but creates a new scale pinning, through the
asymmetry of the promise, which will further stifle scaling should
further super-agency

\subsection{Namespaces and hierarchies as multi-tenancy in identity space}

A namespace is a cooperative tenancy in which members are isolated
from their surroundings by a gateway host. The host promises to make
names of its tenants unique, usually be extending the names
hierarchically under the common umbrella of its own name. This is now
sometimes called `disambiguation' in software.
It is essentially the same as adding more lines to place addresses
so bound location by containment within.  It is a different strategy
opt uniqueness than tuple-coordinatization.

\begin{example}
  The same street names appear in many
  of the local towns, often based on names of historical figures, but this is not a
  problem, because each exemplar can make a unique address by adding
  the name of the local town to disambiguate from the others.
  
  Similarly, the `High Street' is a common fixture in British towns.
  Since most towns have a High Street, the name of the town and
  district can be added for uniqueness.
\end{example}
The term `namespace' is a popular concept in computing, but the concept
of namespaces are clearly in common and widespread use.

\begin{definition}[Namespace]
  A namespace $N$ is an isolated subspace of a semantic
space, formed from a collection of agents.  Inside a
  namespace, all agent coordinates and names are unique by mutual
  cooperation. 
\beq
A_i &\promise{+name_i | name_i \not= name_j}& A_j ~~~~~~ \forall i,j \in N\\
A_i &\promise{\pm C(name)}& A_j
\eeq
  Outside the namespace, names can be made unique by transforming the names
according to some bijective function, e.g. either
  extending the name with a boundary prefix identifier, or performing a name
translation.
\beq
A_i &\promise{+f(name_i, N) }& A_k ~~~~~~ \forall i \in N, k \not\in N\\
A_k &\promise{-f(name_i, N) }& A_i
\eeq
\end{definition}

The hierarchy implicit in nesting namespaces allows us to view these as tenancy relationships,
where a host is appointed as the gatekeeper mediating adjacency between the namespace and the
outside:
\begin{example}
Namespace examples:
\begin{itemize}
\item A family surname labels a namespace in which Christian names can be made unique
relative to other families. Today, this is not a very successful scheme, as it has not scaled
well to modern populations.
\item Filesystem trees, like the Unix or Windows hierarchical directories name sub-directories and
files tenants of their parents, in a forest graph structure.
\item Recursion with local variables in functions and subroutines uses the functional closure as
the host agent.
\item Subnets and networks form a two or more level hierarchy of
  attachment.  A layer 2 broadcast domain is a namespace, in which an
  IP router is the hosting gateway prefixing addresses with their
  subnet prefix.
\item Taxonomy and classification hierarchies use subject categories as hosting
agencies which contain sub-categories.
\item Programming class hierarchies, class member functions etc are hosted
within named objects, much like a file-tree.
\end{itemize}
\end{example}

\subsection{Topology and the indexing of coordinate-spaces in solid-state}

As we saw in section \ref{addresses}, addressing and tenancy are
related through the notion of routing between autonomous agents. The
basic asymmetry of tenancy is what allows message sorting by address
to implemented by cooperation, and the ability to scale this is therefore
connected to how we scale tenancy.

Branching hierarchies are the most common form of classification
sorting (`disambiguation') in information technology. They follow in
the Aristotelean tradition of taxonomy, widely embraced during the
19th century. It is not uncommon to shoehorn models into a tree
structure out of habit. This should not be necessary, however.  The
challenge of any coordinate system, for address encoding, is to mimic the
structure of spacetime in the naming conventions of the points. This
enables predictability and emergent routing from local autonomy.  The Cartesian
tuple-based coordinates discussed in paper I are flexible, and are not
tied to any particular origin.

\begin{figure}[ht]
\begin{center}
\includegraphics[width=8cm]{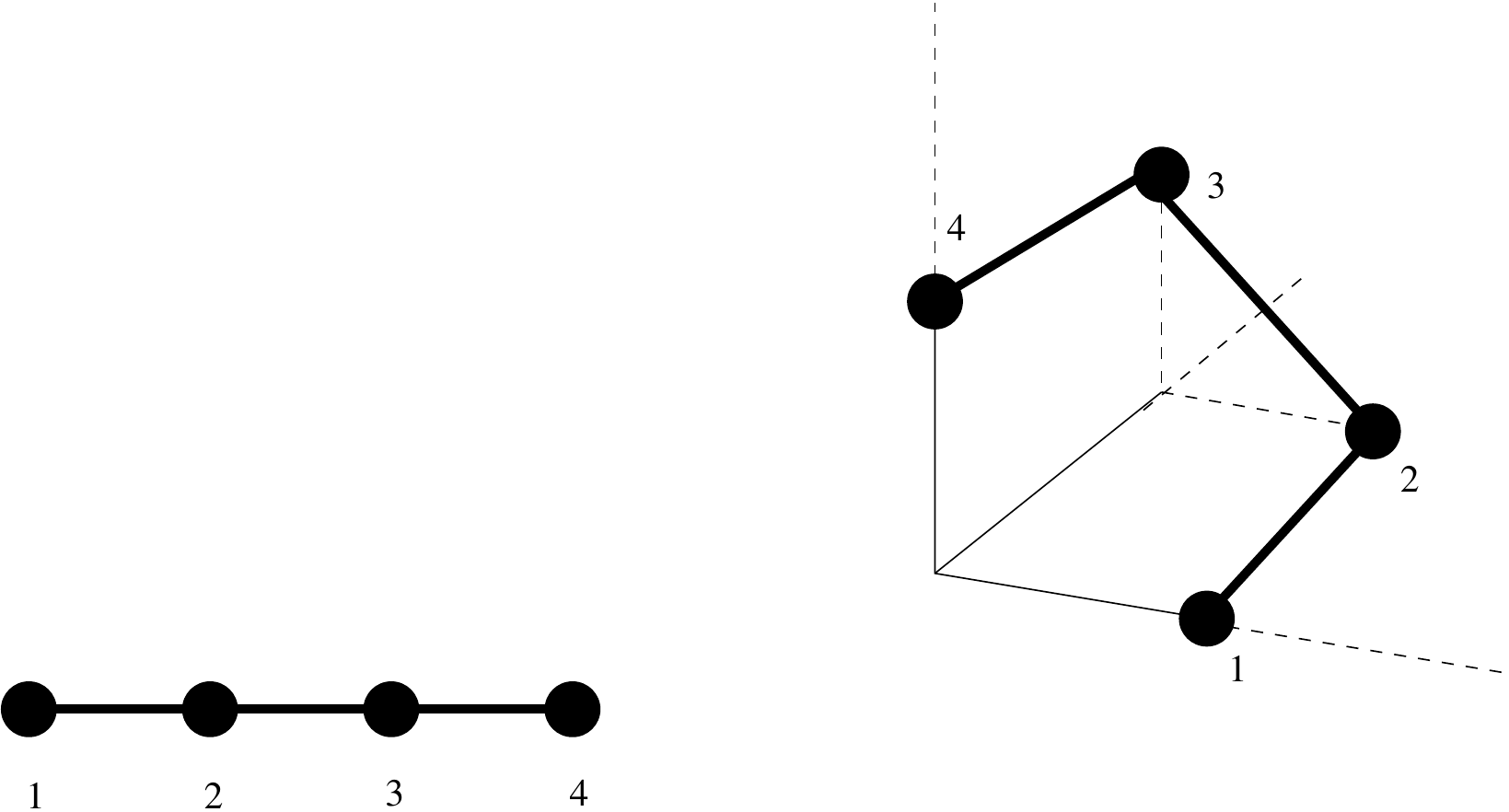}
\caption{\small Even though the actual occupancy of the Cartesian lattice is
sparse, we can use tuple coordinates with a convenient dimensionality. In this example, 
the only occupied points are (1,0,0), (0,1,0), (1,1,0) and (1,1,1). Using
any other values is simply defined to be disallowed.\label{line}}
\end{center}
\end{figure}

Since sub-agent names are interior (local) to a given super-agency, they can be
organized as that agency sees fit. If the size of the namespace is not too
large, almost any naming scheme is workable. However, as the number of
internal sub-agents grows, the efficacy of tuples depends on the internal
connectivity. This need not be a hindrance to using tuples for the following
reason: tuples can always be fitted to a spacetime as a covering, with a possibly
complex boundary. Put simply, even if the tuple space is not fully populated
with points in a Cartesian lattice, one can use the points that are available
and ignore the others (see figure \ref{line}).

A namespace can be covered by an arbitrary set of coordinate tuples of
the right size so as to span all agents in a spacetime uniquely,
giving each agent its own unique ID. For, example, a tree
reference
\begin{quote}
/tenant/container/sub-component
\end{quote}
is trivially written as a tuple:
\begin{quote}
(tenant, container, sub-component)
\end{quote}
The principal different between these two forms is that the order of the tuple components
does not imply any particular ordering of dominance or containment. An agent can approach
the identification of points by iterating over the tuple members in any order to parse the
space according to the desired semantics.

\section{Applications of multi-tenancy}

With a number of tools, principles, and concepts for multi-tenancy
under our belts, we are now approaching a vision for how one could
design and operationalize environments, both as isolated `organisms',
and as fully connected ecosystems of autonomous agencies.

Let's consider some examples to illustrate these points, focusing
particularly on the world of information infrastructure, or what is
now called `cloud computing'. These cases have immediate utility.
We can summarize the basic principles covered so far as a number of
points to cover for each analysis:
\begin{enumerate}
\item Determine language of promise bodies.
\item Determine the conceptual head of the organism, from input/output flow.
\item Describe segmentation or tenancy/sharing relationships.
\item Explain the role of spacetime phase (solid, gas, hybrid, etc).
\end{enumerate}
In these examples, I will sketch out some outlines only, leaving the details
as an exercise to the reader. Completing a full analysis of a
comprehensive system would be a significant undertaking.

\subsection{Example: Processing element in IT infrastructure}

\begin{example}
  Modern IT infrastructure (what is now termed cloud computing) is
  built from arrays of processing nodes, storage devices, and network
  switching devices. There is a plethora of terminology which I will
  try to avoid. In this example, I want to focus only on the
  processing nodes. A processing node (sometimes called a `compute
  node' or `machine') generally consists of a physical computer (also
  called a `server' for historical reasons). It acts as a host which
  promises an operating system ($+f(C,R)$) providing tenancy to {\em processes}.
  These sometimes consist of virtual machines or `containers' ($+R$).

  The purpose of hosting these machines and containers is to support
  the running of software applications within the tenants. For now,
  I'll disregard the details of the applications and simply assume
  that there are isolated processes (see figure \ref{appinfra}).

\begin{figure}[ht]
\begin{center}
\includegraphics[width=11cm]{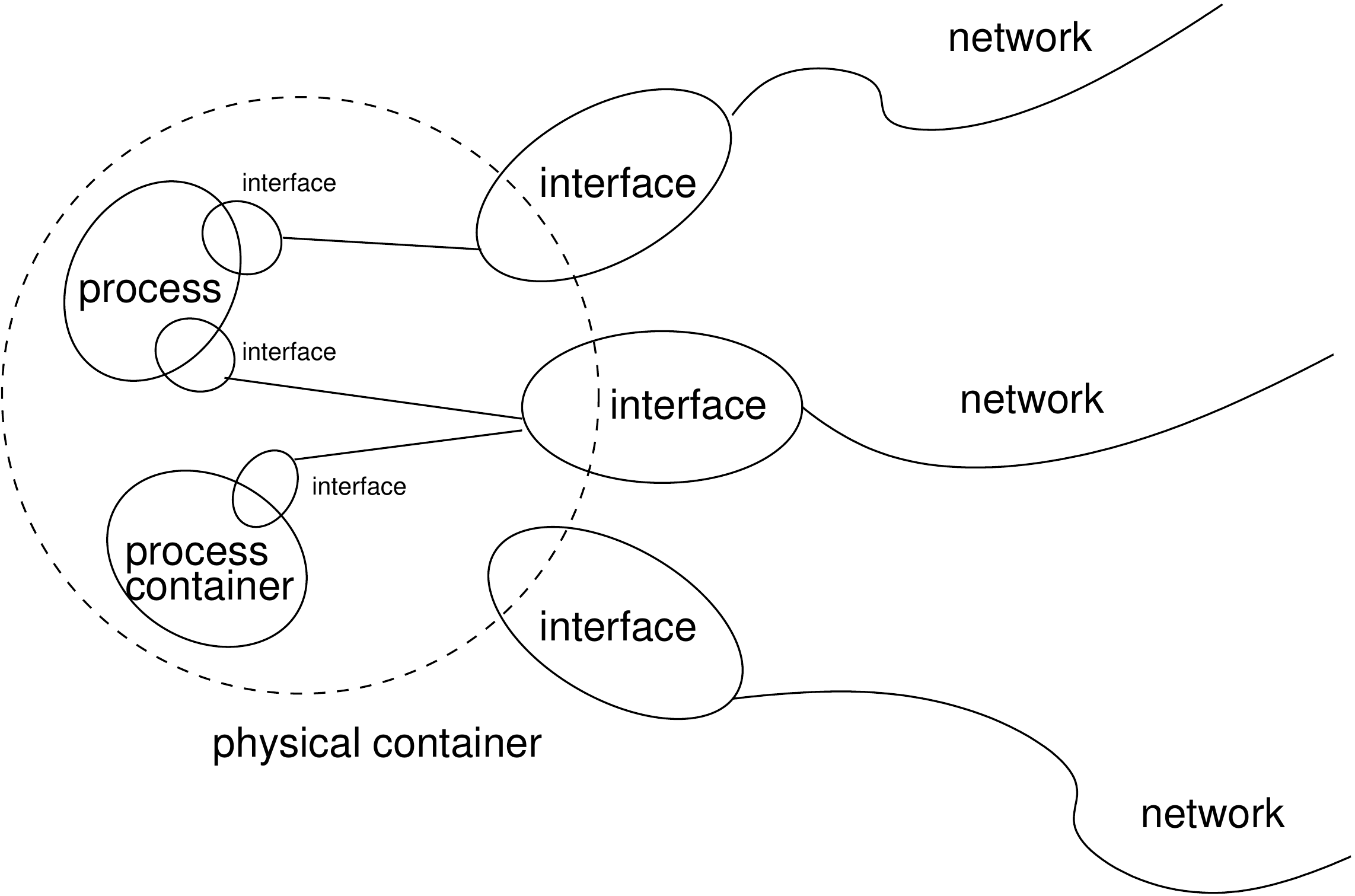}
\caption{\small A tenant-oriented infrastructure. Notice the limited self-similarity
between the process container scale and the physical container scale.\label{appinfra}}
\end{center}
\end{figure}

An array of computing machines can be thought of as a semantic
spacetime, with each host representing a location.  Taking the view
that the hosting service is a `front' for the internal tenants, each
host may be modelled as a super-agency, with internal structure
consisting of the tenants. Each host has a name and an address, and
the processes inside them have names and addresses too.

Let's consider the four points:
\begin{enumerate}
\item {\em The language of promise bodies.}

  We always start by expressing promises that we know we can keep.
  Promising something vague or undefinable is an act of self-sabotage.
  Simple promises, close to the capabilities of the agents are
  preferred.  Here, the alphabet of promises may include promises to
  execute programs at a certain rate (based on the CPU clock speed), transmit data, isolate processes
  from one another, etc.

\item {\em Determine the conceptual head of the organism, from input/output flow.}

  The head of a processing node is where the interfaces to the outside
  world enter (see figure \ref{appinfra}).  These are connected
  directly to the kernel, and the kernel (represented by the dotted line) acts
  as a kind of brain for the hosting of the tenants. The tenants are
  the processes, which have virtual interfaces connecting them to the
  actual network interfaces, via the kernel.

\item {\em Describe segmentation or tenancy/sharing relationships.}

Segmentation of the host's body is in terms of the process containers. The isolation
is maintained by two-mode operation at the hardware level, and thereafter mediated
by the kernel.

\item {\em Explain the role of spacetime phase (solid, gas, hybrid, etc).}

  The phase of the spacetime is undefined here.  Inside a super-agent,
  processes come and go, addresses typically change, looking like a
  gas.  Outside the host, the Internet at large has the structure of a
  gas, for much the same reason. However, inside a datacentre, fabric
  design with regular symmetrical arrays (e.g. in leaf-spine Clos
  networks) and fixed networking has the structure of a solid.
This looks not unlike a biological cell, with structure floating around inside
and outside a boundary.
\end{enumerate}
How does this scale up? Scaling can be done in one of two ways: either
on the {\em interior} of the super-agent boundary of the host, or on the {\em exterior}:
\begin{enumerate}
\item Interior agents can be scaled by parallelizing each sub-agency
  of the super-agent into a larger sub-agent. The number of sub-agents
  grows linearly and the super-agents remains constant.  This is how
  we build a mainframe, or NUMA-scaled cluster. Each component is made
  stronger by adding parallel numbers, but the structure of the design
  is constant. It is low level redundant scaling, which is known to
  lead to best effort reliability, according to the reliability folk
  theorem\cite{burgessbook2,hoyland1}.

\item Exterior agents can be scaled by parallelizing entire hosts.
  This increases the total number of super-agents with constant number
  of interior sub-agents.  This is how we build a server
  farm, like a cloud environment.
\end{enumerate}
The main difference between these two is the adjacency structure, or
how communication between the agents is wired.

What can we say about the host-tenant asymmetries? Within each
super-agent, there are actually many host-tenancy relationships
working together.  Referring to figure \ref{appinfra}, we see a
hierarchy of agency at a typical host. The stippled circle illustrates
the super-agent boundary of the host (running the system kernel), and the
solid ellipses demark various sub-agents associated with it. The
sub-agents form a variety of multi-tenant bindings, with different
semantics. These are summarized in this table:
\begin{center}
\begin{tabular}{r|l}
Role of Host &Role of Tenant\\
\hline
Process container & Process interface\\
Physical container & Process container\\
Physical container & Physical interface\\
Physical container interface & Process container interface\\
Physical network & Process network\\
\end{tabular}
\end{center}
A process container houses several process interface tenants, each of
which pay by mediating a connection between the internals of the
application and the process via logical network channels. The process
network channel interfaces are themselves tenants of the physical
network in the manner of figure \ref{sharednet}. 

In a similar way, but at a coarser scale, the physical interfaces are
tenants of the physical host super-agent, mediating physical network
connections by direct analogy.  Ironically, the process interfaces
bind as tenants of the physical interfaces, drawing their agency from
the network resource provided by the physical interface host.  Such
reversals are completely unproblematic, as they are simply semantics
pointing towards a resource in a particular viewpoint of the collaborative ecosystem.

It is habitual in IT modelling to simply impose a hierarchy on these
resource agencies, from a single viewpoint; however, the imposition of
a hierarchy, without reference to the promises they make, leads to
viewpoints that masquerade as authoritative, but are not.  What one
learns from Promise Theory is that, for every asymmetric
relation, there is a complementarity transformation which reverses the
preferred interpretation and the natural ordering.

Let's focus on only one issue: addressability. Can we assign a
coordinate system of names in a way that is faithful to the structure,
and conducive to locating resources from any viewpoint?
The possible varieties of naming of agencies are extensive, though
not all are used in practice:
\begin{center}
\begin{tabular}{r|l}
Agency & Promised identifier(s)\\
\hline
Physical interface & $\lbrace$ MAC-addresses $\rbrace$, $\lbrace$ IP-addresses $\rbrace$\\
Process interface & $\lbrace$ IP-addresses $\rbrace$\\
Namespace & Namespace umbrella name\\
Physical container & Hostname\\
Physical IP-address & $\lbrace$ DNS translation names $\rbrace$\\
Process IP-address & $\lbrace$ Namespace translation names $\rbrace$\\
Process container & Process container name\\
\end{tabular}
\end{center}
Following industry practice, it is normal to coarse grain away several
of the distinctions between these promises. This leads to problems in
tracing the origins of promises made by some of the coarse-grained agencies.

Well-known problems associated with design of the Domain Name Service
(DNS) (especially reverse lookup) can be cited as an example of how
coordinatization based on perceived containers, rather than general
tuples leads to difficulty in tracing inter-agency collaborative
processes. I'll leave it as an exercise to the reader to design an
improved DNS service which acts as an invertible, based on the resolvability
lemma (lemma \ref{resolver}).
\end{example}

\subsection{Example: tenant-oriented hosting infrastructure}\label{finalsec}

In the previous example, the low-level resource container (hardware)
was considered to be a super-agent boundary for processes running
inside. However, we have great freedom to change the nature of a
hierarchy in Promise Theory \footnote{An analogy might be the
  following: do we consider the mind to be a tenant of the body, or
  the body to be a tenant of the mind? Dynamically (physically) the
  former makes sense, but semantically the latter is a highly
  convenient viewpoint.}.

Functional thinking is invariably top-down. Computer science teaches
top-down decomposition of problems through a logical branching
process. Often this leads to trouble at lower levels due to
inconsistency. The bottom-up aggregation process avoids inconsistency,
and permits logical scaling, however we don't interface with systems
from the bottom up.  The `cephalization' of agency around physical
connectivity tends to fix attention on the host and its role as a kind
of manager or brain.  However, the applications (the minds of the
system) are run inside the tenants, and there are more tenants than
hosts. From a human perspective, then, the tenants see their concerns
as a focal point, since they pay rental fees to support the host's
existence as their service provider. The customer service viewpoint thus
weighs in semantically, where as the engineering viewpoint of the previous
example was attractive dynamically.

\begin{example}\label{final}
  In a computing platform, design for hosting software systems,
  multiple applications run as tenants of an infrastructure provided
  for them. This separation allows delegation with cooperation,
  sometimes called DevOps: tenants deal with content development,
  while the host deals with operational delivery.  

  By analogy with the radio example \ref{radioexample} above, the
  agencies, which a user of the application would like to see, are
  shown in figure \ref{tenex2}. The agent, in the role of user, does
  not want to see the internals or even the `battery' that makes the
  thing work, it only cares about the surface boundary of the application,
and its exterior promises.
\begin{figure}[ht]
\begin{center}
\includegraphics[width=9cm]{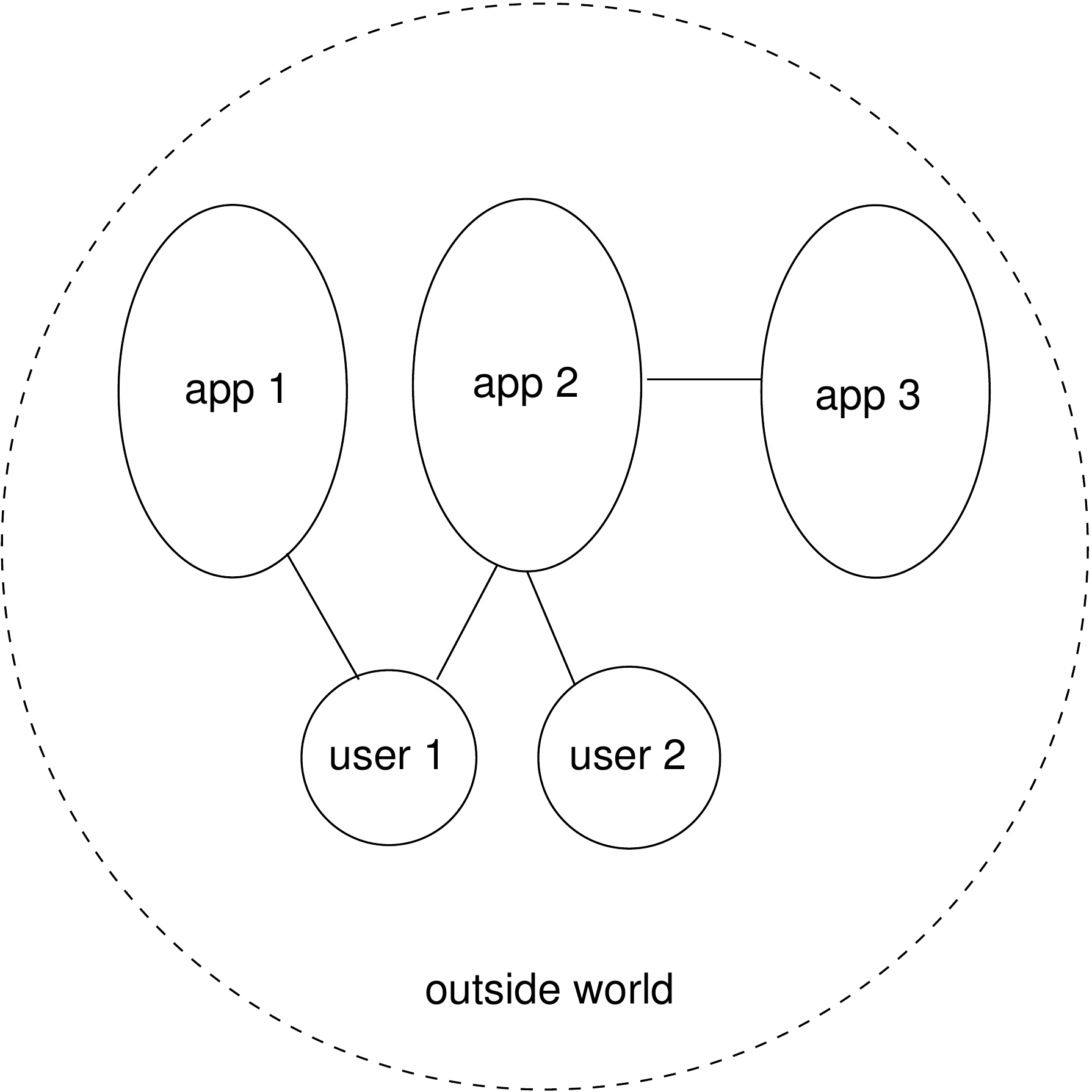}
\caption{\small The user-application interface scale hides all details of what makes the application
tick.\label{tenex2}}
\end{center}
\end{figure}
The illustration in figure \ref{tenex2} shows how tenants would like
to conceptualize spacetime, placing their concerns as the top-most or
outer-most interface. Once again, this is no limitation, since Promise
Theory tells us that we have significant freedom to turn hierarchies
and viewpoints upside down for convenience.  With this viewpoint, the
application tenant is in focus, and, from this user perspective, the
application subsumes its infrastructure within the outer boundary.  It
wears user semantics on the outside.

The four concerns are:
\begin{enumerate}
\item {\em The language of promise bodies}

  This will include the exterior promises made by the application to
  its users, and the interior promises made between the layers of
  hosts and tenants, described below.

  At each scale of agency, one can imagine creating a `compiler' from
  a domain specific body language to a more explicit pattern-generated
  explication of meaning for the lower level components.  \beq \beta_M
  \rightarrow  \beta_{M'} + \beta_M \eeq

\item {\em Determine the conceptual head of the organism, from input/output flow}

The head of the organism is now the application-user interface (see figure \ref{tenex2a}).

\item {\em Describe segmentation or tenancy/sharing relationships}

There are two layers of tenancy:
\begin{enumerate}
\item Application users are tenants of the application platform.
\item The application is a tenant of the infrastructure platform.
\end{enumerate}
The tenancies are based on approximately the following trades. For
the user-application tenancy:
\beq
R &=& \rm application\, account\, login\nonumber\\
C &=& \rm user\, identification\, credentials (money?)\nonumber\\
f(C,R) &=& \rm application\, functionality
\eeq
and for the application-platform tenancy,
\beq
R &=& \rm platform\, login\, by\, application\nonumber\\
C &=& \rm application\, identification\, credentials (money?)\nonumber\\
f(C,R) &=& \rm pay\, by\, use\, computer, storage, network\, resources
\eeq
Maintainers of the application, and maintainers of the infrastructure
on which the application resides will make different choices,
depending on what semantics they wish to expose.

\begin{figure}[ht]
\begin{center}
\includegraphics[width=8cm]{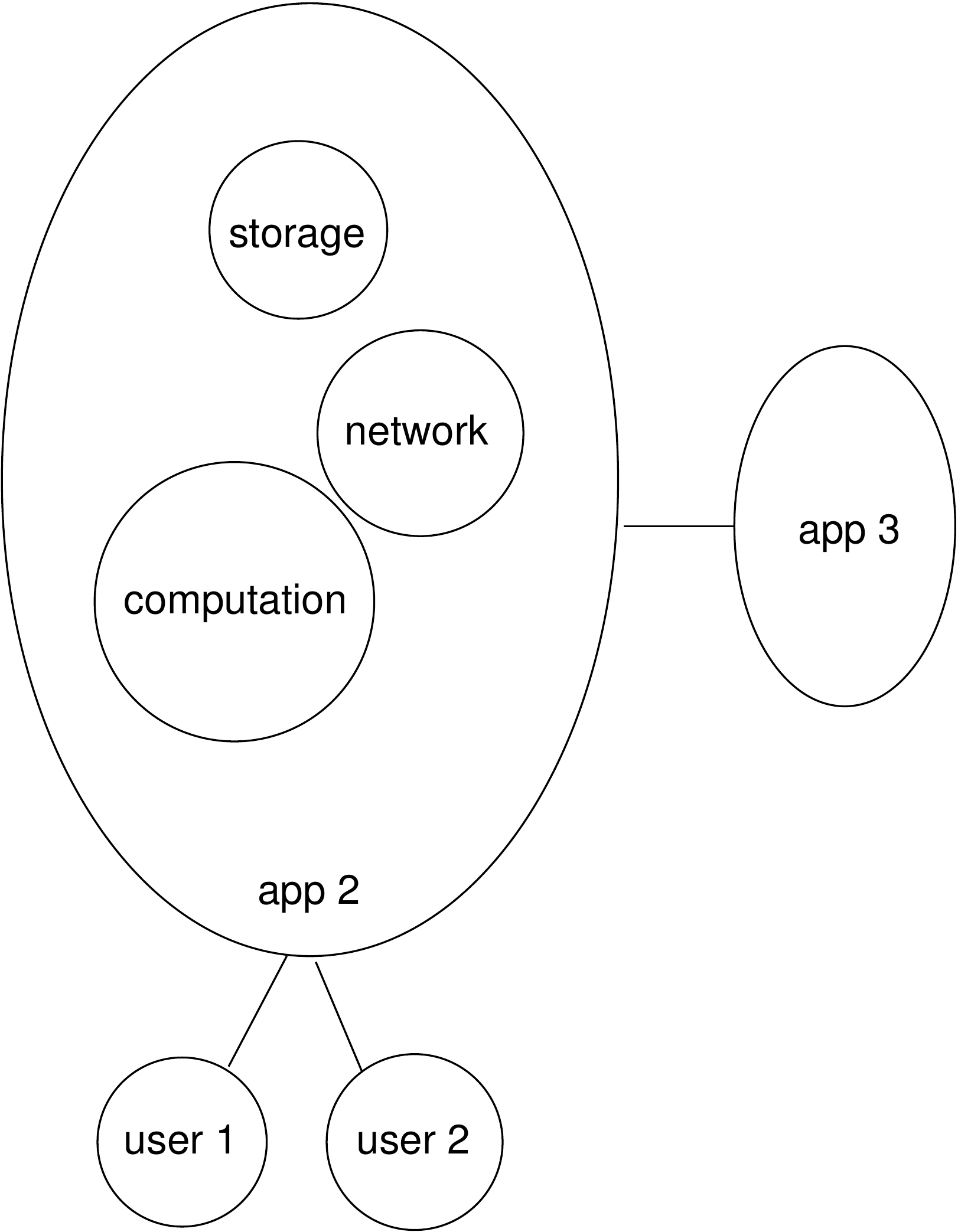}
\caption{\small The application-platform interface exposes the resource knobs
needed by application.\label{tenex2a}}
\end{center}
\end{figure}
At the level of computers, network, and storage, the picture
might look something like figure \ref{ecosys}.
The infrastructure is composed of three types of agent: machine,
storage, and network, which make promises accordingly.
Applications are thus tenants of the platform infrastructure, and
the main resources are represented the infrastructure scale in figure \ref{ecosys}:

\item {\em Explain the role of spacetime phase (solid, gas, hybrid, etc)}

From the application perspective, this is a single atomic agency, ready
to bind to a user in a gaseous state. Inside the application, the state
of the interior depends on the underlying infrastructure that powers the 
application.
\end{enumerate}

The agency scales of interest are:
\beq
{\rm User}: A &=&\rm \{ application, user \}\\
{\rm Application}: T &=&\rm \{ computation, storage, network, namespace, \nonumber\\
                     &~& ~~~~~\rm application, user \}\\
{\rm Infrastructure}: M &=& \rm \{ process\, container, disk, router, application, \nonumber\\ 
                     &~& ~~~~~\rm box, computation, storage, network, namespace, \nonumber\\ 
                     &~& ~~~~~\rm interface, names/addresses, application, user \}
\eeq
Notice how, when increasing levels of detail, the higher levels are not replaced by the lower
levels, since they still contain unique information (see section \ref{irreducible}).
Also, here I have introduced a semantic category scale `box' to symmetrize
over the resource providers that are interconnected by network channels (see figure \ref{tenex3}).
\beq
{\rm Box}: A &=&\rm \{ computer, storage, router, interface \}
\eeq
In addition to these `physical' agencies, there are other more abstractly defined agencies at play
within an information system.
\begin{itemize}
\item Names and addresses of machines and storage (IP/DNS addresses), for use by tenant name-services.
These are private per application.
\item Data from machines and storage, for use by tenant networks.
\item Transport of data, for use by computer machines and storage (also represented as super-agency `box').
\item Directory service lookups, for use by computers and storage, in order to locate others in their private
network.
\end{itemize}

\begin{figure}[ht]
\begin{center}
\includegraphics[width=8cm]{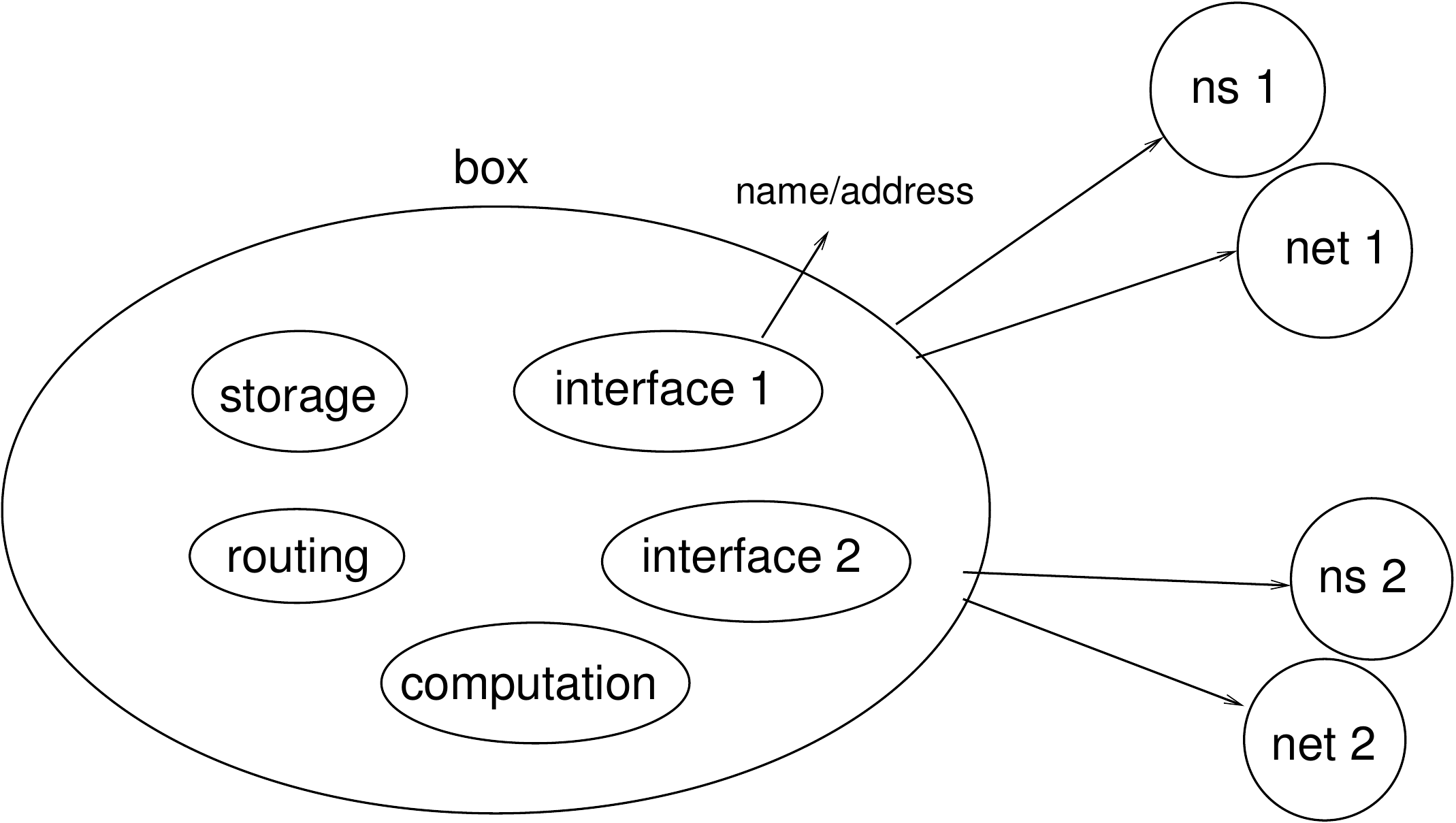}
\caption{\small The network bindings to resource containers\label{tenex3}}
\end{center}
\end{figure}

The network binding to a resource agent has the form of a tenancy also
(see figure \ref{tenex3}), as a process container might be able to
bind to multiple networks, in principle. An application might choose
to branch its logic into multiple private channels, just as it chooses
to branch sub-functions into private branches.
In this reverse viewpoint, like in the previous example, applications
can be considered to form private tenancies on top of the common
spacetime, by using the intermediate agencies of the spacetime as
proxies.

IT applications share a common infrastructure, provided for all
applications by mutual cooperation. This forms a semantic spacetime,
which is entirely interior to what we may call Application Space.
Each application creates its own private world, as a tenant of the
application itself.

\begin{figure}[ht]
\begin{center}
\includegraphics[width=12cm]{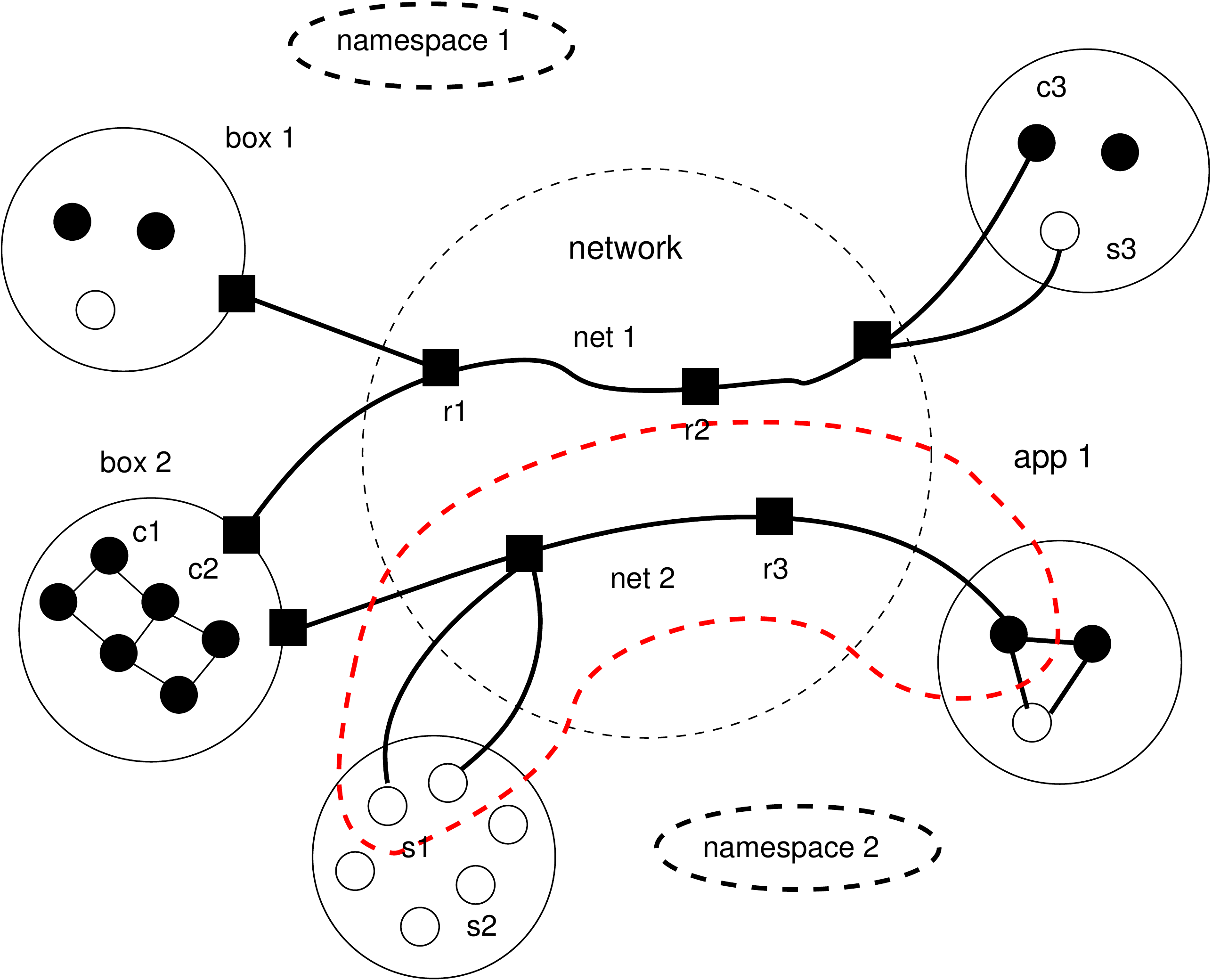}
\caption{\small An IT infrastructure network, supporting
  multi-tenancy.  Encircled clusters of agents are super-agents show
  the tenancy scale. Inside these is the infrastructure scale: filled
  circles are computer nodes, unfilled circles are storage nodes, and
  squares are routing nodes.  Network terminations at the box level
  are distributive to their constituent members.\label{ecosys}}
\end{center}
\end{figure}

Applications may thus be viewed as `atomic' super-agents, either
locally or globally; thus, from a promise perspective, it is
irrelevant whether the architecture is monolithic (implemented inside
a closed space) or distributed (without fixed boundary). The promise
abstractions are identical in both cases; the only difference is one
of scale.

\end{example}

\section{Epilogue: abstract agents, and knowledge modelling}

So far, most of the agencies we've discussed have been real entities,
in a physical spacetime. It is also possible, even common, to
construct entirely virtual `knowledge spaces' of a more abstract kind,
whose very structure is a representation of its semantics.  This
section is a bridge between the notion of tenancy and the treatment
knowledge spaces in paper III.

\subsection{Classification, categorization, and disambiguation of tenants}

The branching of names into taxonomies is closely allied to the way
narratives and storylines branch in logical reasoning.
This notion is extensive, and will be explored in paper III.
\begin{example}
The subsections in this section about tenancy form a hierarchy under the namespace
of multi-tenancy. Trying to untwine and separate the concepts in to identify what is common
and what is independent is a tricky challenge, which can quickly descend into a exercise of
listing every related concept one can imagine. Partially ordering these to account for dependencies
is a further challenge. This is one of the issues we struggle with when organizing information
and representing knowledge.
\end{example}

Classification does not only apply to idea, but also ideas we have about physical entities.
Hence the power of a semantic space lies in making the distinction  between pure information
and information about a separate entity moot: in both cases the information has to be encoded
by something physical, whether in the thing itself or in a proxy thing elsewhere.
\begin{example}
Exclusive clubs are one thing, but categorization of tenants leads to
explosions of new agency:
\begin{itemize}
\item Premium customers
\item Schools for separate sex, race etc
\item Cabin classes on airlines (economy, business, first class, etc)
\item Car/truck parking
\item Taxonomies of species or subject categories.
\item Periodic table of elements.
\item You can share a public bathroom, or you can have a private one
\end{itemize}
Multi-tenancy is a bridge between the understanding of resources and
the differential categorization of knowledge into concepts.
\end{example}

\subsection{The economics of scale}

\begin{figure}[ht]
\begin{center}
\includegraphics[width=10cm]{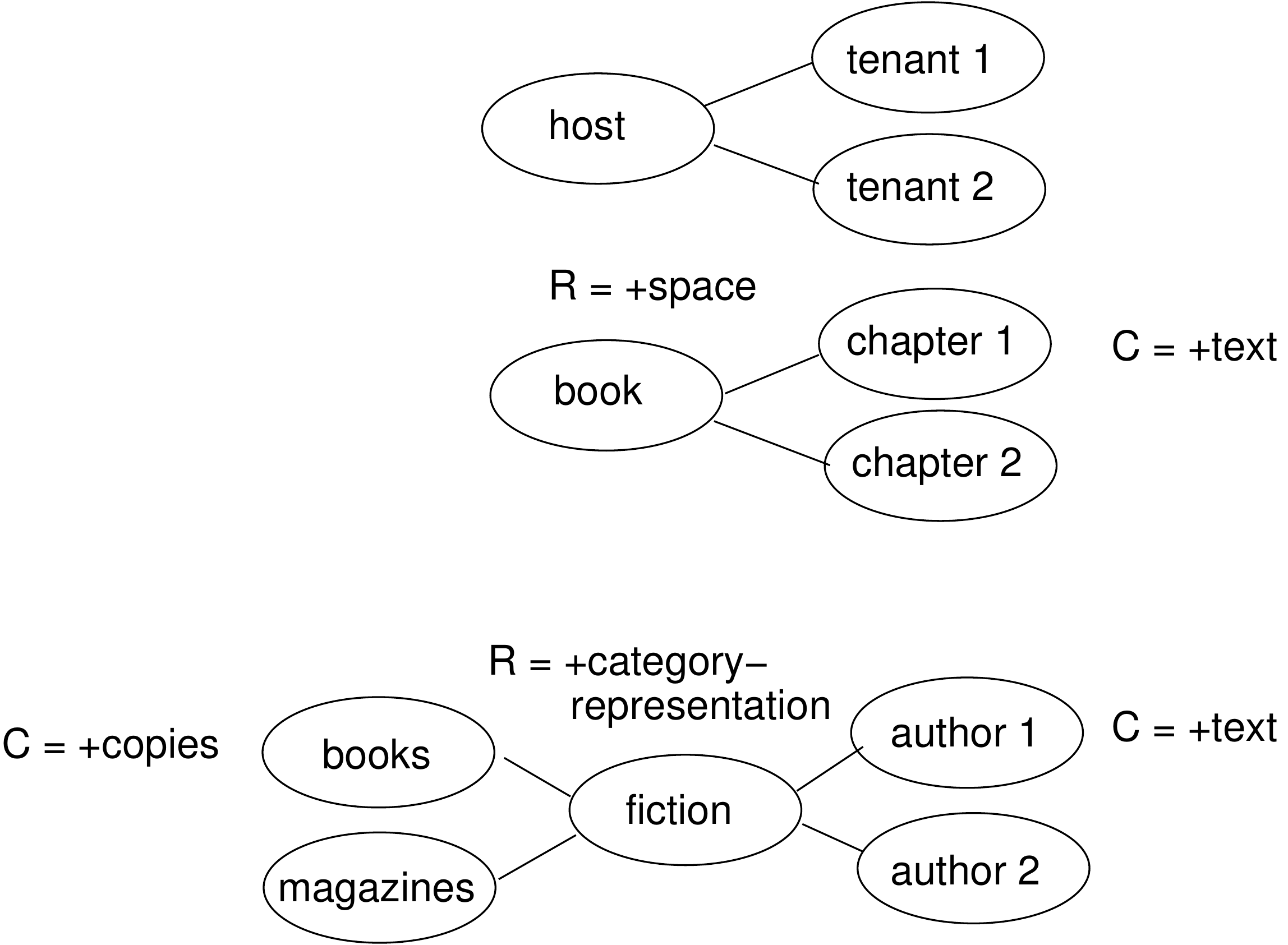}
\caption{\small Orientation of host-tenancy roles. With host-ness to the left and
tenancy increasingly to the right, we see that it is not always possible to
think of tenants as being simple isolated `things', especially when the promises
they represent are more abstract. A tenant shares some space in a host, but what
kind of space? A branching process might result in new tenants (+), or new hosts (-).
\label{ht}}
\end{center}
\end{figure}

Consider the example in figure \ref{ht}, logical reasoning creates a
natural branching process that subdivides agencies into categories,
often from a reductionist standpoint. Conversely, category labels act
as logical hosts that unify similar contributors under a common label.
A book, such as an anthology of articles, is an aggregation of
chapters, where the space $R$ is rented out to chapters for text $C$.
The book provides a vehicle for the chapters to be marketed under a
common brand.  In the same way, authors come together into categories
of fiction, poetry, biography, etc, and the physical exemplars of the
books share the same category, each contributing something to occupy
the hosting of the category as a `market brand'. The brand category
develops relationships with an audience at a different level or scale
than a single book or author can, hence there is a value to the hosting.
This is sometimes called the economics of scale.

\section{Summary and comments}

Continuing on from paper I, we've seen that it is possible to define
semantic spacetimes with cooperative structures that scale in both
semantics and dynamics.  The importance of scale to both semantics and
dynamics is frequently underestimated in system behavioural analysis,
especially when it comes to semantics. The goal of Promise Theory is
to place these two aspects (semantics and dynamics) on the same kind
of footing, for a wide range of spacelike scenarios, to make a physics
of information systems. Even in this deceptively simple topic, and
this cursory sketch of a full treatment, there is a lot to swallow.
Perhaps this is why computer science has yet to answer some of these
basic questions about scaling.

Multi-agent distributed systems delocalize semantics, and potentially
make systems harder to understand. This is the value in coarse
graining and going to a higher level.
As intuited by Milner\cite{milnerbigraph}, the design of a system is
largely about what faces does one wish to expose to whom? But lacking
a proper understanding of spacetime prevents one from implementing 
the interfaces and moving between the abstract and physical viewpoints.

One thing that comes more clearly out of the approach discussed here
is the often-noted misunderstanding about purely reductionist thinking
in cooperative phenomena. One cannot take a system apart and expect to
see all phenomena in the isolated parts.  At each new scale, there are
new promises that cannot be reduced to the promises at lower levels.
Thus we need rules for preserving information during scaling and
coarse-graining.  However, we also find that this information lost to
coarse-graining can be captured as a directory (or index) of the
internal details, so that the complete picture can be represented for
probing downwards. Such services are well known as naming or directory
services in the world of information technology, but here we see that
this is a fundamental requirement for semantic scaling.

Some obvious conclusions drop out of the formalism nicely: central hosted
authority is not scale invariant, for instance.  Separation of agency scales is a
convenience, but one must add to a simple dynamical picture the
answer to the semantic question: why would you ignore or trust the
promises from a finer-grained substrate of agents?  The ability to
rely on the integrity of the substrate implies a long range order, and
data integrity, but also uniform semantics (see \cite{promisebook}
section 11.3).

Lastly, the natural point of departure from this paper II is the issue
of abstract agency, naming and categorization. This is a bridge to
understanding the more abstract issues of relationships and encoded
memory in a functional space. The ultimate goal of this work is to
reach a unified description of semantic spacetimes, where the layout
of space itself is a realization of knowledge relationships and the
comprehension of mind-brain-like structures.  I'll return to these issues
in paper III.

\bibliographystyle{unsrt}
\bibliography{spacetime}

\end{document}